\documentclass[journal]{IEEEtran}
\ifCLASSINFOpdf
\else
\fi

\usepackage{graphicx}
\usepackage{amssymb}
\usepackage{amsmath}
\usepackage{booktabs}
\usepackage{multirow}
\usepackage{enumitem}

\graphicspath{{figures/}}
\usepackage[outdir=./figures_pdf/]{epstopdf}

\usepackage{xcolor}
\definecolor{ECS-Blue}{rgb}{0,0.4392,0.7529}
\definecolor{ECS-Red}{rgb}{0.7529,0,0}
\definecolor{ECS-Green}{rgb}{0.4667,0.5765,0.2353}

\usepackage[para]{threeparttable}
\makeatletter
\def\TPT@doparanotes{\par
   \prevdepth\z@ \TPT@hsize
   \TPTnoteSettings
   \parindent\z@ \pretolerance 8
   \linepenalty 200
   \renewcommand\item[1][]{\relax\ifhmode \begingroup
       \unskip
       \advance\hsize 10em 
       \penalty -45 \hskip\z@\@plus\hsize \penalty-19
       \hskip .15\hsize \penalty 9999 \hskip-.15\hsize
       \hskip .01\hsize\@plus-\hsize\@minus.01\hsize 
       \hskip 0em\@plus .3em
      \endgroup\fi
      \tnote{##1}\,\ignorespaces}%
   \let\TPToverlap\relax
   \def\endtablenotes{\par}%
}

\begin{document}

\renewcommand{\theenumi}{\alph{enumi}}
\setlength{\abovedisplayskip}{4pt plus 1pt minus 1pt}
\setlength{\belowdisplayskip}{4pt plus 0pt minus 1pt}
\setlength{\abovecaptionskip}{0pt plus 1pt minus 1pt}
\setlength{\belowcaptionskip}{0pt plus 0pt minus 1pt}
\setlength{\textfloatsep}{3pt plus 1pt minus 1pt}
\setlength{\floatsep}{3pt plus 0pt minus 1pt}
\setlength{\dbltextfloatsep}{3pt plus 1pt minus 1pt}
\setlength{\dblfloatsep}{3pt plus 0pt minus 1pt}

%
\title{A nA-Range Area-Efficient Sub-100-ppm/$^\circ$C\\
Peaking Current Reference Using Forward Body\\
Biasing in 0.11-$\mu$m Bulk and 22-nm FD-SOI}
%
%
%

\author{Martin~Lefebvre,~\IEEEmembership{Graduate Student Member,~IEEE},
        and David~Bol,~\IEEEmembership{Senior Member,~IEEE}
\vspace{-0.75cm}
\thanks{Manuscript received 9 January 2024; revised 13 April 2024; accepted 24 May 2024. This article was approved by Associate Editor Fabio Sebastiano. This work was supported by the Fonds de la Recherche Scientifique (FRS-FNRS) of Belgium under grant CDR J.0014.20. \textit{(Corresponding author: Martin Lefebvre.)}

The authors are with the Université catholique de Louvain, Institute of Information and Communication Technologies, Electronics and Applied Mathematics, B-1348 Louvain-la-Neuve, Belgium (e-mail: \{martin.lefebvre; david.bol\}@uclouvain.be).

Color versions of one or more figures in this article are available at https://doi.org/10.1109/JSSC.2024.3406423.

Digital Object Identifier 10.1109/JSSC.2024.3406423
}}

%
%

\markboth{IEEE Journal of Solid-State Circuits,~Vol.~xx, No.~xx, xx~2024}%
{Shell \MakeLowercase{\textit{et al.}}: Bare Demo of IEEEtran.cls for IEEE Journals}
%
\IEEEoverridecommandlockouts
\IEEEpubid{\begin{minipage}{\textwidth}\ \\[12pt] \begin{scriptsize}This document is the paper as accepted for publication in JSSC, the fully edited paper is available at https://ieeexplore.ieee.org/document/10555551. \copyright 2024 IEEE. Personal use of this material is permitted. Permission from IEEE must be obtained for all other uses, in any current or future media, including reprinting/republishing this material for advertising or promotional purposes, creating new collective works, for resale or redistribution to servers or lists, or reuse of any copyrighted component of this work in other works.\end{scriptsize}
\end{minipage}}
\maketitle

\begin{abstract} In recent years, the development of the Internet of Things (IoT) has prompted the search for nA-range current references that are simultaneously constrained to a small area and robust to process, voltage and temperature variations. Yet, such references have remained elusive, as existing architectures fail to reach a low temperature coefficient (TC) while minimizing silicon area. In this work, we propose a nA-range constant-with-temperature (CWT) peaking current reference, in which a resistor is biased by the threshold voltage difference between two transistors in weak inversion. This bias voltage is lower than in conventional architectures to cut down the silicon area occupied by the resistor, and is obtained by forward body biasing one of the two transistors with an ultra-low-power voltage reference so as to reduce its threshold voltage. In addition, the proposed reference includes a circuit to suppress the leakage of parasitic diodes at high temperature, and two simple trimming mechanisms for the reference current and its TC. As the proposed design relies on the body effect, it has been validated in both \mbox{0.11-$\mu$m} bulk and \mbox{22-nm} fully-depleted silicon-on-insulator, to demonstrate feasibility across different technology types. In post-layout simulation, the \mbox{0.11-$\mu$m} design generates a \mbox{5-nA} current with a \mbox{65-ppm/$^\circ$C} TC and a \mbox{2.84-$\%$/V} line sensitivity (LS), while in measurement, the \mbox{22-nm} design achieves a \mbox{1.5-nA} current with an \mbox{89-ppm/$^\circ$C} TC and a \mbox{0.51-$\%$/V} LS. As a result of the low resistor bias voltage, the proposed references occupy a silicon area of 0.00954~mm$^2$ in 0.11~$\mu$m (resp. 0.00214~mm$^2$ in 22~nm) at least 1.8$\times$ (resp. 8.2$\times$) smaller than fabricated nA-range CWT references, but with a TC improved by 6.1$\times$ (resp. 4.4$\times$).
\end{abstract}

\begin{IEEEkeywords}
Peaking current reference, temperature coefficient (TC), temperature-independent, constant-with-temperature (CWT), forward body biasing (FBB).
\end{IEEEkeywords}

%
\IEEEpeerreviewmaketitle

\section{Introduction}
\label{sec:1_introduction}
%
%
%
%
\begin{figure}[!t]
	\centering
	\includegraphics[width=.45\textwidth]{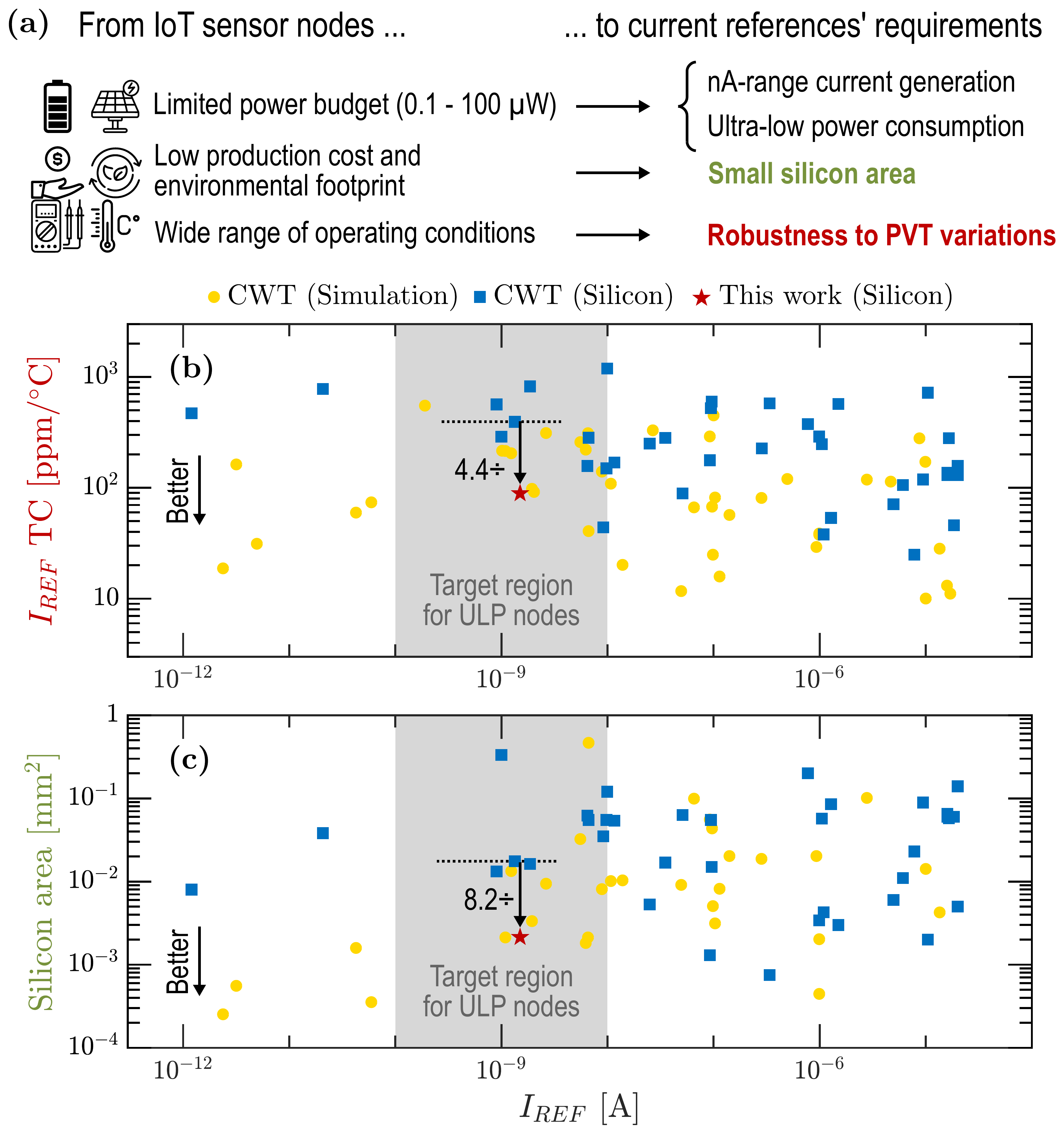}
	\caption{(a) Requirements of current references in the context of the IoT, and landscape of existing CWT current references in terms of (b) temperature coefficient (TC) and (c) silicon area, highlighting the absence of both temperature-independent and area-efficient architectures in the nA range.}
	\label{fig:1_context}
\end{figure}
\IEEEPARstart{T}{he} last decade has seen the development of numerous ultra-low-power (ULP) sensor nodes embedding intelligence at the edge, fostered by the growth of Internet-of-Things (IoT) applications. As is the case for most integrated circuits, the analog blocks constituting these sensor nodes need to be biased by a current, commonly generated by on-chip current references. Nonetheless, designing such sensor nodes poses three main challenges listed in Fig.~\ref{fig:1_context}(a), which result in specific requirements for the current references they embed. First, the power consumption of these sensor nodes is greatly constrained by the fact that they operate from limited-capacity batteries or energy harvesting. The typical average power consumption is thus comprised between 0.1 and 100 $\mu$W \cite{Blaauw_2014}, albeit this value largely varies from one application to another. A reference current in the nA range is thus desirable to bias always-on circuit blocks operating in sleep mode, and to ensure an ultra-low power consumption while providing sufficient performance in active mode. Then, silicon area must be minimized to limit the production cost and the associated direct environmental footprint, typically comprised between 1 and 4~kgCO$_2$-eq/cm$^2$ \cite{Bardon_2020, Pirson_2023}, given the large production volumes projected for the IoT \cite{Pirson_2021}. Therefore, the area occupied by current references should also be limited to leave more space for circuits providing useful functionalities, especially in the case of IoT nodes using small silicon dies. Finally, sensor nodes must be able to cope with a wide range of operating conditions due to the diversity of possible deployment scenarios. \IEEEpubidadjcol
Hence, current references need to be robust to process, voltage and temperature (PVT) variations, to avoid degrading the performance of analog blocks such as real-time clock generators \cite{Wang_2017_ESSCIRC, Liao_2023} and temperature sensors \cite{Jeong_2014, Wang_2017_Nature}.\\
\indent As already observed in \cite{Lefebvre_2023}, the landscape of existing current references [Figs.~\ref{fig:1_context}(b) and (c)] reveals the absence of fabricated references that are simultaneously robust to temperature variations and area-efficient in the nA range, while such references are ideally suited to the needs of IoT sensor nodes \cite{Blaauw_2014}. A common way to generate a current from a voltage is to use a voltage-to-current ($V$-to-$I$) converter. First, resistor-based references \cite{Ji_2017, Wang_2019_TCAS, Huang_2020} rely on a resistor biased by the threshold voltage ($V_T)$ difference between two transistors of different $V_T$ types. Hence, they offer a good TC thanks to the relatively linear temperature characteristics of the resistance, at the cost of a large area necessary to reach the nA range, as it is constituted of several segments in series. Second, references based on gate-leakage transistors \cite{Chang_2022} offer a similar trade-off, except that the large area originates from the necessity to connect numerous gate-leakage devices in parallel. Finally, references based on a self-cascode MOSFET (SCM) \cite{Wang_2019_VLSI, Lefebvre_2023} present a significantly reduced area compared to the two previous categories, but generally feature a poorer TC because of the nonlinear $I$-$V$ characteristics of the SCM, and the difficulty to bias it with a voltage leading to a constant-with-temperature (CWT) current.\\
\indent This work proposes a nA-range CWT peaking current reference (PCR), in which the threshold voltage difference between two subthreshold transistors of the same $V_T$ type, denoted as $\Delta V_T$, biases a resistor. This $\Delta V_T$ results from the forward body biasing (FBB) of one of the transistors by a two-transistor (2T) ULP voltage reference, leading to a reduced bias voltage and thus, to a reduced resistance and area. The proposed design includes a leakage suppression circuit to mitigate the impact of parasitic diodes at high temperature, as well as trimming mechanisms for the reference current and its TC to maintain performance across process corners. As the proposed reference relies on the body effect, it has been validated with post-layout simulations in \mbox{0.11-$\mu$m} bulk and \mbox{22-nm} fully-depleted silicon-on-insulator (FD-SOI) technologies to demonstrate feasibility in these two technology types. In addition, the proposed reference has been fabricated in \mbox{22-nm} \mbox{FD-SOI}, yielding a \mbox{1.5-nA} current with an \mbox{89-ppm/$^\circ$C} TC and a \mbox{0.51-$\%$/V} line sensitivity (LS), while consuming 2.87~nW at 0.75~V and occupying a record silicon area of 0.00214~mm$^2$. This paper is organized as follows. First, Section~\ref{sec:2_operation_principle_and_sizing} presents the operation principle of the proposed reference and discusses its sizing. Section~\ref{sec:3_architectural_optimizations} details architectural optimizations, while Sections~\ref{sec:4_post-layout_simulation_results} and \ref{sec:5_measurement_results} respectively present simulation and measurement results. Finally, Section~\ref{sec:6_comparison_to_the_state_of_the_art} compares this work to the state of the art, while Section~\ref{sec:7_conclusion} offers perspectives for future works.

\section{Operation Principle and Sizing}
\label{sec:2_operation_principle_and_sizing}
\begin{figure}[!t]
	\centering
	\includegraphics[width=.45\textwidth]{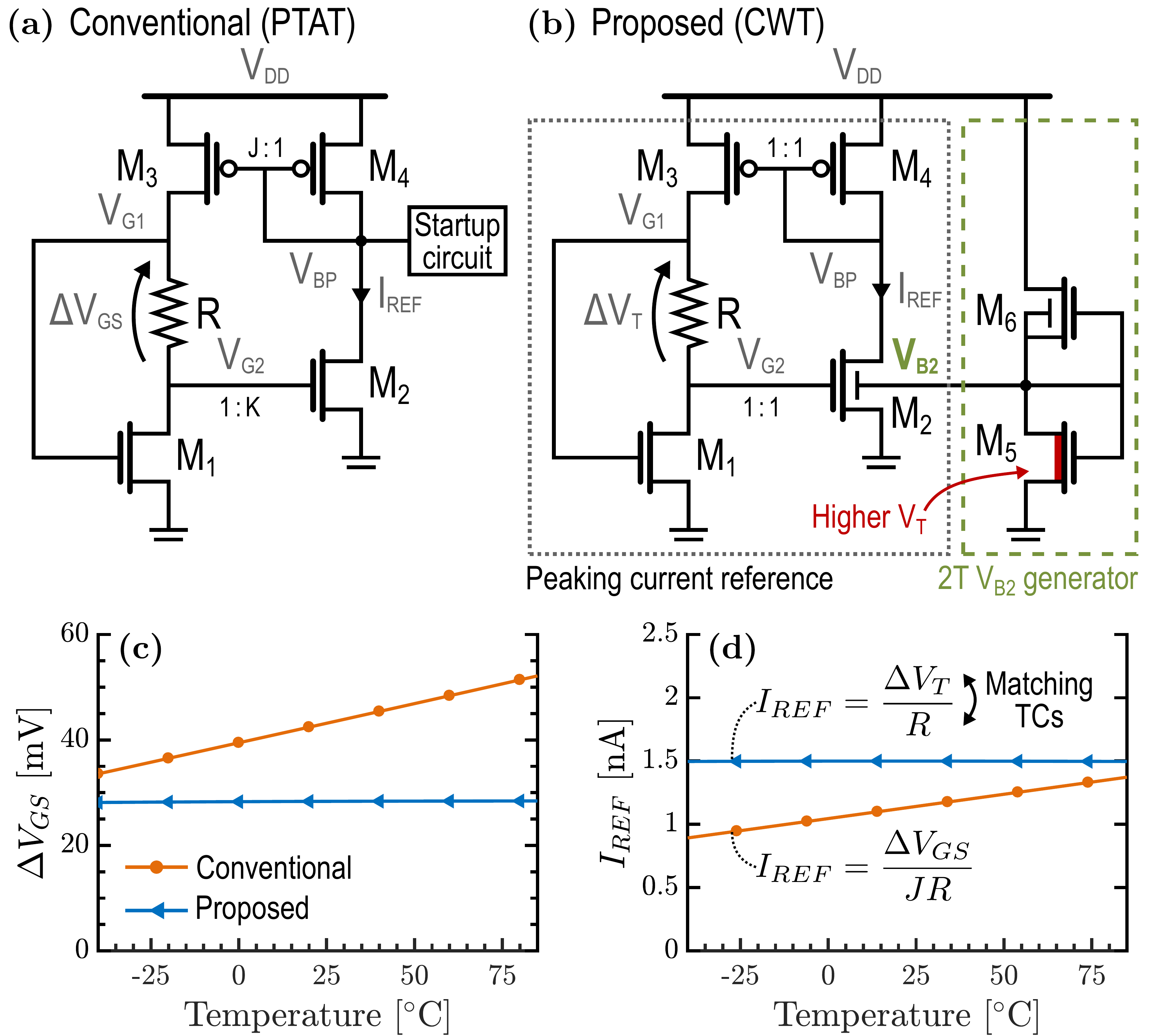}
	\caption{Schematic of (a) the conventional PTAT peaking current reference and (b) the simplified proposed CWT one, which relies on the threshold voltage difference $\Delta V_T$ between $M_{1-2}$, obtained by forward body biasing $M_2$ with a 2T voltage reference ($M_{5-6}$). Operation principle illustrated with pre-layout simulations of (c) $\Delta V_{GS}$ and (d) $I_{REF}$ in \mbox{22-nm} FD-SOI, for the conventional design with $J=K=2$, and the proposed one.}
	\label{fig:2_basic_schematic}
\end{figure}
The main objective of this section is to respectively remind the operation principle of the conventional proportional-to-absolute-temperature (PTAT) PCR, and explain the one of the proposed CWT PCR, based on their governing equations in Sections~\ref{subsec:2A_conventional_PTAT_peaking_current_reference} and \ref{subsec:2B_proposed_CWT_peaking_current_reference}. Then, Section~\ref{subsec:2C_design_and_sizing_methodology} discusses the design and sizing methodology of the proposed CWT PCR in \mbox{0.11-$\mu$m} bulk and \mbox{22-nm} \mbox{FD-SOI} technologies.\\
\indent Before getting into the heart of the governing equations, we need to remind two important concepts. First, the drain-to-source current in subthreshold is given by
\begin{equation}
	I_{DS} = I_{SQ}S\exp\left(\frac{V_{GS}-V_T}{nU_T}\right)\textrm{,}\label{eq:ids_subthreshold}
\end{equation}
\noindent for $V_{DS} > 4U_T$, with $I_{SQ} = \mu C_{ox}^{'}(n-1)U_T^2$ the specific sheet current, $\mu$ the carrier mobility, $C_{ox}^{'}$ the gate oxide capacitance per unit area, $n$ the subthreshold slope factor, $U_T$ the thermal voltage, $S = W/L$ the transistor's aspect ratio, and $V_T$ the threshold voltage at any $V_{BS}$, as opposed to $V_{T0}$, which refers to the threshold voltage at zero $V_{BS}$. Consequently, the gate-to-source voltage is described by
\begin{equation}
	V_{GS} = V_T + nU_T\ln\left(\frac{I_{DS}}{I_{SQ}S}\right)\textrm{.}\label{eq:vgs_subthreshold}
\end{equation}
The second important concept to be introduced is the the body effect, which is captured by
\begin{IEEEeqnarray}{RCL}
	\Delta V_T = V_T - V_{T0} & = & \gamma_b\left(\sqrt{2\phi_{fp}-V_{BS}} - \sqrt{2\phi_{fp}}\right)\label{eq:body_effect_bulk}\\
	& \approx & -\gamma_b^{*} V_{BS}\IEEEnonumber
\end{IEEEeqnarray}
in a bulk technology, where $\phi_{fp}$ denotes Fermi's potential, $\gamma_b$ the body factor, $\gamma_b^{*}$ its linearization around $V_{BS}$ = 0, and by
\begin{equation}
	\Delta V_T = V_T - V_{T0} = -\gamma_b^{*} V_{BS}\label{eq:body_effect_FD-SOI}
\end{equation}
in an \mbox{FD-SOI} technology. $\gamma_b^{*}$ is temperature-dependent in bulk, and CWT at first order in \mbox{FD-SOI} \cite{daSilva_2021}.

\subsection{Conventional PTAT Peaking Current Reference}
\label{subsec:2A_conventional_PTAT_peaking_current_reference}
In the conventional PTAT architecture [Fig.~\ref{fig:2_basic_schematic}(a)], we consider that $M_{1-2}$ have the same technological parameters and do not use any kind of body biasing, i.e., $V_{T01} = V_{T02} = V_T$ and $n_1 = n_2 = n$, excluding mismatch. In addition, the current mirror ratios $J$ and/or $K$ need to be larger than one, to have a positive voltage drop across resistor $R$. This results in
\begin{equation}
	\Delta V_{GS} = V_{GS1} - V_{GS2} = nU_T \ln\left(JK\right)\label{eq:dvgs_conventional}
\end{equation}
using (\ref{eq:vgs_subthreshold}), and
\begin{equation}
	I_{REF} = \frac{\Delta V_{GS}}{JR} = \frac{nU_T \ln\left(JK\right)}{JR}\textrm{.}\label{eq:iref_conventional}
\end{equation}
The TC of $I_{REF}$ at a given temperature can be computed as
\begin{equation}
	\textrm{TC} = \frac{1}{I_{REF}}\frac{dI_{REF}}{dT} = -\frac{1}{R}\frac{dR}{dT} + \frac{1}{T}\textrm{.}\label{eq:iref_TC_conventional}
\end{equation}
Eq. (\ref{eq:iref_TC_conventional}) shows that the TC of $I_{REF}$ cannot be equal to zero unless the temperature coefficient of resistance (TCR) is equal to $1/T$. But in practice, the TCR is always smaller than $1/T$ and the TC of $I_{REF}$ is thus positive. In Figs.~\ref{fig:2_basic_schematic}(c) and (d), we observe that the PTAT behavior of $\Delta V_{GS}$ results in a PTAT behavior of $I_{REF}$, which is consistent with the positive TC.\\
\indent Another quantity of interest is the LS of $I_{REF}$, which can be computed from a small-signal analysis of the PCR. Based on the small-signal schematic in Fig.~\ref{fig:3_small_signal_schematic}(a), we find that
\begin{equation}
	v_{bp} = \frac{(g_{m4}+g_{d4})v_{dd} - g_{m2}v_{g2}}{g_{m4}+g_{d4}+g_{d2}}\approx v_{dd} - \left(\frac{g_{m2}}{g_{m4}}\right)v_{g2}\label{eq:vbp_LS_conventional}
\end{equation}
with the simplification $g_m \gg g_d$, and we can thus compute the LS of $I_{REF}$ as
\begin{equation}
	\frac{i_{ref}}{v_{dd}} = \frac{1}{J}\:\frac{g_{d3}}{g_{m1}}\left(\frac{1}{R}-g_{m1}\right) + g_{d2} \approx g_{d2}\textrm{,}\label{eq:iref_LS_conventional}
\end{equation}
to be evaluated around the operation point of the current reference. Interestingly, the LS is predominantly related to the output conductance of $M_2$.

\subsection{Proposed CWT Peaking Current Reference}
\label{subsec:2B_proposed_CWT_peaking_current_reference}
\begin{figure}[!t]
	\centering
	\includegraphics[width=.45\textwidth]{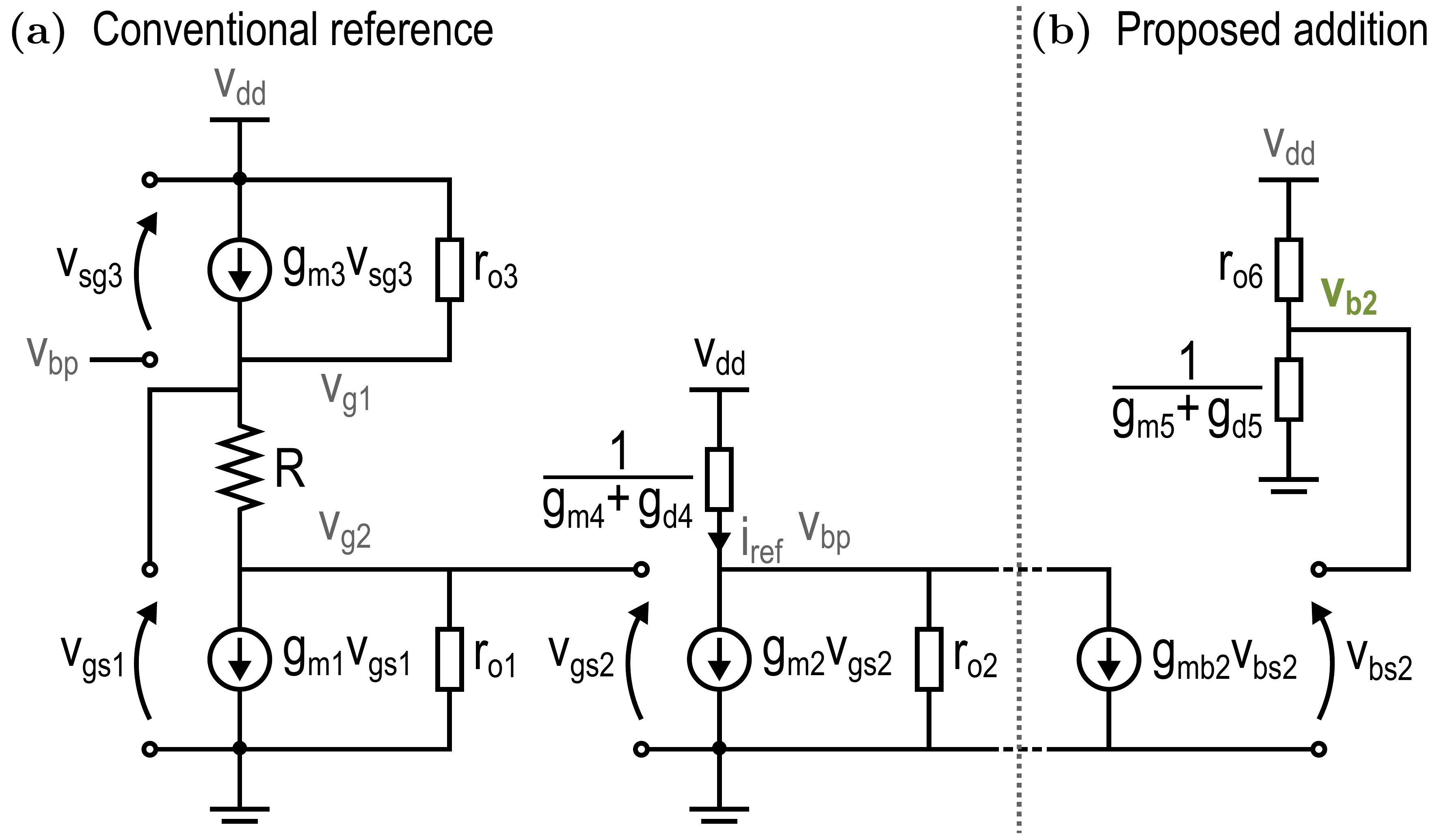}
	\caption{Small-signal schematic of (a) the conventional PTAT PCR and (b) the proposed FBB addition to obtain a CWT reference current.}
	\label{fig:3_small_signal_schematic}
\end{figure}
In the proposed CWT PCR [Fig.~\ref{fig:2_basic_schematic}(b)], $M_1$ has its body connected to its source and is therefore not impacted by the body effect, hence $V_{T1} = V_{T01}$, while $M_2$ is biased with $V_{BS2} > 0$ and undergoes a forward body biasing lowering its threshold voltage, i.e., $V_{T2} < V_{T1}$. In addition, unitary current mirror ratios $J = K = 1$ are used. In this architecture, the body voltage of $M_2$ is generated by the 2T voltage reference formed by $M_{5-6}$, in which the equilibrium of subthreshold currents $I_{DS5} = I_{DS6}$, given by (\ref{eq:ids_subthreshold}), leads to
\begin{equation}
	V_{B2} = \left(V_{T05} - \frac{n_5}{n_6}V_{T06}\right) + n_5U_T\ln\left( \frac{I_{SQ6}S_6}{I_{SQ5}S_5} \right)\textrm{.}\label{eq:vb2_proposed}
\end{equation}
The voltage drop across resistor $R$ is therefore equal to the difference of threshold voltages between $M_1$ and $M_2$ expressed by (\ref{eq:body_effect_bulk}) or (\ref{eq:body_effect_FD-SOI}), i.e.,
\begin{equation}
	\Delta V_{GS} = \Delta V_T = V_{T01} - V_{T2} \approx \gamma_b^{*} V_{B2}\textrm{,}\label{eq:dvgs_proposed}
\end{equation}
and results in a reference current
\begin{equation}
	I_{REF} = \frac{\Delta V_{T}}{R} = \frac{\gamma_b^{*} V_{B2}}{R}\textrm{.}\label{eq:iref_proposed}
\end{equation}
The TC of $I_{REF}$ at a given temperature can be computed as
\begin{equation}
	\textrm{TC} = \frac{1}{I_{REF}}\frac{dI_{REF}}{dT} = -\frac{1}{R}\frac{dR}{dT} + \frac{1}{\gamma_b^{*}}\frac{d\gamma_b^{*}}{dT} + \frac{1}{V_{B2}}\frac{dV_{B2}}{dT}\textrm{.}\label{eq:iref_TC_proposed}
\end{equation}
Contrary to (\ref{eq:iref_TC_conventional}), in which there is no tuning knob to set the TC of $I_{REF}$, the proposed architecture allows to zero out the TC by properly selecting the TC of $V_{B2}$, which will compensate for the temperature dependence of $R$ and $\gamma_b^{*}$. $V_{B2}$ TC can indeed be tuned by changing the ratio of aspect ratios $S_6/S_5$ in (\ref{eq:vb2_proposed}). This will be the basis for the TC trimming circuit later discussed in Section~\ref{subsec:3B_temperature_coefficient_and_reference_current_trimming}, which allows to mitigate the impact of $V_{B2}$ process variations on $I_{REF}$ TC. In Figs.~\ref{fig:2_basic_schematic}(c) and (d), we notice that a close-to-CWT $\Delta V_{GS}$ gives a CWT $I_{REF}$, by matching the TC of $\Delta V_{GS}$ with the TCR. Hence, the operation principle of the proposed reference is to bias the resistor with the $\Delta V_T$ between two subthreshold transistors of the same $V_T$ type, one of them being forward body biased to create a $V_T$ imbalance. Interestingly, no startup circuit is required as the proposed reference is unstable at zero $I_{REF}$. The 2T voltage reference indeed generates a positive voltage $V_{B2}$ even if the rest of the reference is stuck at zero. This leads to a non-zero $I_{DS2}$ which starts the reference. Other CWT PCRs employing slightly different ideas exist in the literature, such as \cite{Hu_2021, Ballo_2022}, biasing the resistor with the $\Delta V_T$ between transistors of the same $V_T$ type but different lengths, and \cite{Eslampanah_2018}, making use of an additional degeneration resistor to achieve temperature compensation. Yet, these works are either limited to simulations and/or generate a larger $I_{REF}$.\\
\indent Similarly to the conventional PCR, the LS can be obtained from a small-signal analysis. Based on the small-signal schematic in Figs.~\ref{fig:3_small_signal_schematic}(a) and (b), we first find
\begin{equation}
	v_{b2} = \frac{g_{d6}}{g_{m5}+g_{d5}+g_{d6}}v_{dd} \approx \left(\frac{g_{d6}}{g_{m5}}\right)v_{dd}\textrm{,}
\end{equation}
and consequently,
\begin{IEEEeqnarray}{RCL}
	\frac{i_{ref}}{v_{dd}} & = & \frac{g_{d3}}{g_{m1}}\left(\frac{1}{R}-g_{m1}\right) + g_{d2} + g_{mb2}\left(\frac{g_{d6}}{g_{m5}}\right)\textrm{,}\label{eq:iref_LS_proposed}\\
	& \approx & g_{d2} + g_{mb2}\left(\frac{g_{d6}}{g_{m5}}\right)\textrm{.}\label{eq:iref_LS_proposed_simplified}
\end{IEEEeqnarray}
The 2T bias generator thus affects the LS of $I_{REF}$ through the second term of (\ref{eq:iref_LS_proposed_simplified}).
\begin{table}[!t]
\centering
\caption{Resistor properties in \mbox{0.11-$\mu$m} bulk and \mbox{22-nm} \mbox{FD-SOI} technologies. The density is measured at 25$^\circ$C and the TCR is characterized from -40 to 85$^\circ$C, and both characteristics are normalized to the lowest obtained value. All the resistors have a width and length of 0.5~$\mu$m and 1~$\mu$m in 0.11~$\mu$m, and 0.36~$\mu$m and 0.4~$\mu$m in 22~nm.}
\label{table:resistor_properties}
\resizebox{.475\textwidth}{!}{%
\begin{threeparttable}
\begin{scriptsize}
\begin{tabular}{llcccccc}
	\toprule
	\multirow{3}{*}{Techno.} & \multirow{3}{*}{Properties} & \multicolumn{3}{c}{N+} & \multicolumn{3}{c}{P+}\\
	\cmidrule(lr){3-5} \cmidrule(lr){6-8}
	& & diff. & poly. & poly. & diff. & poly. & poly.\\
	& & & & (high res.) & & & (high res.)\\
	\midrule
	\multirow{2}{*}{0.11~$\mu$m} & Normalized density & \textcolor{ECS-Red}{\textbf{1}} & 1.28 & - & 1.06 & 2.88 & \textcolor{ECS-Blue}{\textbf{4.12}}\\
	& Normalized TCR & \textcolor{ECS-Red}{\textbf{14.98}}\tnote{$\ast$} & 1.69\tnote{$\ast$} & - & 7.87\tnote{$\ast$} & \textcolor{ECS-Blue}{\textbf{1}}\tnote{$\ast$} & 5.34\tnote{$\dagger$}\\
    \midrule
	\multirow{2}{*}{22~nm} & Normalized density & 1.13 & 2.51 & 3.98/\textcolor{ECS-Blue}{\textbf{4.07}} & \textcolor{ECS-Red}{\textbf{1}} & - & -\\
	& Normalized TCR & 5.55\tnote{$\ast$} & \textcolor{ECS-Blue}{\textbf{1}}\tnote{$\ast$} & 1.36\tnote{$\dagger$}$\;$/1.12\tnote{$\dagger$} & \textcolor{ECS-Red}{\textbf{5.72}}\tnote{$\ast$} & - & -\\
	\bottomrule
\end{tabular}%
\end{scriptsize}
\begin{footnotesize}
\begin{tablenotes}
	\item[$\ast$] Positive TCR.
	\item[$\dagger$] Negative TCR.
\end{tablenotes}
\end{footnotesize}
\end{threeparttable}
}
\end{table}

\subsection{Design and Sizing Methodology}
\label{subsec:2C_design_and_sizing_methodology}
In what follows, the LS and TC are computed using the conventional box method, as is done in \cite{Lefebvre_2023}. The design and sizing methodology of the proposed CWT PCR can be brought down to three main steps:
\begin{enumerate}[label={\arabic*)}]
	\item The selection of the resistor, based on its density in $\Omega/\square$ and its TCR in ppm/$^\circ$C.
	\item The sizing of the 2T body bias generator, which consists in selecting the transistor type and dimensions for $M_{5-6}$.
	\item The sizing of all remaining transistors $M_{1-4}$, focusing on maintaining these transistors saturated in all PVT corners, and keeping mismatch-induced variability under an acceptable threshold.
\end{enumerate}
\begin{figure}[!t]
	\centering
	\includegraphics[width=.45\textwidth]{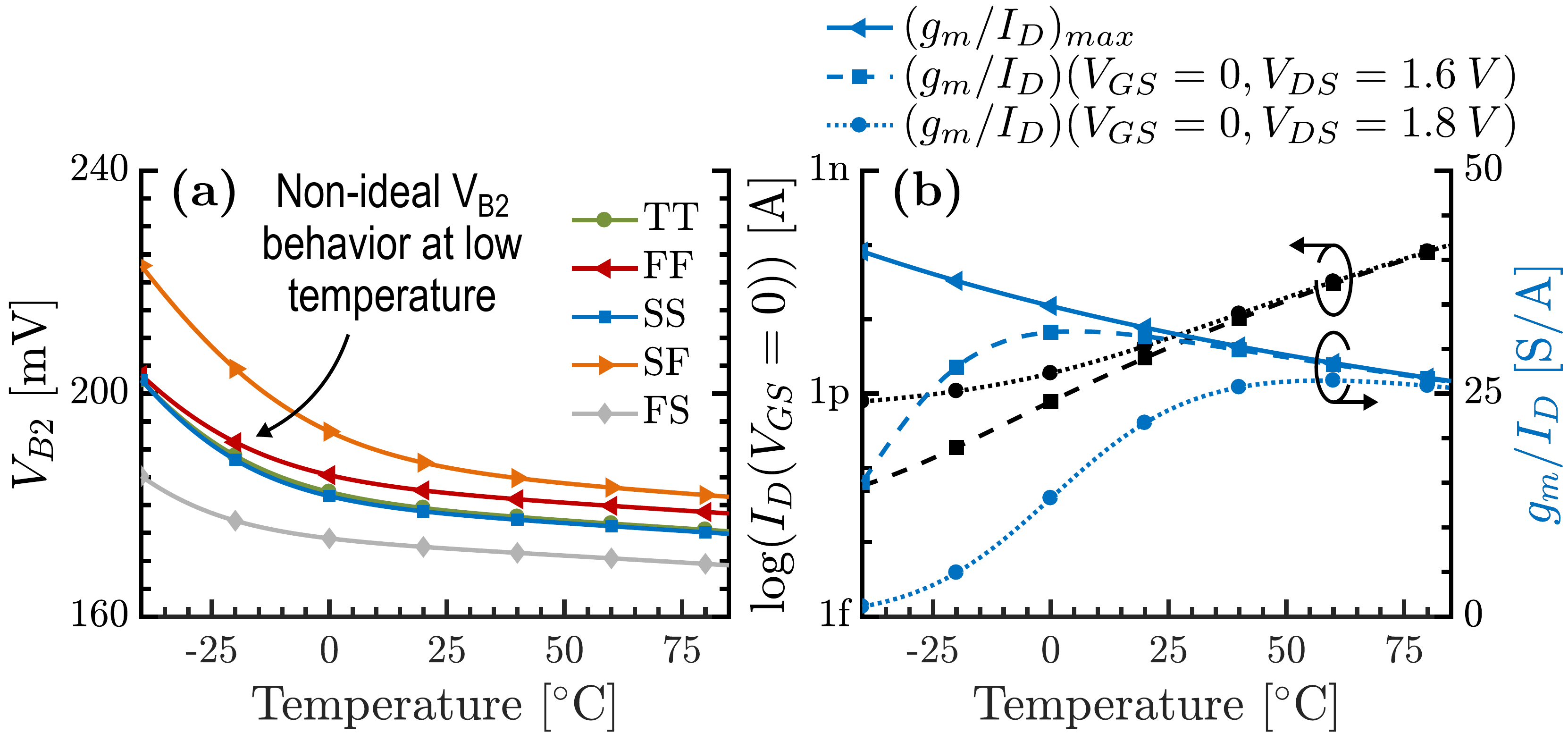}
	\caption{\textbf{The 2T voltage reference presents a nonideal behavior at low temperature, due to gate leakage, or to GIDL at high $\boldsymbol{V_{DS}}$.} All figures correspond to \mbox{22-nm} \mbox{FD-SOI}. Temperature dependence of (a) $V_{B2}$ in all process corners for a 2T voltage reference supplied at 1.8~V and implemented with \mbox{1-$\mu$m-long} I/O nMOS, and (b) $\log(I_D)$ and $(g_m/I_D)$ at $V_{GS}$ = 0, and $\left(g_m/I_D\right)_{\textrm{max}}$, for two different $V_{DS}$ and for an I/O SLVT nMOS with $W$ = 3.89~$\mu$m and $L$ = 1~$\mu$m in the SS process corner.}
	\label{fig:4_problem_2T_vref}
\end{figure}
\begin{figure}[!t]
	\centering
	\includegraphics[width=.48\textwidth]{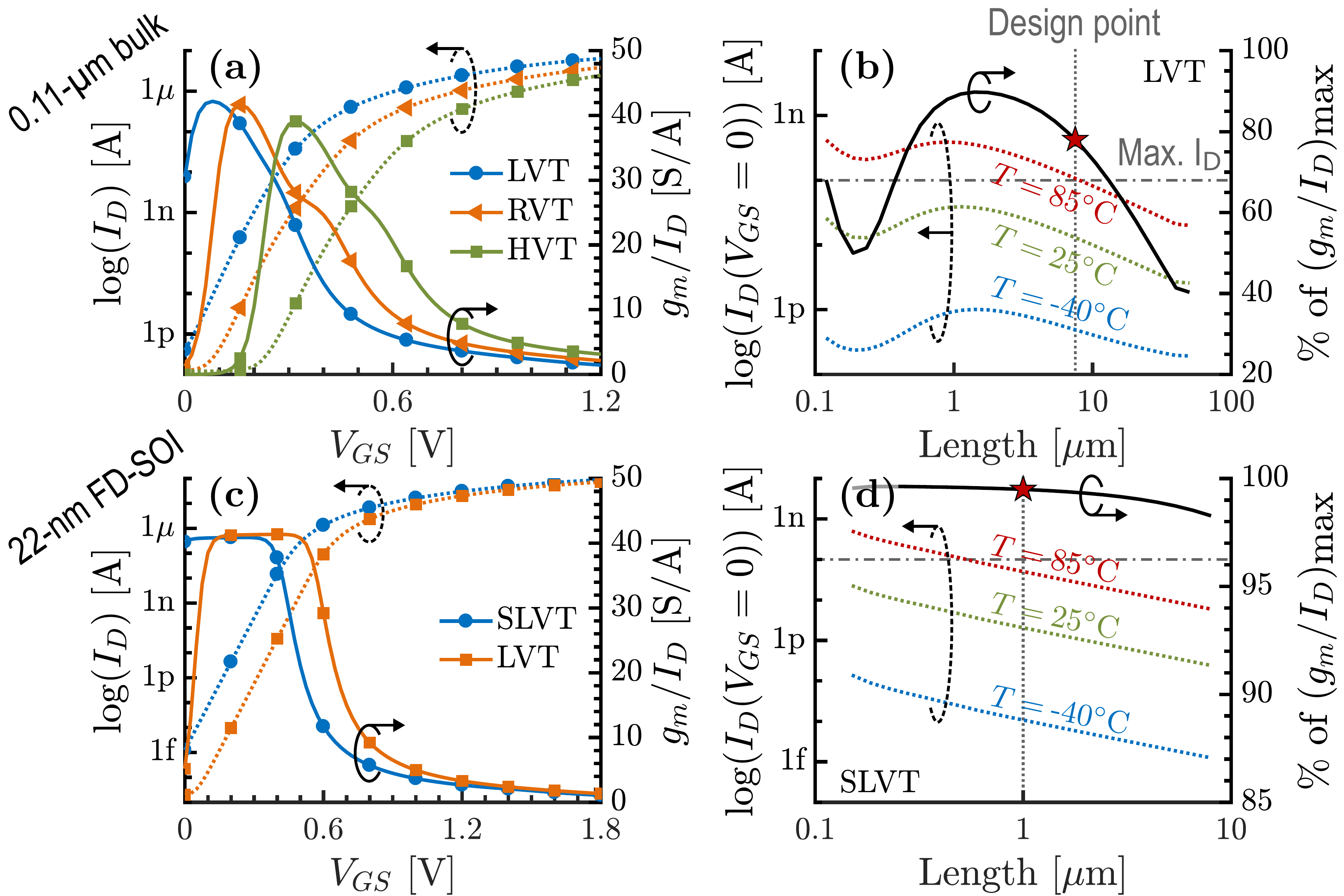}
	\caption{All figures correspond to the SS process corner with a fixed $V_{DS} = V_{GS,\textrm{max}}/2$. $\log(I_D)$, $g_m/I_D$ vs. $V_{GS}$ curves at -40$^\circ$C (a) in 0.11~$\mu$m for core LVT, RVT and HVT nMOS with $W$ = 0.5~$\mu$m and $L$ = 10.45~$\mu$m, and (c) in 22~nm for I/O SLVT and LVT nMOS with $W$ = 2~$\mu$m and $L$ = 8~$\mu$m. $\log(I_D)$ at $V_{GS}$ = 0 and -40, 25 and 85$^\circ$C, and $g_m/I_D$ at $V_{GS}$ = 0 and -40$^\circ$C as a percentage of $\left(g_m/I_D\right)_{\textrm{max}}$ (b) in 0.11~$\mu$m for a core LVT nMOS with $W$ = 0.5~$\mu$m and $L$ ranging from 0.12 to 50~$\mu$m, and (d) in 22~nm for an I/O SLVT nMOS with $W$ = 2~$\mu$m and $L$ ranging from 0.15 to 8~$\mu$m.}
	\label{fig:5_sizing_type_length}
\end{figure}

\begin{figure}[!t]
	\centering
	\includegraphics[width=.45\textwidth]{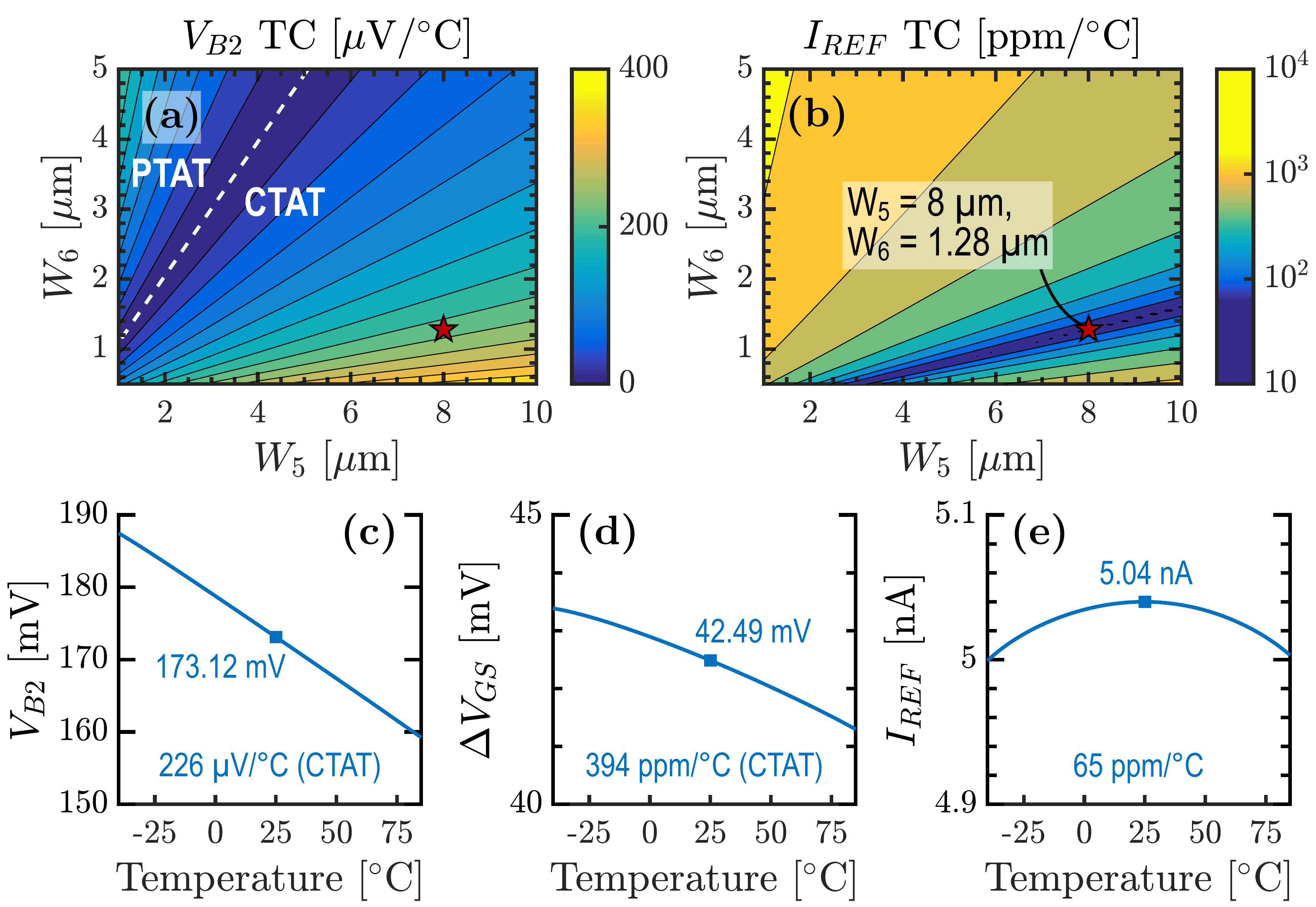}
	\caption{\textbf{In bulk technologies such as \mbox{0.11-$\mu$m}, a CTAT $\boldsymbol{V_{B2}}$ is required to achieve a CWT $\boldsymbol{I_{REF}}$ due to the temperature dependence of the linearized body factor $\boldsymbol{\gamma_b^{*}}$.} Temperature coefficient of (a) $V_{B2}$ and (b) $I_{REF}$, as a function of transistor widths $W_5$ and $W_6$. Temperature dependence of (c) $V_{B2}$, (d) $\Delta V_{GS}$ and (e) $I_{REF}$ at the chosen design point.}
	\label{fig:6_sizing_0p11um}
\end{figure}

\begin{figure}[!t]
	\centering
	\includegraphics[width=.45\textwidth]{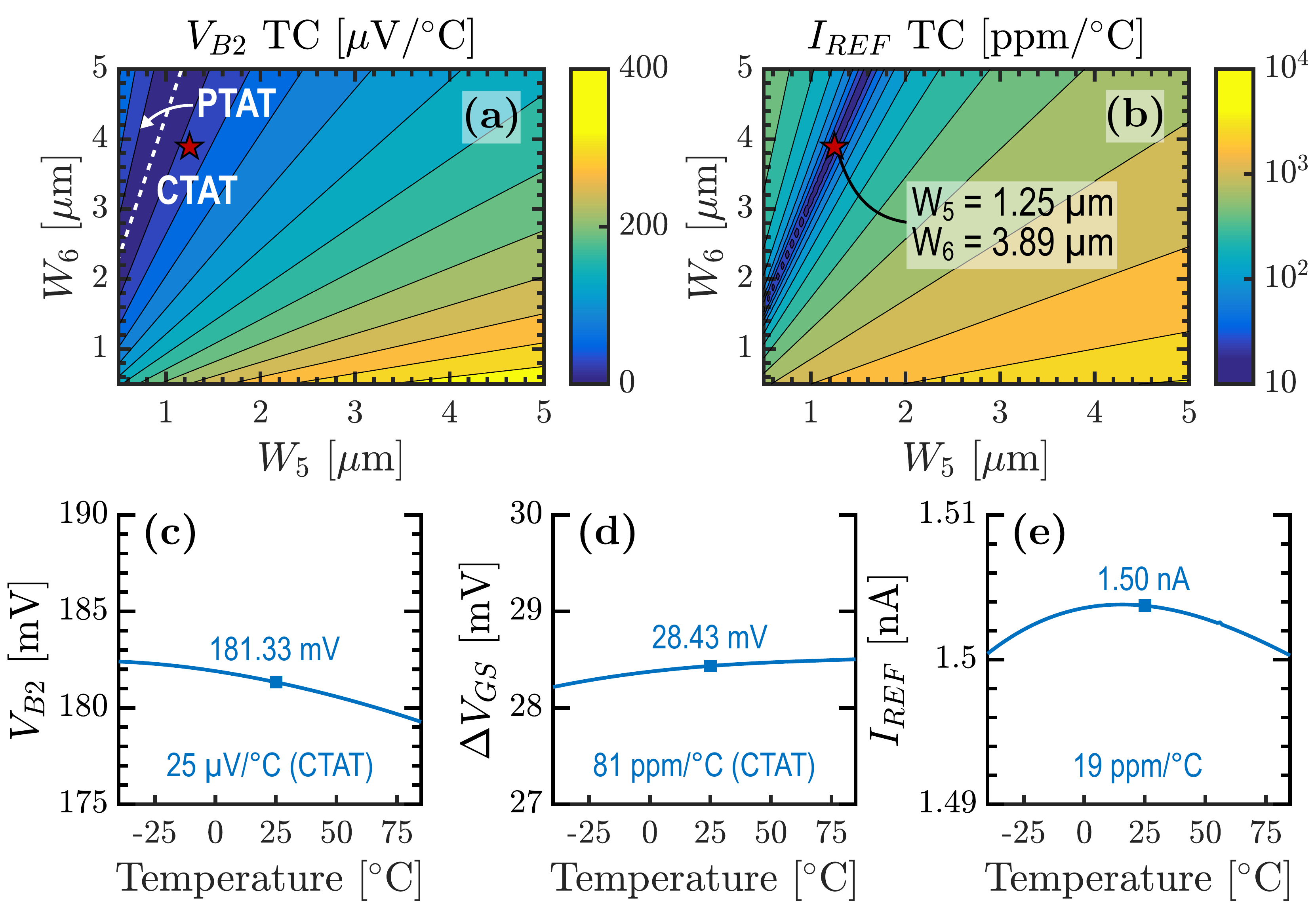}
	\caption{\textbf{In \mbox{FD-SOI} technologies such as \mbox{22-nm}, a close-to-CWT $\boldsymbol{V_{B2}}$ is required to achieve a CWT $\boldsymbol{I_{REF}}$ due to the first-order temperature independence of the linearized body factor $\boldsymbol{\gamma_b^{*}}$.} Temperature coefficient of (a) $V_{B2}$ and (b) $I_{REF}$, as a function of transistor widths $W_5$ and $W_6$. Temperature dependence of (c) $V_{B2}$, (d) $\Delta V_{GS}$ and (e) $I_{REF}$ at the chosen design point.}
	\label{fig:7_sizing_22nm}
\end{figure}

\indent First, step~1) relies on Table~\ref{table:resistor_properties} to select an appropriate resistor type. We notice that diffusion resistors are typically a poor choice for nA-range CWT current references, as they present both a low density and a large TCR. High-resistivity (high-res.) polysilicon (poly) resistors are a better fit as they usually have a large density and a low TCR. In \mbox{22-nm} \mbox{FD-SOI}, the high-res. poly resistor can even have its width narrowed down to 150 and 40~nm, thereby effectively increasing its density by 2.4 and 9$\times$ compared to the values shown in Table~\ref{table:resistor_properties}. In 0.11~$\mu$m, a P+ high-res. poly resistor is selected, while in 22~nm, an N+ high-res. narrow poly resistor is chosen. Because this work focuses on area reduction, we select the resistor with the highest density, but a better TC could potentially be attained by selecting the resistor with the lowest TCR. Section~\ref{sec:4_post-layout_simulation_results} however demonstrates that a sub-100-ppm/$^\circ$C TC can be reached with the proposed choice.\\
\indent Next, step~2) is presented in Figs.~\ref{fig:4_problem_2T_vref} to \ref{fig:7_sizing_22nm}. The main problem that can appear in an inappropriately-sized 2T body bias generator is a divergence of the generated $V_{B2}$ from its ideal behavior at low temperature, as depicted in Fig.~\ref{fig:4_problem_2T_vref}(a).
\begin{figure}[!t]
	\centering
	\includegraphics[width=.45\textwidth]{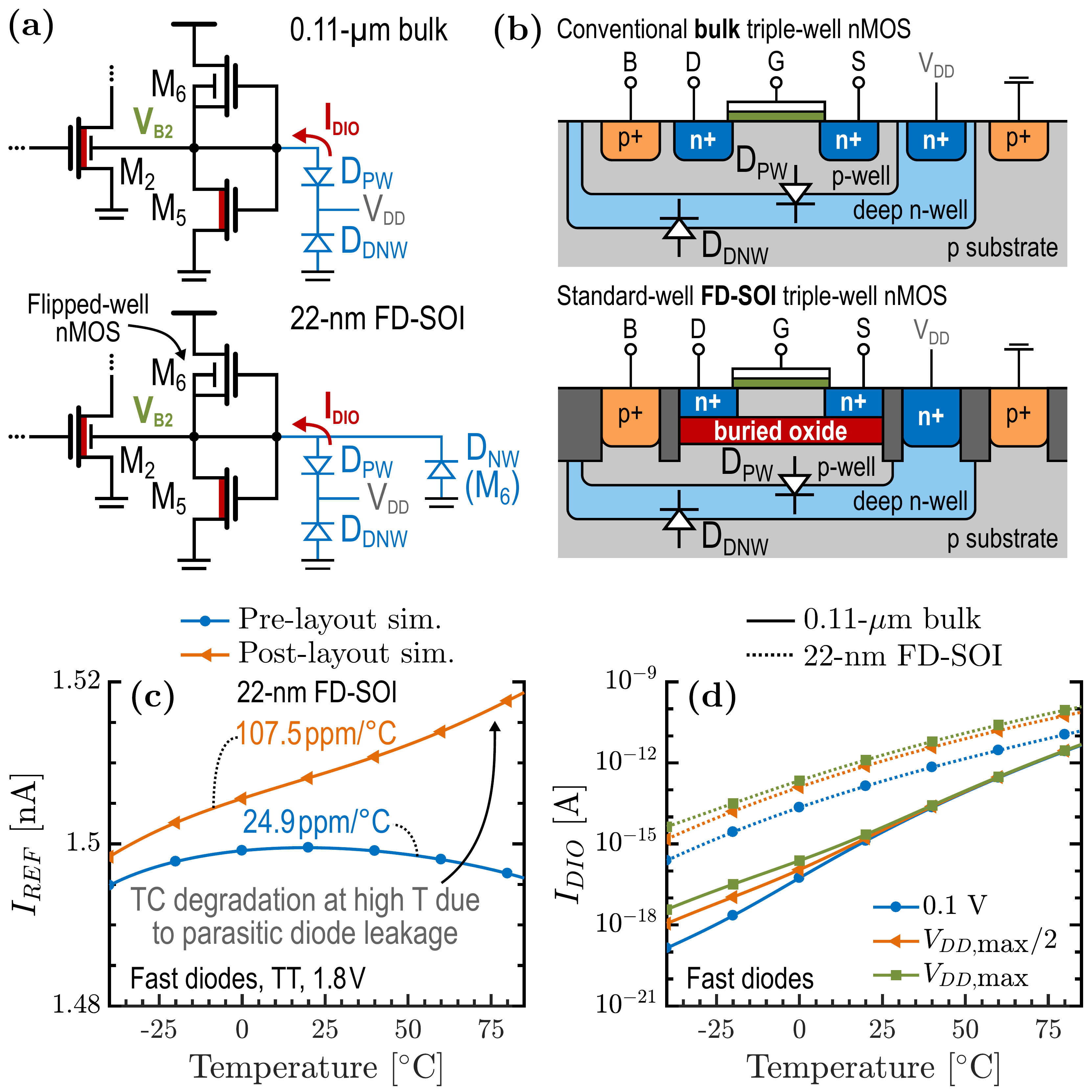}
	\caption{\textbf{$\boldsymbol{I_{REF}}$ TC is degraded by the leakage of the parasitic p-well/deep n-well diode $\boldsymbol{D_{PW}}$.} All figures are for \mbox{0.11-$\mu$m} and \mbox{22-nm} technologies except for (c). (a) Schematic of the 2T body bias generator with parasitic diodes. (b) Cross-section of triple-well nMOS devices. (c) Temperature dependence of $I_{REF}$ in TT 1.8~V and fast diodes, based on pre- and post-layout simulations in \mbox{22-nm}. (d) p-well/deep n-well diode leakage in a fast corner, as a function of temperature, for different reverse voltages, and for a diode area of 968~$\mu$m$^2$ in 0.11~$\mu$m (resp. 370~$\mu$m$^2$ in 22~nm).}
	\label{fig:8_parasitic_diode_leakage}
\end{figure}
This discrepancy either stems from gate leakage, or from gate-induced drain leakage (GIDL), whose relative impact is exacerbated at low temperature due to an increased $V_T$. In the considered technologies, gate leakage is unlikely as \mbox{0.11-$\mu$m} core devices still have a relatively thick oxide, and the devices used in 22~nm are I/O ones. Fig.~\ref{fig:4_problem_2T_vref}(b) reveals that, at low temperature, $I_{DS}$ at $V_{GS}$ = 0 flattens instead of decreasing exponentially, while $(g_m/I_D)$ at $V_{GS}$ = 0, i.e., the slope of the $\log(I_D)$ vs. $V_{GS}$ curve at $V_{GS}$ = 0, plummets. This sharp drop is even more pronounced for a high $V_{DS}$, which is consistent with the fact that GIDL is exacerbated at high $V_{DS}$. This issue can nevertheless be alleviated by properly selecting the transistor type and length of the transistors used in the 2T body bias generator, as shown in Fig.~\ref{fig:5_sizing_type_length}. In 0.11~$\mu$m, a low-$V_T$ (LVT) transistor is preferable to a regular- or high-$V_T$ (RVT or HVT) one, as these latter types present a strong degradation of $(g_m/I_D)$ at $V_{GS}$ = 0 [Fig.~\ref{fig:5_sizing_type_length}(a)]. Regarding the transistor length, $L$ = 7.5~$\mu$m is selected, thereby yielding a $(g_m/I_D)$ at $V_{GS}$ = 0 equal to 78.1~$\%$ of $(g_m/I_D)_{\textrm{max}}$ in the SS -40$^\circ$C corner, and a \mbox{107.9-pA} leakage per transistor in the SS 85$^\circ$C corner which is close to the selected \mbox{100-pA} target [Fig.~\ref{fig:5_sizing_type_length}(b)]. In 22~nm, super-low-$V_T$ (SLVT) I/O transistors are selected as they feature a nearly-perfect behavior at $V_{GS}$ = 0 [Fig.~\ref{fig:5_sizing_type_length}(c)]. A length of 1~$\mu$m is chosen, leading to a $(g_m/I_D)$ at $V_{GS}$ = 0 equal to 99.5~$\%$ of $(g_m/I_D)_{\textrm{max}}$ in SS -40$^\circ$C, and a \mbox{50.7-pA} leakage per transistor in SS 85$^\circ$C. Note that, as the supply voltage can go up to 1.8~V thanks to the use of I/O devices, an additional zero-$V_{GS}$ transistor $M_7$ identical of $M_6$ is added on top of it in Fig.~\ref{fig:12_final_implementation_schematic}(b) to reduce the $V_{DS}$ to values similar to Fig.~\ref{fig:5_sizing_type_length}, thereby limiting the impact of GIDL while simultaneously improving LS. Now that the transistor types and lengths have been selected, a two-dimensional sweep of $W_{5-6}$ allows to determine the ratio $S_6/S_5$ leading to a CWT $I_{REF}$, as represented in Figs.~\ref{fig:6_sizing_0p11um} and \ref{fig:7_sizing_22nm} for the \mbox{0.11-$\mu$m} and \mbox{22-nm} designs, respectively.
In 0.11~$\mu$m, a complementary-to-absolute-temperature (CTAT) $V_{B2}$ is required to compensate the temperature dependence of $\gamma_b^{*}$ in bulk [Fig.~\ref{fig:6_sizing_0p11um}(a)]. The chosen design point corresponds to $W_5$ = 8~$\mu$m and $W_6$ = 1.28~$\mu$m [Fig.~\ref{fig:6_sizing_0p11um}(b)], and leads to a \mbox{226-$\mu$V/$^\circ$C} CTAT slope for $V_{B2}$ [Fig.~\ref{fig:6_sizing_0p11um}(c)] and a 65-ppm/$^\circ$C TC for $I_{REF}$ [Fig.~\ref{fig:6_sizing_0p11um}(e)]. In 22~nm, a close-to-CWT $V_{B2}$ is required because of the first-order temperature independence of $\gamma_b^{*}$ in FD-SOI \cite{daSilva_2021} [Fig.~\ref{fig:7_sizing_22nm}(a)]. The chosen design point corresponds to $W_5$ = 1.25~$\mu$m and $W_6$ = 3.89~$\mu$m [Fig.~\ref{fig:7_sizing_22nm}(b)], and leads to a \mbox{25-$\mu$V/$^\circ$C} CTAT slope for  $V_{B2}$ [Fig.~\ref{fig:7_sizing_22nm}(c)] compensating the CTAT TCR of the chosen high-res. N+ poly. resistor, and a 19-ppm/$^\circ$C TC for $I_{REF}$ [Fig.~\ref{fig:7_sizing_22nm}(e)].\\
\indent Finally, step 3) must ensure that $M_{1-4}$ remain saturated across PVT corners, and that variability constraints are met. On the one hand, $M_{1-2}$ are operated in weak inversion and their aspect ratio $S_{1-2}$ must be chosen so that $V_{GS2} = V_{DS1} > 4U_T$ in the FF 85$^\circ$C corner. $M_{3-4}$ should be biased in moderate inversion to minimize the current mirror mismatch within a reasonable silicon area, and their aspect ratio $S_{3-4}$ must be such that $V_{GS4} = V_{DS4} > 4U_T$. In practice, this limit on $V_{GS4}$ should be larger than $4U_T$ because of the operation in moderate inversion. On the other hand, variability considerations can also come into play, but this requires to establish an analytical expression linking the variability of $I_{REF}$ to the transistor dimensions. Eq.~(\ref{eq:var_iref}), developed in Appendix~\ref{sec:8_appendix}, informs us that the variability of $I_{REF}$ stems from the $V_T$ mismatch between the transistor pairs $M_{1-2}$ and $M_{5-6}$, as well as the current imbalance between $M_{3-4}$, the latter being the dominant source of mismatch as later discussed in Section~\ref{sec:4_post-layout_simulation_results}. Special care must therefore be taken in sizing the pMOS current mirror.

\section{Architectural Optimizations}
\label{sec:3_architectural_optimizations}
This section details the features added to the simplified design in Fig.~\ref{fig:2_basic_schematic}(b), namely a circuit to mitigate the impact of parasitic diode leakage in Section~\ref{subsec:3A_parasitic_diode_leakage_mitigation}, and another to trim $I_{REF}$ and its TC in Section~\ref{subsec:3B_temperature_coefficient_and_reference_current_trimming}. Then, all the details regarding the final implementation of the proposed PCR in \mbox{0.11-$\mu$m} bulk and \mbox{22-nm} \mbox{FD-SOI} are provided in Section~\ref{subsec:3C_final_implementation}.

\subsection{Parasitic Diode Leakage Mitigation}
\label{subsec:3A_parasitic_diode_leakage_mitigation}
The remaining issue with the 2T voltage reference is related to parasitic diode leakage, which becomes significant at high temperature and leads to a distortion of $V_{B2}$, consequently impacting the TC of $I_{REF}$. It stems from the fact that the 2T voltage reference biases the body of the nMOS transistor $M_2$ with a voltage $V_{B2} > 0$ [Fig.~\ref{fig:8_parasitic_diode_leakage}(a)]. Therefore, $M_2$ needs to be implemented as a triple-well device, whose cross-section is represented for bulk and \mbox{FD-SOI} technologies in Fig.~\ref{fig:8_parasitic_diode_leakage}(b). Its p-well is biased at $V_{B2} > 0$, while its deep n-well and p-substrate are respectively connected to $V_{DD}$ and ground. The connections of the parasitic diodes resulting from this triple-well implantation are shown in Fig.~\ref{fig:8_parasitic_diode_leakage}(a), and these parasitic diodes are located in the cross-section in Fig.~\ref{fig:8_parasitic_diode_leakage}(b). The leakage of the parasitic p-well/deep n-well diode leads to a current $I_{DIO}$ flowing from $V_{DD}$ to $V_{B2}$. Fig.~\ref{fig:8_parasitic_diode_leakage}(c) displays the degradation of $I_{REF}$ TC from 107.5 to 24.9~ppm/$^\circ$C between pre- and post-layout simulations in \mbox{22-nm} \mbox{FD-SOI}. Furthermore, Fig.~\ref{fig:8_parasitic_diode_leakage}(d) reveals that this leakage increases exponentially with temperature and is also impacted by the reverse voltage drop $V_D$ across the diode, with a higher $V_D$ inducing a higher leakage. At 85$^\circ$C, for a p-well corresponding to the dimensions of $M_2$ with its surrounding dummies, $I_{DIO}$ is around 4.8~pA in 0.11~$\mu$m, and ranges from 15.4 to 120.4~pA in 22~nm. These results suggest that $I_{DIO}$ is largely underestimated in \mbox{0.11-$\mu$m} models, as at 25$^\circ$C and $V_D$ = 0.1~V, the leakage per unit area is 2.83~aA/$\mu$m$^2$ in 0.11~$\mu$m compared to 565.96~aA/$\mu$m$^2$ in 22~nm, so a \mbox{200-$\times$} difference. At 85$^\circ$C, this difference narrows down to 8.9$\times$, with a leakage per unit area worth 4.70~fA/$\mu$m$^2$ in 0.11~$\mu$m and 41.73~fA/$\mu$m$^2$ in 22~nm. If we compare the leakage in 0.11~$\mu$m bulk to the dark current of a P+/n-well photodiode in 0.18~$\mu$m bulk at 25$^\circ$C \cite{Koklu_2013}, the measured value is 7.16~fA for a \mbox{20$\times$20-$\mu$m$^2$} diode, corresponding to a \mbox{17.9-aA/$\mu$m$^2$} leakage per unit area 6.3$\times$ larger than the simulated value in Fig.~\ref{fig:8_parasitic_diode_leakage}(d).\\
\indent The proposed solution to mitigate the impact of this leakage, referred to as \textit{leakage suppression}, consists in zeroing out the leakage by biasing the parasitic p-well/deep n-well diode with a reverse voltage $V_D$ close to zero.
\begin{figure}[!t]
	\centering
	\includegraphics[width=.45\textwidth]{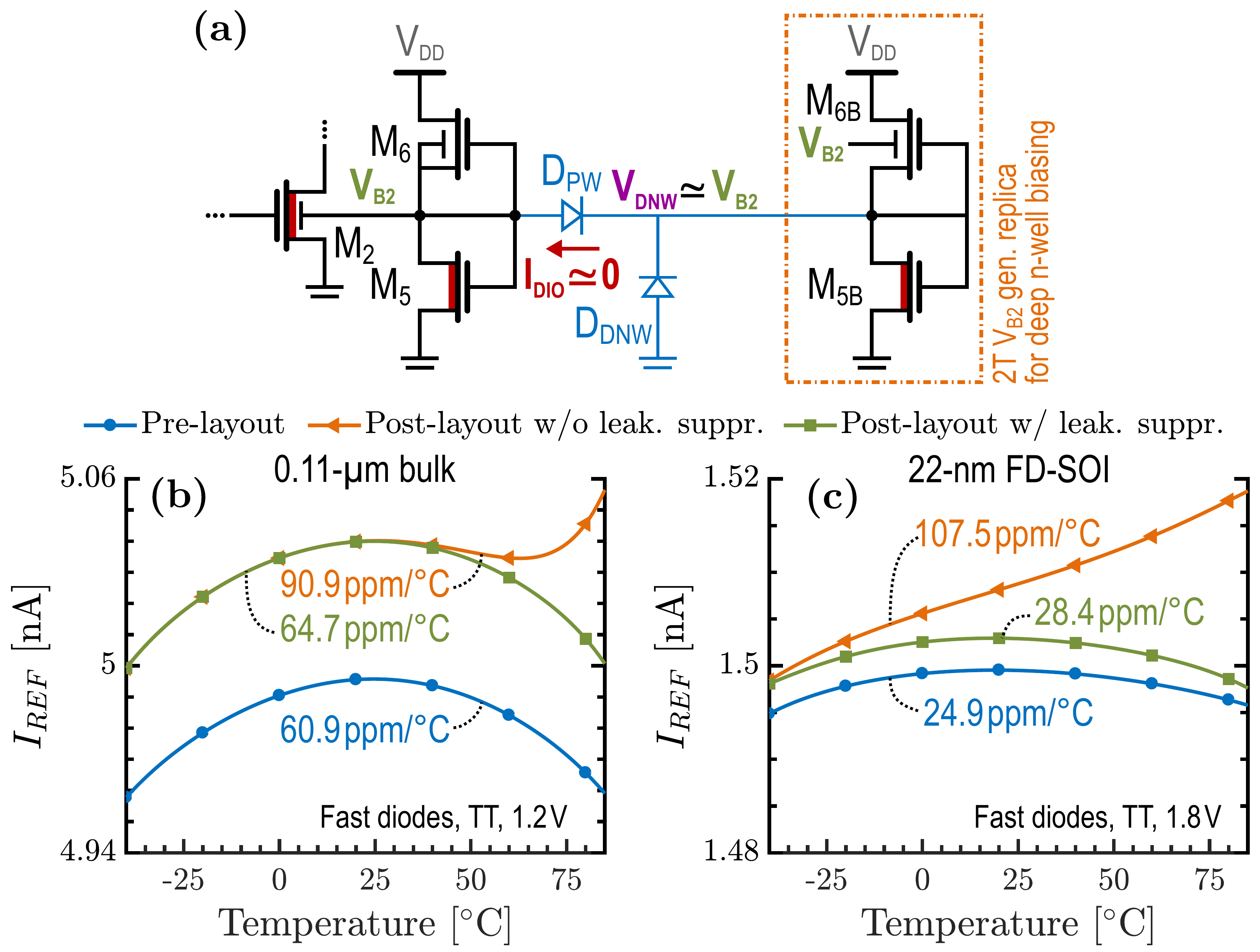}
	\caption{\textbf{$\boldsymbol{I_{REF}}$ TC degradation is mitigated by leakage suppression, relying on a replica of the 2T body bias generator to bias the deep n-well.} (a) Schematic of the 2T body bias generator with parasitic diodes and a replica for leakage suppression. Temperature dependence of $I_{REF}$, based on pre-layout simulations, and post-layout simulations without and with leakage suppression, (b) in \mbox{0.11-$\mu$m} bulk, with a \mbox{100-$\times$} scaling factor on the p-well/deep n-well diode leakage, and (c) in \mbox{22-nm} \mbox{FD-SOI}.}
	\label{fig:9_leakage_suppression}
\end{figure}
This is achieved by biasing the deep n-well with a voltage $V_{DNW}$ close to $V_{B2}$, generated by a replica of the 2T body bias generator, as shown in Fig.~\ref{fig:9_leakage_suppression}(a). This replica has the same $S_6/S_5$ ratio as the body bias generator, but not necessarily the same widths $W_{5-6}$, thereby allowing to adjust its power consumption. Note that, to increase the temperature range up to 125$^\circ$C, a larger $S_6/S_5$ in the replica compared to the body bias generator might be needed, to ensure that the parasitic p-well/deep n-well diode remains reverse-biased and that its leakage current remains negligible. The leakage suppression technique was initially proposed in the context of voltage references \cite{Fassio_2021}, and used to generate a CTAT \mbox{191-nA} current reference in \cite{Fassio_2021_current}, with both works providing measurement results in bulk technologies.
For the \mbox{0.11-$\mu$m} design, because the parasitic diode model underestimates the leakage, we observe no significant degradation of $I_{REF}$ TC between pre- and post-layout simulations. However, when we apply a \mbox{100-$\times$} scaling factor to this leakage as in Fig.~\ref{fig:9_leakage_suppression}(b), $I_{REF}$ TC increases from 60.9 to 90.9~ppm/$^\circ$C, and the leakage suppression circuit cuts down this degradation to 64.7~ppm/$^\circ$C. For the \mbox{22-nm} design in Fig.~\ref{fig:9_leakage_suppression}(c), $I_{REF}$ TC is degraded from 24.9 to 107.5~ppm/$^\circ$C due to the parasitic diode leakage, but this degradation is nearly entirely recovered by the leakage suppression circuit, leading to a post-layout \mbox{28.4-ppm/$^\circ$C} TC. In the complete schematic in Fig.~\ref{fig:12_final_implementation_schematic}, the careful reader may notice that in \mbox{0.11-$\mu$m} bulk, $M_{6-6B}$ have their body connected to $V_{B2}$ [Fig.~\ref{fig:12_final_implementation_schematic}(a)] and are located in the same p-well as $M_2$. On the contrary, in \mbox{22-nm} \mbox{FD-SOI}, $M_{6-6B}$ have their body connected to $V_{DNW}$ [Fig.~\ref{fig:12_final_implementation_schematic}(b)], because they are implemented with SLVT flipped-well devices required to have a lower $V_T$ than $M_{5-5B}$. They are thus located in an n-well, which is biased by the replica to keep the leakage of diode $D_{NW}$ [Fig.~\ref{fig:8_parasitic_diode_leakage}(a)] from impacting $V_{B2}$.
\setcounter{figure}{10}
\begin{figure}[!t]
	\centering
	\includegraphics[width=.5\textwidth]{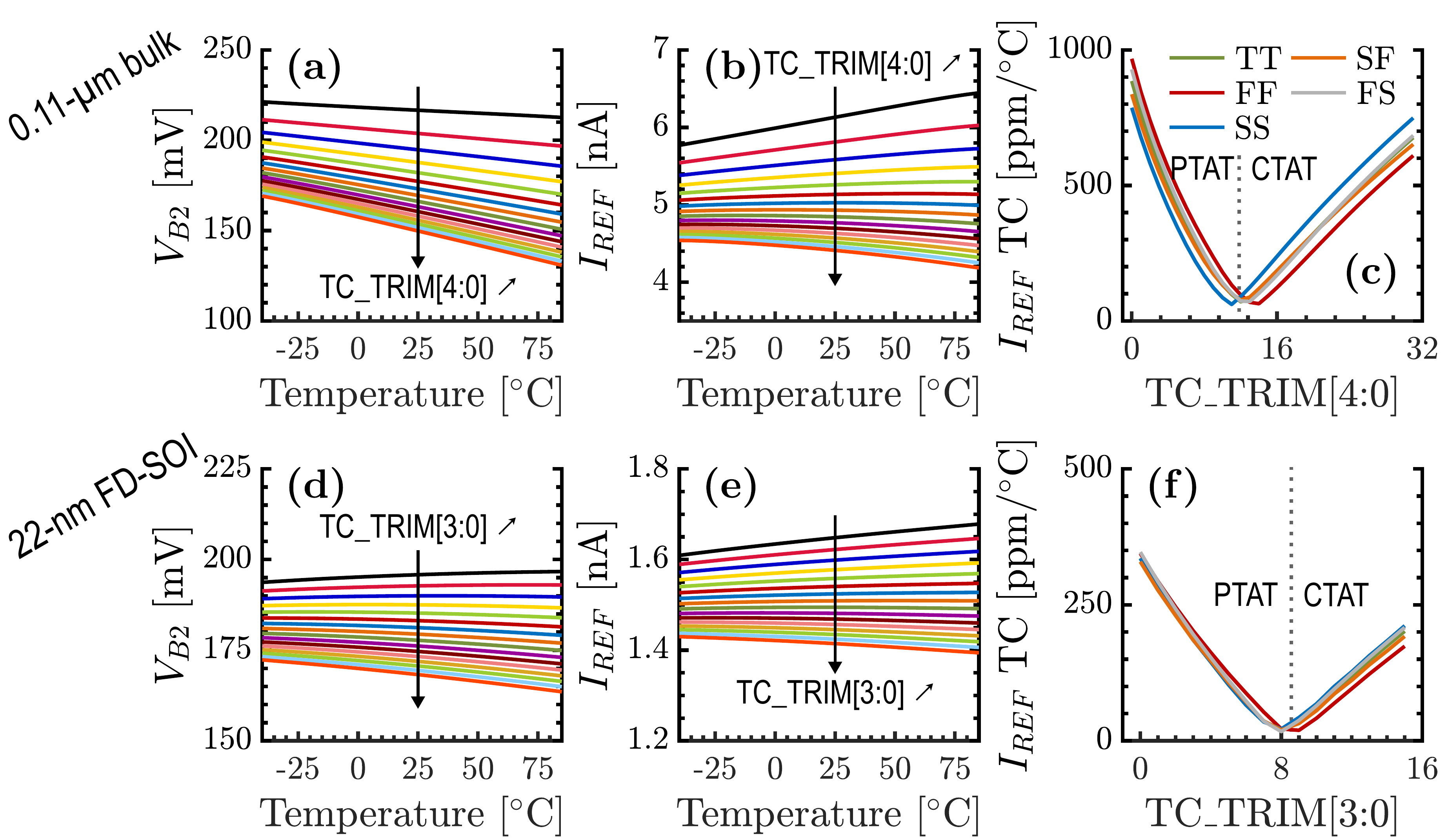}
	\caption{\textbf{$\boldsymbol{I_{REF}}$ TC is trimmed by changing the effective width of $\boldsymbol{M_5}$ in the 2T body bias generator.} In TT 1.2~V (resp. 1.8~V), temperature dependence of $V_{B2}$ [(a) and (d)] and $I_{REF}$ [(b) and (e)] as a function of the trimming code in 0.11~$\mu$m (resp. 22~nm). (c)(f) $I_{REF}$ TC as a function of the trimming code in conventional process corners.}
	\label{fig:10_TC_trimming}
\end{figure}
\begin{figure}[!t]
	\centering
	\includegraphics[width=.5\textwidth]{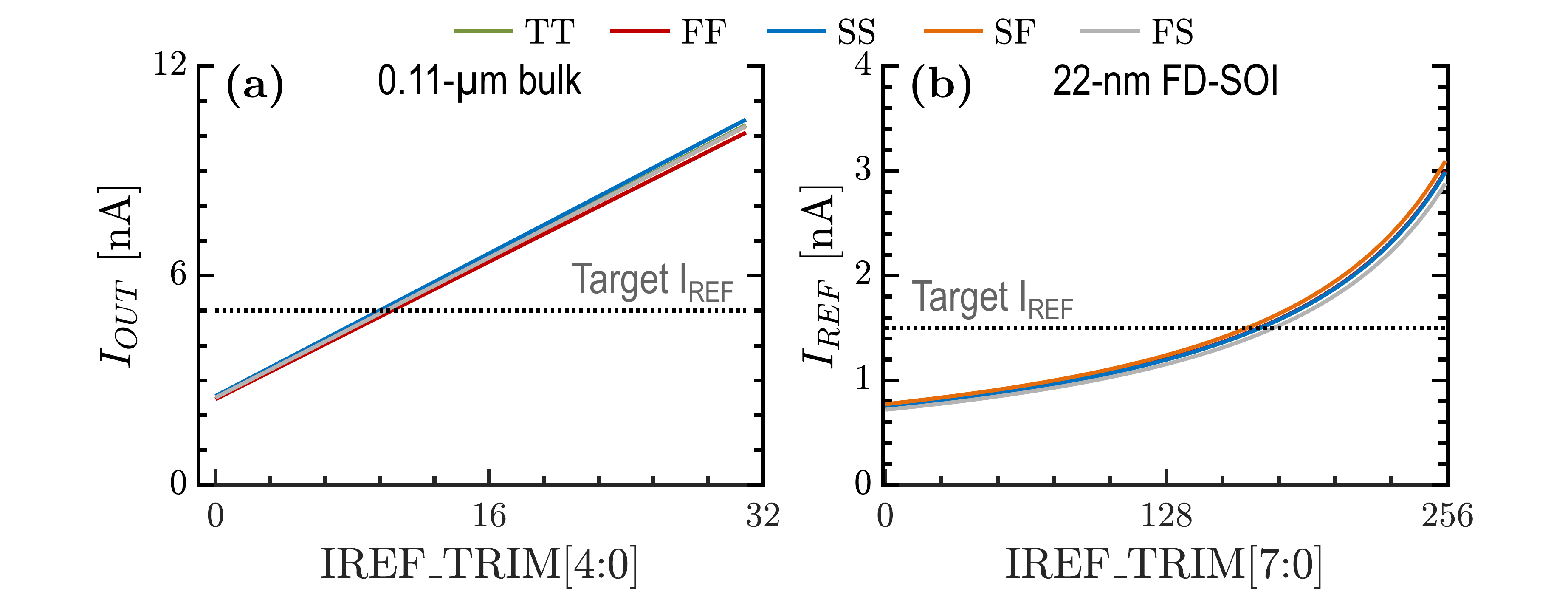}
	\caption{\textbf{$\boldsymbol{I_{OUT}}$ is trimmed by means of a binary-weighted current mirror in bulk, and $\boldsymbol{I_{REF}}$ by a trimmable binary-weighted resistance in FD-SOI.} At 1.2~V (resp. 1.8~V), trimmed current at 25$^\circ$C as a function of the trimming code in (a) 0.11~$\mu$m and (b) 22~nm.}
	\label{fig:11_iref_trimming}
\end{figure}
\setcounter{figure}{9}
\begin{figure*}[!t]
	\centering
	\includegraphics[width=.9\textwidth]{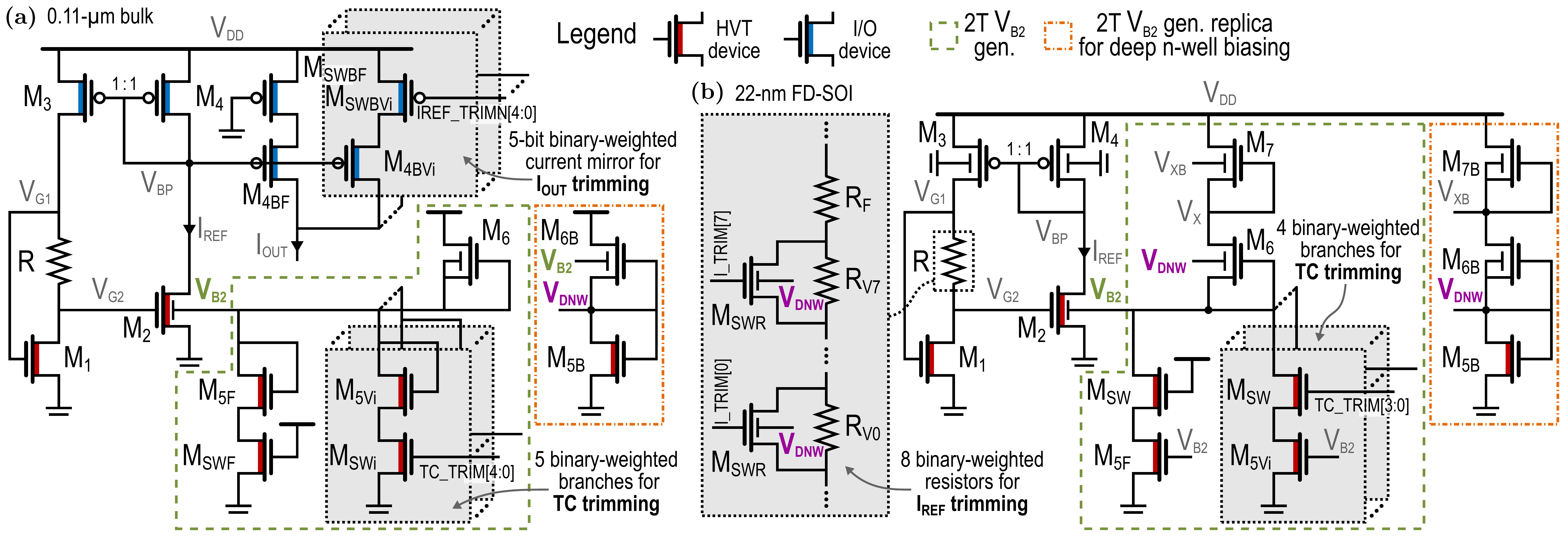}
	\caption{Complete schematic of the proposed nA-range CWT reference in (a) \mbox{0.11-$\mu$m} bulk and (b) \mbox{22-nm} FD-SOI. The schematic includes trimming circuits for TC and $I_{REF}$, as well as a replica of the 2T body bias generator to suppress the leakage of the parasitic p-well/deep n-well diode.}
	\label{fig:12_final_implementation_schematic}
\end{figure*}
\setcounter{figure}{12}
\begin{table*}[!t]
\centering
\caption{Sizing of the proposed nA-range CWT current references.}
\label{table:final_implementation_sizes}
\scalebox{.8}{%
\begin{threeparttable}
\begin{scriptsize}
\begin{tabular}{lcccccclcccccc}
	\toprule
	& \multicolumn{6}{c}{\mbox{0.11-$\mu$m} bulk\tnote{$\ast$}} & & \multicolumn{6}{c}{\mbox{22-nm} FD-SOI}\\
	\cmidrule(lr){2-7} \cmidrule(lr){9-14}
	& \multicolumn{3}{c}{w/o trimming} & \multicolumn{3}{c}{w/ TC and $I_{REF}$ trimming} & & \multicolumn{3}{c}{w/o trimming} & \multicolumn{3}{c}{w/ TC and $I_{REF}$ trimming}\\
	\cmidrule(lr){2-4} \cmidrule(lr){5-7} \cmidrule(lr){9-11} \cmidrule(lr){12-14}
	& Type\tnote{$\star$} & $W$ [$\mu$m] & $L$ [$\mu$m] & Type\tnote{$\star$} & $W$ [$\mu$m] & $L$ [$\mu$m] & & Type\tnote{$\star$} & $W$ [$\mu$m] & $L$ [$\mu$m] & Type\tnote{$\star$} & $W$ [$\mu$m] & $L$ [$\mu$m]\\
	\midrule
	$M_{1-2}$ & LL & 2$\times$8 & 4$\times$5 & LL & 2$\times$8 & 4$\times$5 & $M_{1-2}$ & LVT & 8 & 4$\times$8 & LVT & 8 & 4$\times$8\\
	$M_{3-4}$ & HS & 2$\times$2 & 2$\times$25 & I/O & 20$\times$2.5 & 5 & $M_{3-4}$ & SLVT & 0.32 & 10$\times$8 & SLVT & 0.32 & 10$\times$8\\
	$M_5$ & LL & 16$\times$4 & 7.5 & - & - & - & $M_5$ & LVT & 8$\times$1.25 & 1 & - & - & -\\
	$M_{6-7}$ & HS & 8$\times$1.28 & 7.5 & HS & 8$\times$1.28 & 7.5 & $M_{6-7}$ & SLVT & 8$\times$3.89 & 1 & SLVT & 8$\times$3.9 & 1\\
	$M_{5B}$ & LL & 4$\times$4 & 7.5 & LL & 4$\times$4 & 7.5 & $M_{5B}$ & LVT & 8$\times$1.25 & 1 & LVT & 16$\times$0.625 & 1\\
	$M_{6B-7B}$ & HS & 2$\times$1.28 & 7.5 & HS & 2$\times$1.28 & 7.5 & $M_{6B-7B}$ & SLVT & 8$\times$3.89 & 1 & SLVT & 8$\times$3.9 & 1\\
	$R$ & P+ poly & 0.5 & 256$\times$15.565 & P+ poly & 0.5 & 212$\times$18.705 & $R$ & N+ poly & 0.04 & 100$\times$9.765 & - & - & -\\
	& (high res.) & & & (high res.) & & & & (high res.) & & & & & \\
	\midrule
	$M_{5F}$ & - & - & - & LL & 4$\times$4 & 7.5 & $M_{5F}$ & - & - & - & LVT & 8$\times$0.625 & 1\\
	$M_{5Vi}$ & - & - & - & LL & 1 to 16$\times$4 & 7.5 & $M_{5Vi}$ & - & - & - & LVT & 1 to 8$\times$0.625 & 1\\
	$M_{SWF}$ & - & - & - & LL & 4$\times$0.25 & 2 & $M_{SW}$ & - & - & - & LVT & 0.16 & 8\\
	$M_{SWVi}$ & - & - & - & LL & 1 to 16$\times$0.25 & 2 & $M_{SWR}$ & - & - & - & SLVT & 10 & 0.15\\
	$M_{4BF}$ & - & - & - & I/O & 10$\times$2.5 & 5 & $R_F$ & - & - & - & N+ poly & 0.04 & 85$\times$5.634\\
	$M_{4BVi}$ & - & - & - & I/O & 1 to 16$\times$2.5 & 5 & & & & & (high res.) & & \\
	$M_{SWBF}$ & - & - & - & I/O & 10$\times$0.25 & 1.25 & $R_{Vi}$ & - & - & - & N+ poly & 0.04 & 1 to 128$\times$5.634\\
	$M_{SWBVi}$ & - & - & - & I/O & 1 to 16$\times$0.25 & 1.25 & & & & & (high res.) & & \\
	\bottomrule
\end{tabular}
\end{scriptsize}
\begin{footnotesize}
\begin{tablenotes}
	\item[$\ast$] Dimensions reported for \mbox{0.11-$\mu$m} bulk are pre-shrink ones, and must be scaled by a factor 0.9$\times$ to obtain silicon dimensions.
	\item[$\star$] In 0.11~$\mu$m, HS refers to high-speed, i.e., LVT, and LL to low-leakage, i.e., HVT. In 22~nm, SLVT refers to super-low-$V_T$, and LVT to low-$V_T$.
\end{tablenotes}
\end{footnotesize}
\end{threeparttable}
}
\end{table*}
\begin{figure}[!t]
	\centering
	\includegraphics[width=.5\textwidth]{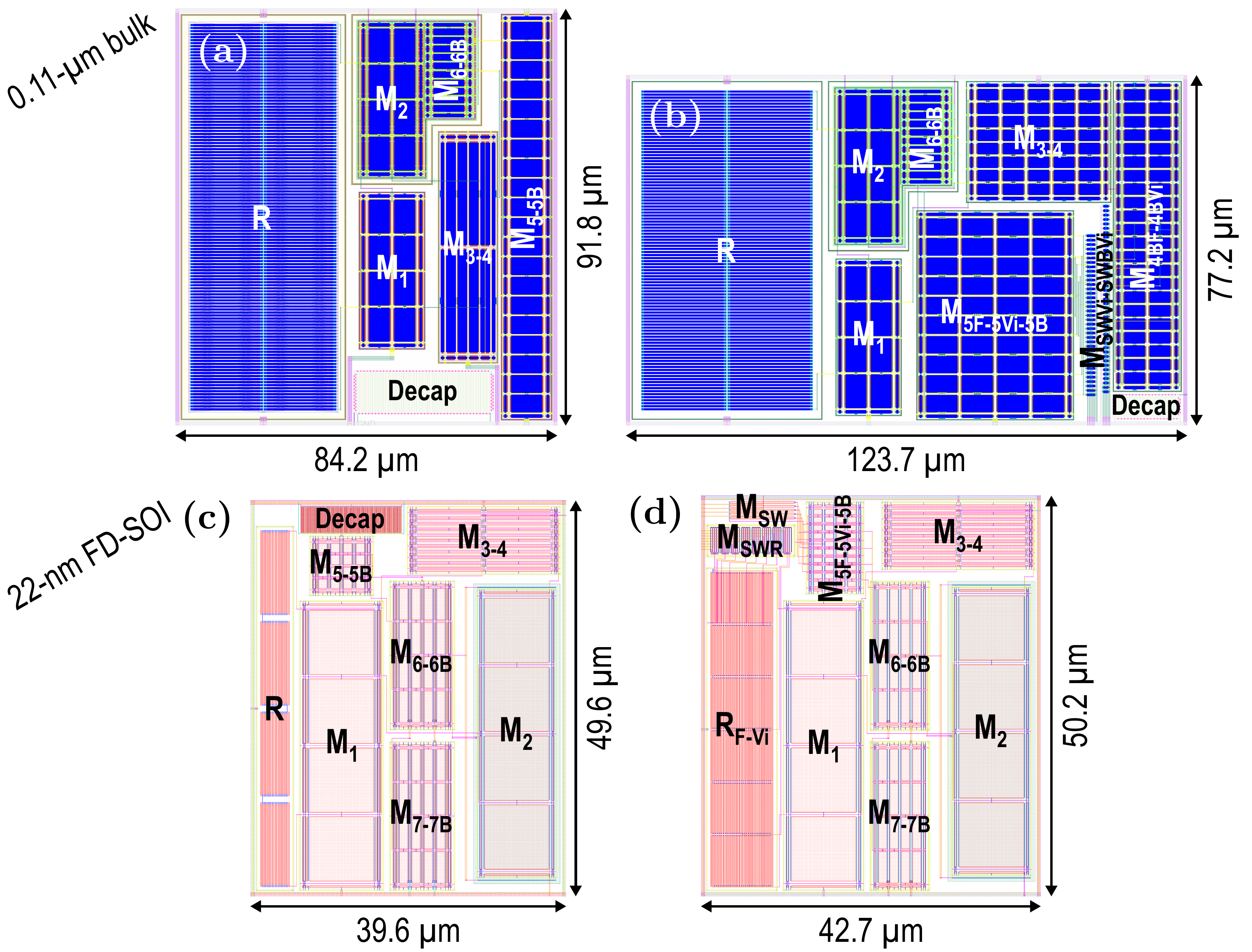}
	\caption{Layout of the proposed reference (a)(c) without and (b)(d) with TC and $I_{REF}$ trimming, in \mbox{0.11-$\mu$m} bulk and \mbox{22-nm \mbox{FD-SOI}}.}
	\label{fig:13_layout}
\end{figure}

\vspace{-0.5cm}
\subsection{Temperature Coefficient and Reference Current Trimming}
\label{subsec:3B_temperature_coefficient_and_reference_current_trimming}
\begin{table}[!t]
\centering
\caption{Summary of the simulated and measured performance of the proposed nA-range CWT references.}
\label{table:summary_performance}
\scalebox{.8}{%
\begin{threeparttable}
\begin{scriptsize}
\begin{tabular}{lccccc}
	\toprule
	& \multicolumn{2}{c}{\mbox{0.11-$\mu$m} bulk} & \multicolumn{3}{c}{\mbox{22-nm} FD-SOI}\\
	\cmidrule(lr){2-3} \cmidrule(lr){4-6}
	& w/o trimming & w/ trimming & w/o trimming & \multicolumn{2}{c}{w/ trimming}\\
	& Sim. & Sim. & Sim. & Sim. & \textbf{Meas.}\\
	\midrule
	$I_{REF}$ [nA] & 5.04 & 5.03 & 1.50 & 1.50 & 1.21/1.50\tnote{$\dagger$}\\
	Power [nW] & 5.03 & 13.88 & 2.04 & 2.35 & 2.87\\
	& $@$0.45V & $@$0.85V & $@$0.65V & $@$0.75V & $@$0.75V\\
	Area [mm$^2$] & 0.00773 & 0.00954 & 0.00197 & \multicolumn{2}{c}{0.00214}\\
	\midrule
	Supply range [V] & 0.45 -- 1.2 & 0.85 -- 1.2 & 0.65 -- 1.8 & \multicolumn{2}{c}{0.75 -- 1.8}\\
	LS [$\%$/V] & 2.22 & 2.84 & 0.44 & 0.41 & 0.51\\
	PSRR\tnote{$\star$}$\quad$[dB] & -35.4 & -35.8 & -38.4 & -35.9 & N/A\\ 
	\midrule
	Temp. range [$^\circ$C] & -40 -- 85 & -40 -- 85 & -40 -- 85 & \multicolumn{2}{c}{-40 -- 85}\\
	TC [ppm/$^\circ$C] & 121.3 & 64.3 & 52.2 & 21.7 & 168.0/89.2\tnote{$\dagger$}\\
	\midrule
	$I_{REF}$ var. & \multirow{2}{*}{11.60} & \multirow{2}{*}{9.05} & \multirow{2}{*}{10.48} & \multirow{2}{*}{9.88} & \multirow{4}{*}{4.10/0.61\tnote{$\dagger$}}\\
	(process) [$\%$] & & & & & \\
	$I_{REF}$ var. & \multirow{2}{*}{0.71} & \multirow{2}{*}{1.32} & \multirow{2}{*}{2.55} & \multirow{2}{*}{2.98} & \\
	(mismatch) [$\%$] & & & & & \\
	\midrule
	$t_{start}$ [ms] & 0.65 & 1.82 & 0.96 & 1.05 & 11.96\\
	\bottomrule
\end{tabular}
\end{scriptsize}
\begin{footnotesize}
\begin{tablenotes}
	\item[$\dagger$] Before and after trimming.
	\item[$\star$] PSRR = $20\log_{10}\left[(i_{ref}/v_{dd})/(I_{REF}/V_{DD})\right]$ is characterized at 10~Hz and $V_{DD,\textrm{min}}$, with a \mbox{1-pF} capacitor between $V_{BP}$ and $V_{DD}$.
\end{tablenotes}
\end{footnotesize}
\end{threeparttable}
}
\end{table}
Depending on the application, it might be necessary to calibrate the TC of $I_{REF}$ and/or $I_{REF}$ itself, to maintain performance across process corners. The proposed design includes two trimming circuits fulfilling this function. First, the TC trimming circuit is presented in Fig.~\ref{fig:12_final_implementation_schematic} and relies on a change of $S_6/S_5$ in (\ref{eq:vb2_proposed}) to tune the TC of $V_{B2}$, and therefore, the TC of $I_{REF}$. It is implemented by changing the effective width of $M_5$ using binary-weighted branches connected in parallel. The switches are either located on top of or below transistors $M_{5Vi}$. We find more effective to place switches below $M_{5Vi}$ in technologies with a large $n$ as it results in a low $I_{on}/I_{off}$ ratio. This switch arrangement alleviates the problem by providing a larger $V_{GS}$ and therefore $I_{DS}$ when the switch is on. Figs.~\ref{fig:10_TC_trimming}(a) to (c) illustrate the effect of this trimming circuit in 0.11~$\mu$m, whereas Figs.~\ref{fig:10_TC_trimming}(d) to (f) depict it in 22~nm. We observe in Figs.~\ref{fig:10_TC_trimming}(a) and (d) that the change in $V_{B2}$ TC allows to make $I_{REF}$ either PTAT or CTAT [Figs.~\ref{fig:10_TC_trimming}(b) and (e)] across the trimming range, and that the minimum $I_{REF}$ TC is easily reached in the five conventional process corners [Figs.~\ref{fig:10_TC_trimming}(c) and (f)].
Section~\ref{subsec:4A_designs_in_0.11-µm_bulk_CMOS_technology} will later highlight that skewed process corners, i.e., different process corners for the devices of different $V_T$ types used for $M_5$ and $M_6$, are more critical and should be adequately considered when defining the trimming range.\\
\indent Moreover, a trimming of $I_{REF}$ can be required in applications relying on an accurate value of $I_{REF}$. In 0.11~$\mu$m, this trimming is performed by a \mbox{5-bit} binary-weighted current mirror [Fig.~\ref{fig:12_final_implementation_schematic}(a)], allowing for a linear control of the output current $I_{OUT}$ [Fig.~\ref{fig:11_iref_trimming}(a)]. Besides, \mbox{3.3-V} I/O devices are used for the mirror formed by $M_3$, $M_4$ and $M_{4BF-4BVi}$, and for the switches $M_{SWBF-SWBVi}$, as they feature a lower $I_{off}$ than core devices. In 22~nm, the $I_{REF}$ trimming circuit is integrated inside the reference, and is implemented by eight binary-weighted resistors which can be shorted by nMOS switches [Fig.~\ref{fig:12_final_implementation_schematic}(b)]. This trimming scheme is possible as the $I_{on}/I_{off}$ ratio is larger in 22~nm than in 0.11~$\mu$m thanks to a lower $n$, and is also more power-efficient as it avoids the overhead associated with $I_{OUT}$. In addition, it enables a finer trimming resolution, but leads to a nonlinear relationship between the calibration code and $I_{REF}$ as it changes the value of resistor $R$ in (\ref{eq:iref_proposed}) [Fig.~\ref{fig:11_iref_trimming}(b)].
\begin{figure}[!t]
	\centering
	\includegraphics[width=.45\textwidth]{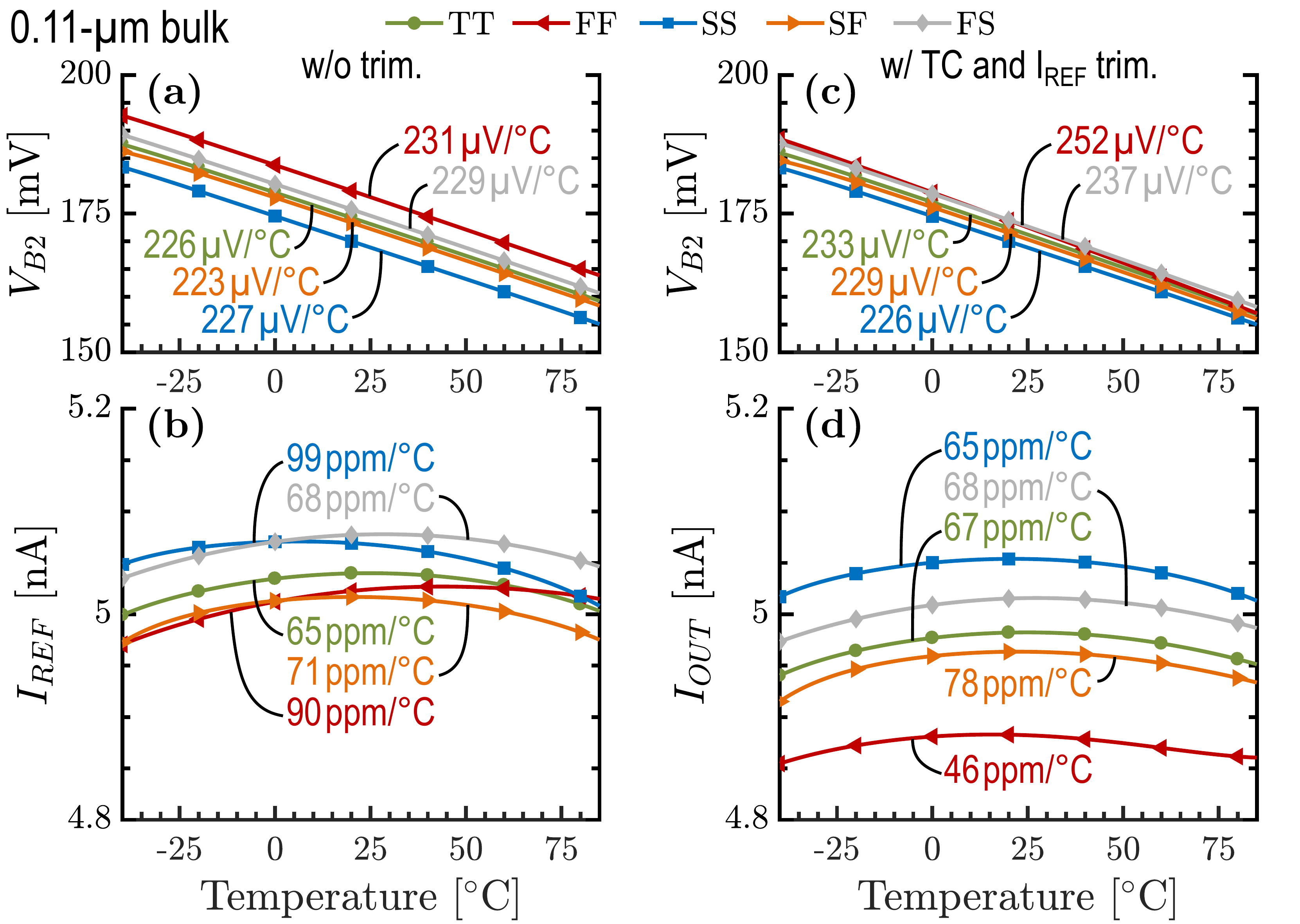}
	\caption{In \mbox{0.11-$\mu$m} bulk, post-layout simulation of the temperature dependence of $V_{B2}$ and $I_{REF}$, in all process corners and at 1.2~V, without trimming [(a) and (b)], and with TC and $I_{REF}$ trimming [(c) and (d)].}
	\label{fig:14_sim_iref_vs_T_0p11um}
\end{figure}
\begin{figure}[!t]
	\centering
	\includegraphics[width=.45\textwidth]{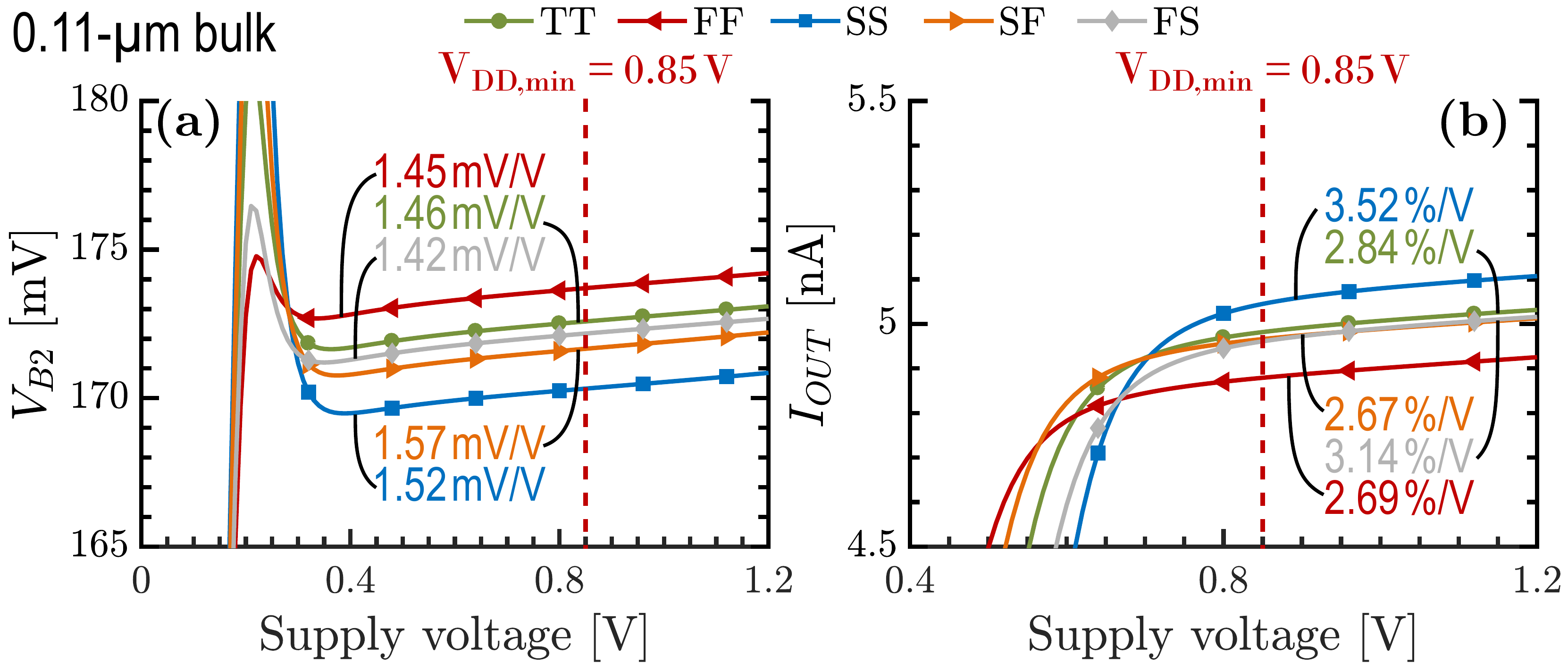}
	\caption{In \mbox{0.11-$\mu$m} bulk, post-layout simulation of the supply voltage dependence of $V_{B2}$ and $I_{REF}$ in all process corners and at 25$^\circ$C, with TC and $I_{REF}$ trimming.}
	\label{fig:15_sim_iref_vs_vdd_0p11um}
\end{figure}

\vspace{-0.25cm}
\subsection{Final Implementation}
\label{subsec:3C_final_implementation}
Two current references, respectively without trimming, and with TC and $I_{REF}$ trimming, have been implemented in each of the \mbox{0.11-$\mu$m} bulk and \mbox{22-nm} \mbox{FD-SOI} technologies. The complete schematic of the proposed reference with leakage suppression, and TC and $I_{REF}$ trimming, is presented in Fig.~\ref{fig:12_final_implementation_schematic}, while the layout and sizing of the four proposed references can be found in Fig.~\ref{fig:13_layout} and Table~\ref{table:final_implementation_sizes}, respectively. Interestingly, Fig.~\ref{fig:13_layout} reveals that the area overhead of the trimming circuit, with respect to the area of the reference without trimming, corresponds to 23.5~$\%$ in 0.11~$\mu$m and can be attributed equally to the TC and $I_{REF}$ trimming circuits, while it is only 9.1~$\%$ in 22~nm and is predominantly due to the binary-weighted resistors used to trim $I_{REF}$.
\begin{figure}[!t]
	\centering
	\includegraphics[width=.45\textwidth]{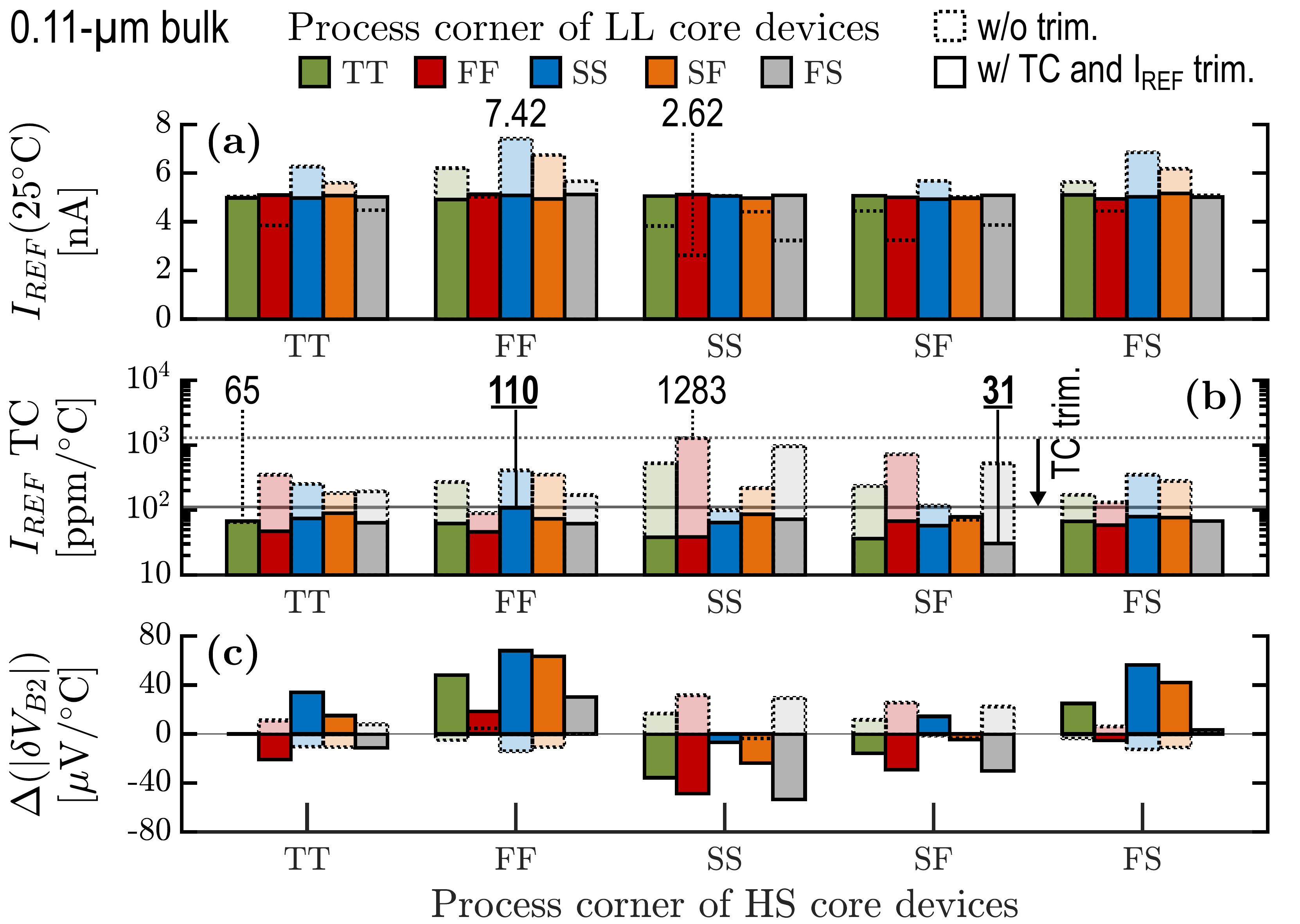}
	\caption{In \mbox{0.11-$\mu$m} bulk and at 1.2~V, post-layout simulation of (a) $I_{REF}$ at 25$^\circ$C, (b) $I_{REF}$ TC from -40 to 85$^\circ$C, and (c) the change in CTAT slope of $V_{B2}$, denoted as $\delta V_{B2}$, with respect to its nominal value, without trimming, and with TC and $I_{REF}$ trimming.}
	\label{fig:16_sim_iref_vs_T_skewed_process_0p11um}
\end{figure}
\begin{figure}[!t]
	\centering
	\includegraphics[width=.5\textwidth]{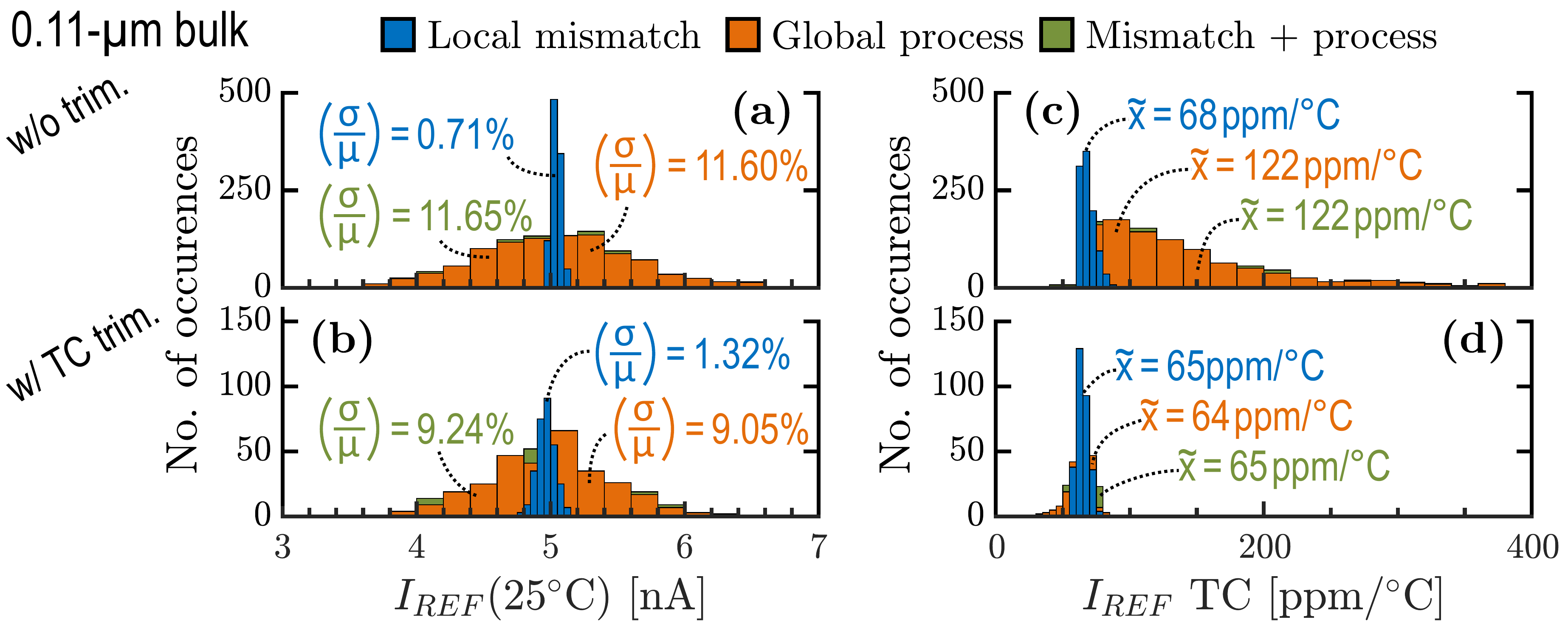}
	\caption{In \mbox{0.11-$\mu$m} bulk, for post-layout MC simulations (10$^3$ without trimming, and 3$\times$10$^2$ with TC trimming) in TT at 1.2~V, histograms of $I_{REF}$ at 25$^\circ$C [(a) and (b)] and of $I_{REF}$ TC from -40 to 85$^\circ$C [(c) and (d)], without trimming [(a) and (c)], and with TC trimming [(b) and (d)]. $\tilde{x}$ denotes the median of the distribution.}
	\label{fig:17_sim_iref_vs_T_mc_0p11um}
\end{figure}

\section{Post-Layout Simulation Results}
\label{sec:4_post-layout_simulation_results}
In this section, we discuss the post-layout simulation results of the four references introduced in Section~\ref{subsec:3C_final_implementation}. More specifically, the temperature dependence is discussed in conventional and skewed process corners, and for Monte-Carlo (MC) simulations. The interest of this section is thus to demonstrate the performance of the proposed references and the robustness of the trimming mechanisms under the impact of local mismatch and global process variations, which cannot be as extensively and thoroughly covered in measurement. Additionally, for the design with TC and $I_{REF}$ trimming, we characterize the supply voltage dependence and startup. Table~\ref{table:summary_performance} summarizes the performance of the designs.

\vspace{-0.25cm}
\subsection{Designs in 0.11-$\mu$m Bulk CMOS Technology}
\label{subsec:4A_designs_in_0.11-µm_bulk_CMOS_technology}
The temperature dependence of $V_{B2}$ and $I_{REF}$ is shown in Fig.~\ref{fig:14_sim_iref_vs_T_0p11um}. In the design without trimming, there is a slight change of $V_{B2}$ [Fig.~\ref{fig:14_sim_iref_vs_T_0p11um}(a)], hence deteriorating $I_{REF}$ TC in FF and SS to 90 and 99~ppm/$^\circ$C, compared to 65~ppm/$^\circ$C in TT [Fig.~\ref{fig:14_sim_iref_vs_T_0p11um}(b)]. Trimming the TC and $I_{REF}$ modifies the CTAT slope of $V_{B2}$ [Fig.~\ref{fig:14_sim_iref_vs_T_0p11um}(c)], and thereby reduces the TC to at most 78~ppm/$^\circ$C in SF [Fig.~\ref{fig:14_sim_iref_vs_T_0p11um}(d)]. While the trimming mechanism in Fig.~\ref{fig:14_sim_iref_vs_T_0p11um} may not yield obvious benefits in conventional process corners, its interest in dealing with different process corners for the two $V_T$ types employed for $M_{5-6}$ is paramount, as will be discussed herebelow in relation to Fig.~\ref{fig:16_sim_iref_vs_T_skewed_process_0p11um}.
Next, the supply voltage dependence is depicted in Fig.~\ref{fig:15_sim_iref_vs_vdd_0p11um}, with the minimum supply voltage given by
\begin{figure}[!t]
	\centering
	\includegraphics[width=.45\textwidth]{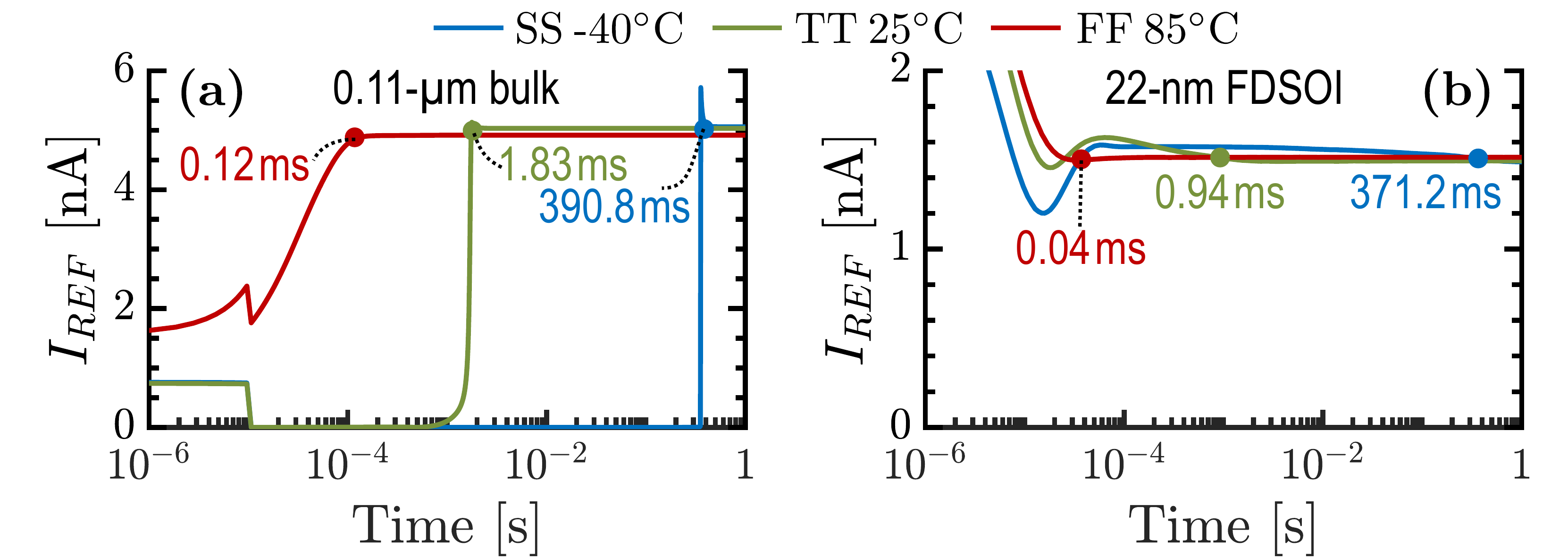}
	\caption{Post-layout startup waveforms of the reference with TC and $I_{REF}$ trimming, in the fastest (FF 85$^\circ$C), typical (TT 25$^\circ$C), and slowest (SS \mbox{-40$^\circ$C}) corners, (a) at 1.2~V in \mbox{0.11-$\mu$m} bulk and (b) at 1.8~V in \mbox{22-nm} FDSOI, using a voltage source with a \mbox{100-$\mu$s} rise time.}
	\label{fig:18_sim_startup}
\end{figure}
\begin{figure}[!t]
	\centering
	\includegraphics[width=.45\textwidth]{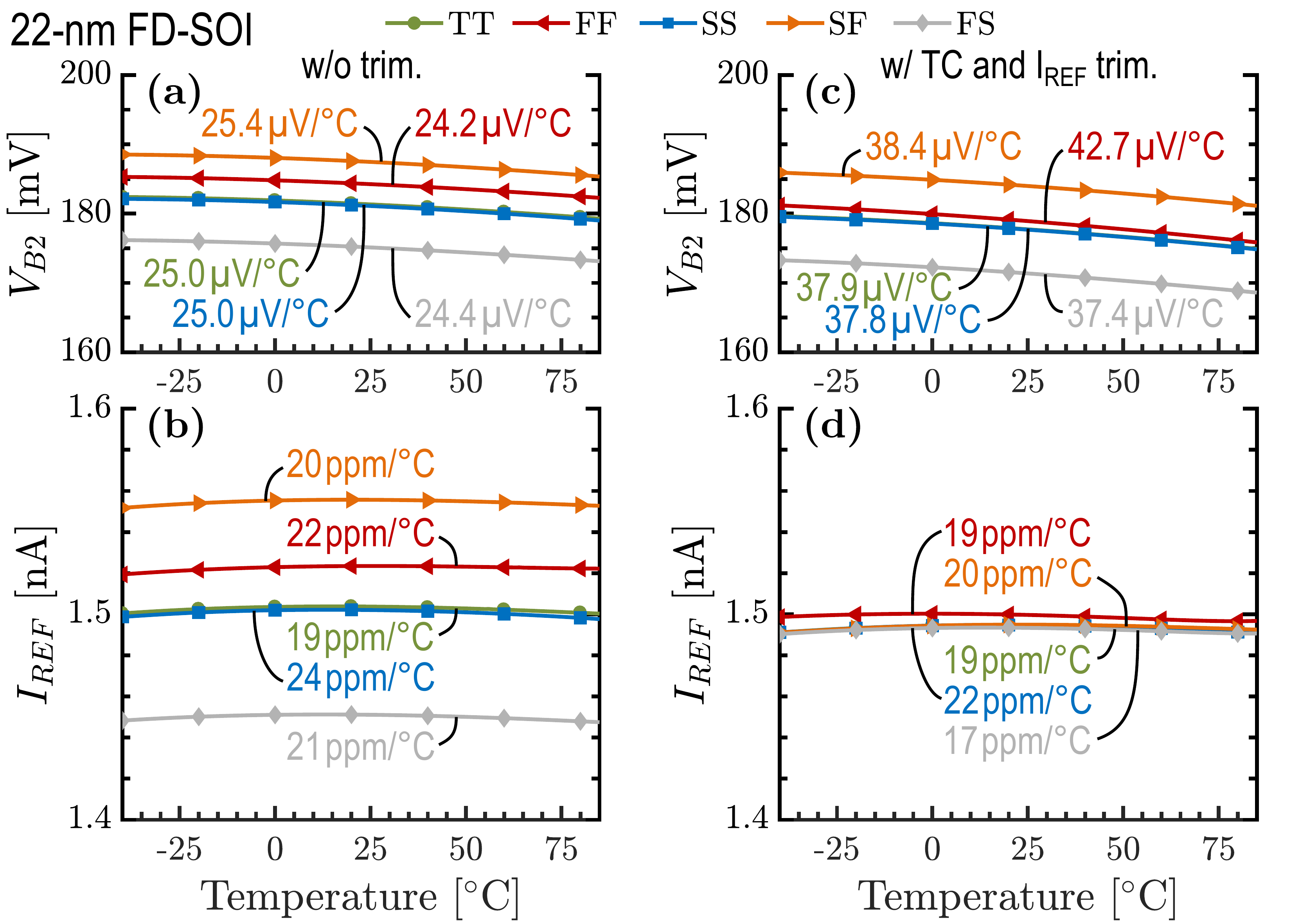}
	\caption{In \mbox{22-nm} \mbox{FD-SOI}, post-layout simulation of the temperature dependence of $V_{B2}$ and $I_{REF}$, in all process corners and at 1.8~V, without trimming [(a) and (b)], and with TC and $I_{REF}$ trimming [(c) and (d)].}
	\label{fig:19_sim_iref_vs_T_22nm}
\end{figure}
\begin{figure}[!t]
	\centering
	\includegraphics[width=.45\textwidth]{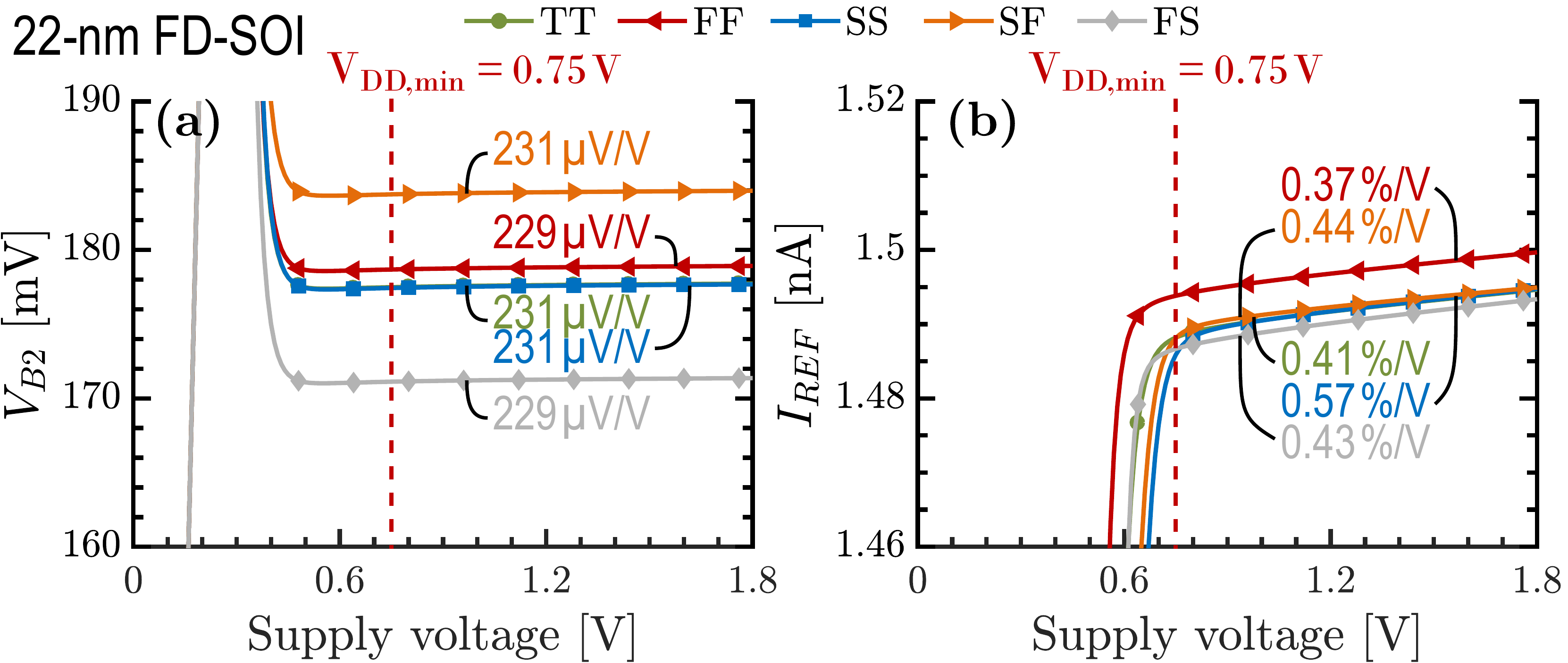}
	\caption{In \mbox{22-nm} \mbox{FD-SOI}, post-layout simulation of the supply voltage dependence of $V_{B2}$ and $I_{REF}$ in all process corners and at 25$^\circ$C, with TC and $I_{REF}$ trimming.}
	\label{fig:20_sim_iref_vs_vdd_22nm}
\end{figure}
\begin{equation}
	V_{DD,\textrm{min}} \approx 4U_T +  \max\left( V_{GS1}, V_{SG4}, V_{B2}\right)\textrm{.}\label{eq:vddmin}
\end{equation}
For this design, $V_{DD,\textrm{min}}$ is limited to 0.85~V by the large $V_{SG4}$ due to the use of I/O devices in the pMOS current mirror. Yet, the 2T body bias generator already produces a valid $V_{B2}$ above 0.4~V, explaining why the design without trimming can operate down to 0.45~V in Table~\ref{table:summary_performance}. $V_{B2}$ LS is comprised between 1.42 and 1.57~mV/V from 0.85 to 1.2~V [Fig.~\ref{fig:15_sim_iref_vs_vdd_0p11um}(a)], while $I_{REF}$ LS ranges from 2.67 to 3.52~$\%$/V [Fig.~\ref{fig:15_sim_iref_vs_vdd_0p11um}(b)]. Eq. (\ref{eq:iref_LS_conventional}) indicates that this LS predominantly stems from the variation of $M_2$ $V_{DS}$.
In addition, Fig.~\ref{fig:16_sim_iref_vs_T_skewed_process_0p11um} considers the impact of HS/LL skewed process corners, i.e., different process corners for the high-speed (HS) and low-leakage (LL) core devices used for $M_{5-6}$. Fig.~\ref{fig:16_sim_iref_vs_T_skewed_process_0p11um}(a) shows that, without trimming, $I_{REF}$ ranges from 7.42~nA in (FF, SS) down to 2.62~nA in (SS, FF), resulting in process variations of \mbox{+47.2$\%$}$\:$/$\:$\mbox{-48.0$\%$}. The remaining variations after trimming are \mbox{+3.6$\%$}$\:$/$\:$\mbox{-1.4$\%$}, related to the finite resolution of the trimming circuit. Regarding $I_{REF}$ TC [Fig.~\ref{fig:16_sim_iref_vs_T_skewed_process_0p11um}(b)], it lies between 65 and 1283~ppm/$^\circ$C without trimming, and is reduced to a range from 31 to 110~ppm/$^\circ$C after trimming. Moreover, $I_{REF}$ TC is below 100~ppm/$^\circ$C in all process corners except (FF, SS), which would necessitate a slightly larger TC trimming range. The effect of the TC trimming scheme is to increase the CTAT slope of $V_{B2}$ in fast nMOS HS corners, and to diminish it in slow ones, on top of smaller variations related to the LL corner [Fig.~\ref{fig:16_sim_iref_vs_T_skewed_process_0p11um}(c)].
Then, MC simulations highlight that, without trimming [Fig.~\ref{fig:17_sim_iref_vs_T_mc_0p11um}(a)], the impact of mismatch is limited, with a $(\sigma/\mu)$ of 0.71~$\%$.This value is relatively close to the 1.18~$\%$ predicted by (\ref{eq:var_iref}), from which 1$\%$ can be attributed to the mismatch of the pMOS mirror. The impact of process variations is more significant, with a $(\sigma/\mu)$ of 11.60~$\%$, as a result of the process variations of $M_{5-6}$ but more critically, of the resistance. In Fig.~\ref{fig:17_sim_iref_vs_T_mc_0p11um}(b), the distribution of the TC due to local mismatch presents a \mbox{68-ppm/$^\circ$C} median and an \mbox{84-ppm/$^\circ$C} 99th percentile, while global process and combined effects both have a \mbox{122-ppm/$^\circ$C} median, with 99th percentiles at 422 and 446~ppm/$^\circ$C. With TC trimming and no $I_{REF}$ trimming, $I_{REF}$ $(\sigma/\mu)$ slightly rises to 1.32~$\%$ with mismatch and shrinks to 9.05~$\%$ with process variations [Fig.~\ref{fig:17_sim_iref_vs_T_mc_0p11um}(c)], due to a different sizing and to the trimming itself. In terms of TC, the median is cut down to 65~ppm/$^\circ$C, with a 99th percentile at 75~ppm/$^\circ$C for mismatch, and 80~ppm/$^\circ$C for process and combined effects [Fig.~\ref{fig:17_sim_iref_vs_T_mc_0p11um}(d)].
\begin{figure}[!t]
	\centering
	\includegraphics[width=.45\textwidth]{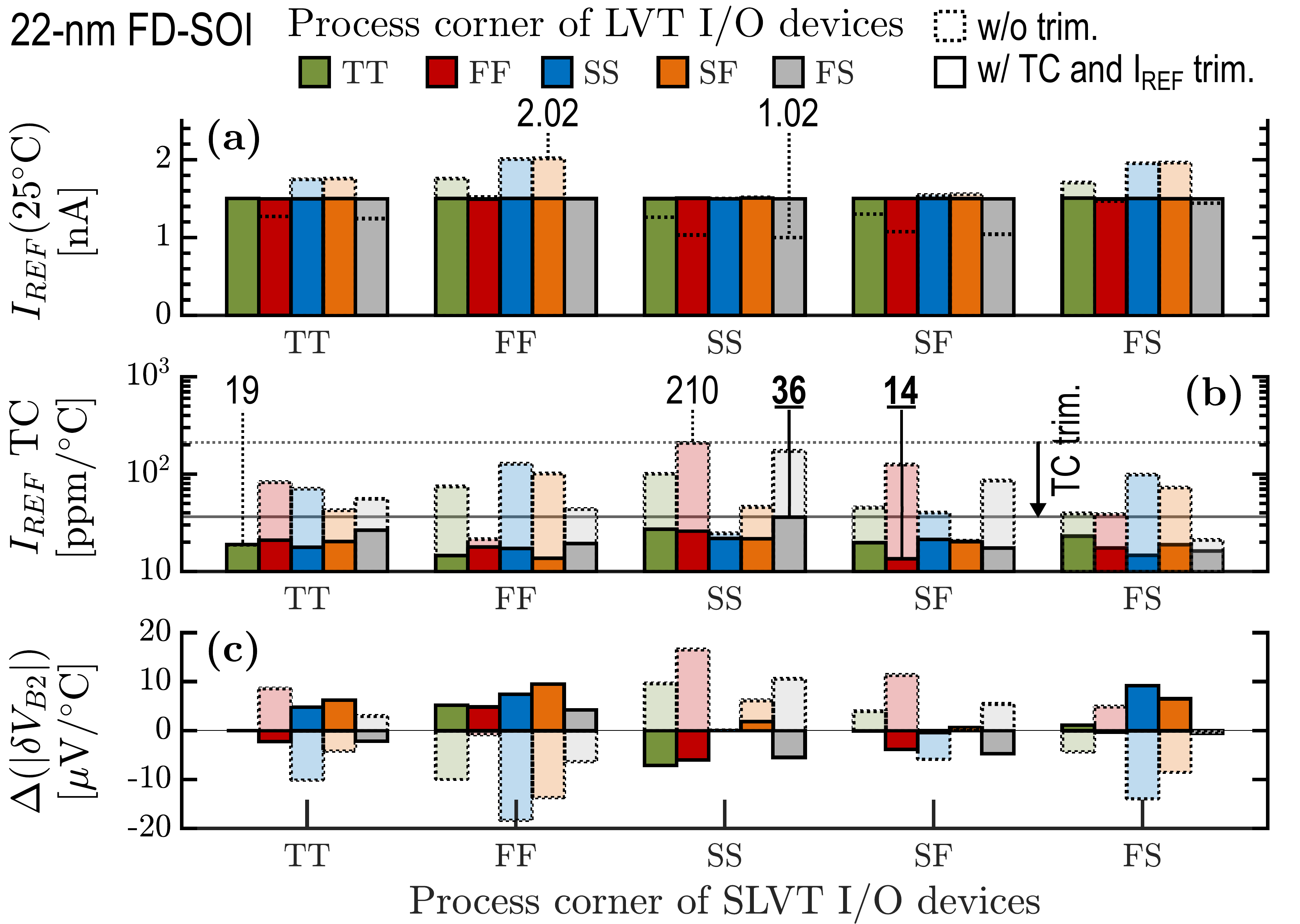}
	\caption{In \mbox{22-nm} \mbox{FD-SOI} and at 1.8~V, post-layout simulation of (a) $I_{REF}$ at 25$^\circ$C, (b) $I_{REF}$ TC from -40 to 85$^\circ$C, and (c) the change in CTAT slope of $V_{B2}$, denoted as $\delta V_{B2}$, with respect to its nominal value, without trimming, and with TC and $I_{REF}$ trimming.}
	\label{fig:21_sim_iref_vs_T_skewed_process_22nm}
\end{figure}
\begin{figure}[!t]
	\centering
	\includegraphics[width=.5\textwidth]{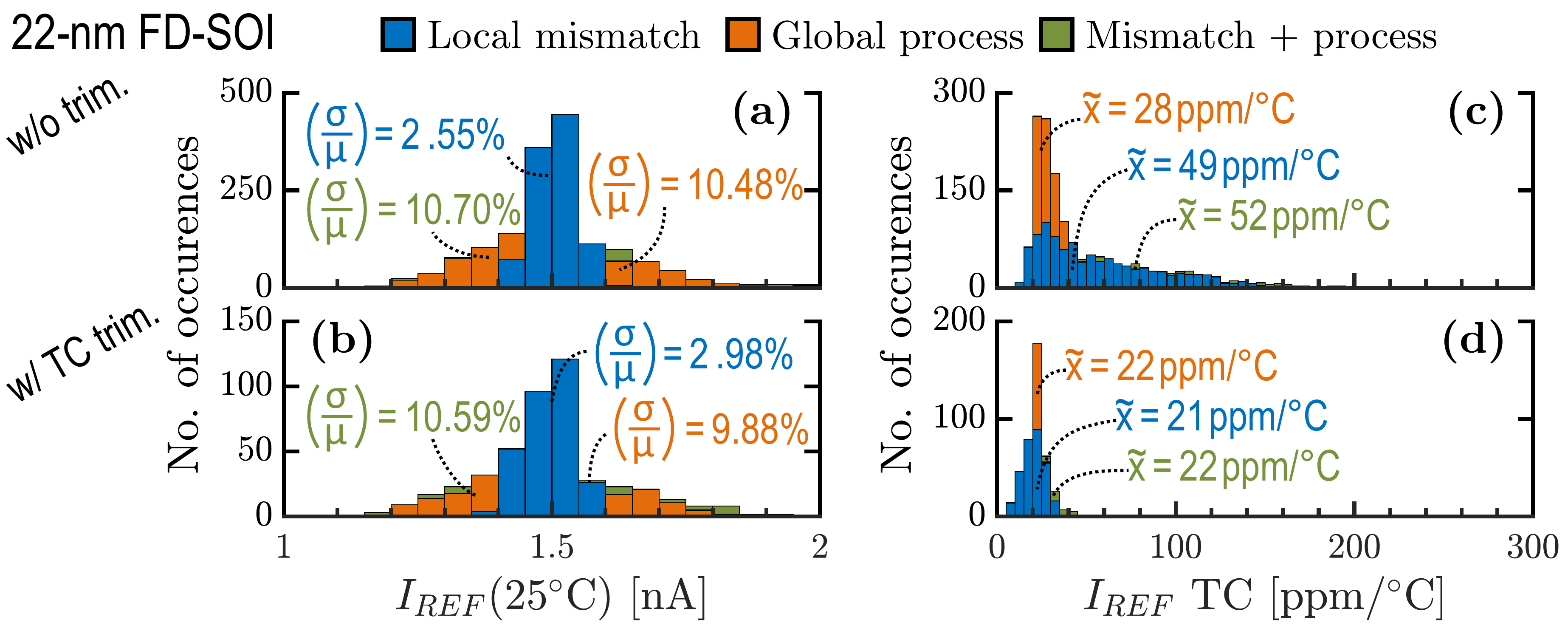}
	\caption{In \mbox{22-nm} \mbox{FD-SOI}, for post-layout MC simulations (10$^3$ without trimming, and 3$\times$10$^2$ with TC trimming) in TT at 1.8~V, histograms of $I_{REF}$ at 25$^\circ$C [(a) and (b)] and of $I_{REF}$ TC from -40 to 85$^\circ$C [(c) and (d)], without trimming [(a) and (c)], and with TC and $I_{REF}$ trimming [(b) and (d)]. $\tilde{x}$ denotes the median of the distribution.}
	\label{fig:22_sim_iref_vs_T_mc_22nm}
\end{figure}
Finally, the \mbox{x-$\%$} startup time corresponds to the time at which $I_{REF}$ remains within (100-x)~$\%$ of its steady-state value. The nominal \mbox{99-$\%$} startup time is equal to 1.83~ms in Fig.~\ref{fig:18_sim_startup}(a), and ranges from 0.12 to 390.8~ms in the best- and worst-case corners, respectively FF 85$^\circ$C and SS \mbox{-40$^\circ$C}.

\subsection{Designs in 22-nm FD-SOI CMOS Technology}
\label{subsec:4B_designs_in_22-nm_FD-SOI_CMOS_technology}
\begin{figure}[!t]
	\centering
	\includegraphics[width=.469\textwidth]{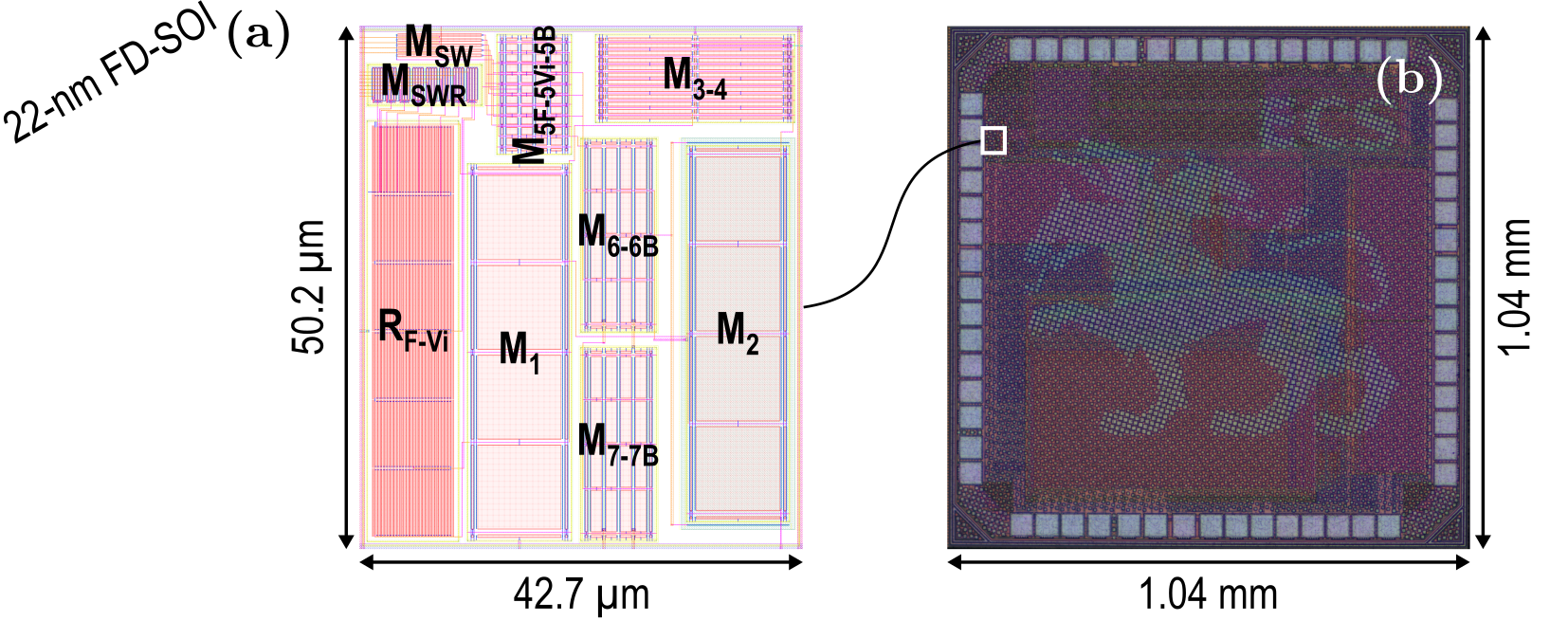}
	\caption{In \mbox{22-nm} \mbox{FD-SOI}, (a) layout of the proposed nA-range CWT current reference and (b) chip microphotograph with overlaid layout of the \mbox{1.08-mm$^2$} CERBERUS MCU.}
	\label{fig:23_microphotograph_layout}
\end{figure}
\begin{figure}[!t]
	\centering
	\includegraphics[width=.48\textwidth]{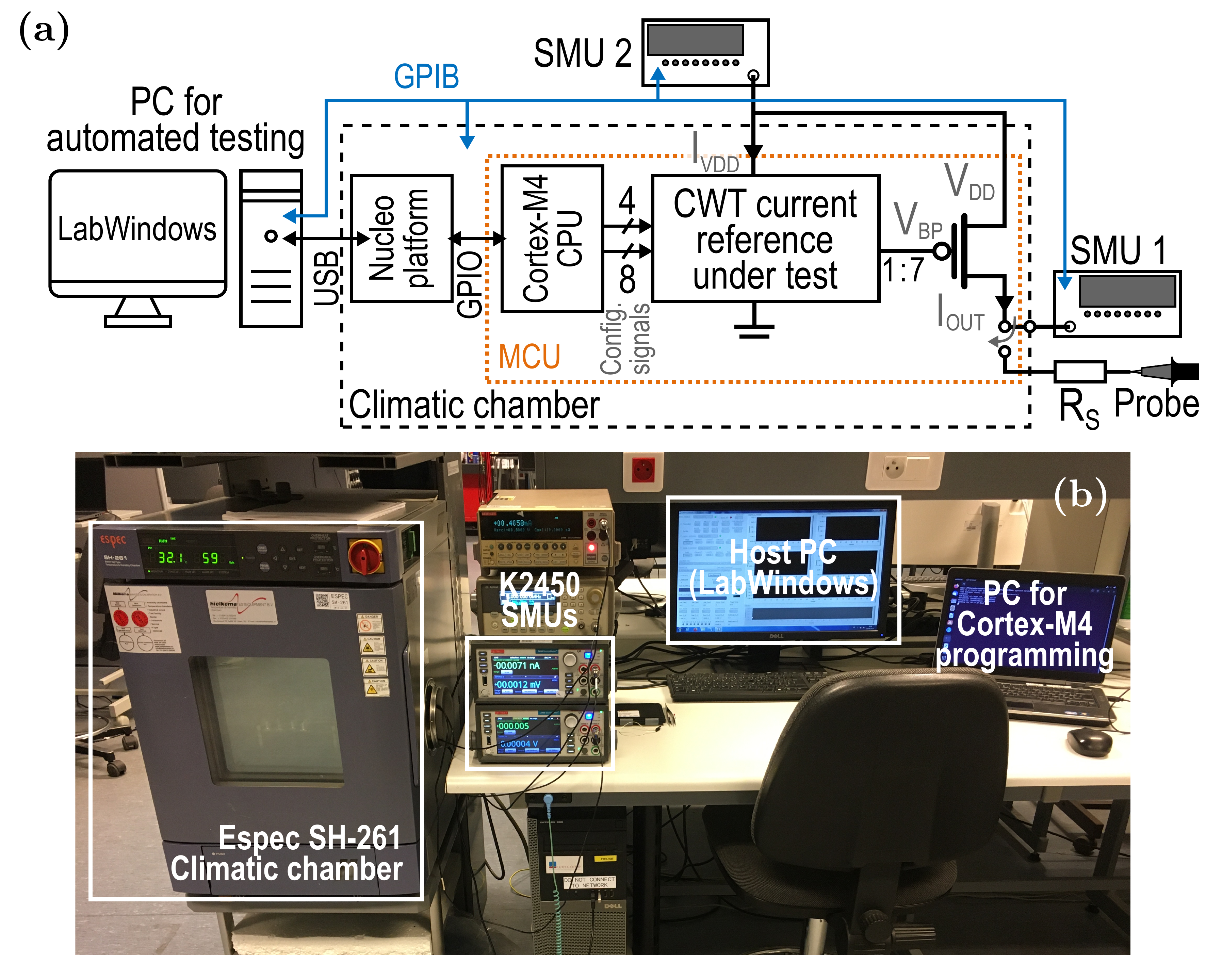}
	\caption{Measurement testbench for the characterization of the dependence to supply voltage and temperature, as well as the startup time.}
	\label{fig:24_meas_setup}
\end{figure}
In Fig.~\ref{fig:19_sim_iref_vs_T_22nm}, depicting the temperature dependence of $I_{REF}$, the design without trimming shows slim variations of $V_{B2}$ CTAT slope between 24.4 and 25.4~$\mu$V/$^\circ$C [Fig.~\ref{fig:19_sim_iref_vs_T_22nm}(a)], translating into negligible variations of $I_{REF}$ TC from 19 to 24~ppm/$^\circ$C [Fig.~\ref{fig:19_sim_iref_vs_T_22nm}(b)]. After trimming, $V_{B2}$ and its TC are slightly different [Fig.~\ref{fig:19_sim_iref_vs_T_22nm}(c)], yet $I_{REF}$ TC remains in a range similar to the design without trimming, comprised between 17 and 22~ppm/$^\circ$C, while a clear effect of the $I_{REF}$ trimming can be observed [Fig.~\ref{fig:19_sim_iref_vs_T_22nm}(d)].
Then, regarding the supply voltage dependence in Fig.~\ref{fig:20_sim_iref_vs_vdd_22nm}, $V_{DD,\textrm{min}}$ is equal to 0.75~V and is not dominated by $V_{SG4}$ as in 0.11~$\mu$m, but rather by the minimum $V_{DD}$ required by the switches in the $I_{REF}$ trimming circuit to avoid distortion. Fig.~\ref{fig:20_sim_iref_vs_vdd_22nm}(a) reveals that $V_{B2}$ is generated at 0.5~V, and presents an LS around 230~$\mu$V/V. In Fig.~\ref{fig:20_sim_iref_vs_vdd_22nm}(b), $I_{REF}$ features an LS between 0.37 and 0.57~$\%$/V, which is a reduction of nearly 6$\times$ compared to the \mbox{0.11-$\mu$m} design, thanks to the larger $(g_m/g_d)$ in FD-SOI.
Moreover, Fig.~\ref{fig:21_sim_iref_vs_T_skewed_process_22nm} illustrates the variations in SLVT/LVT skewed process corners, i.e., different process corners for the SLVT and LVT I/O devices used for $M_{5-6}$. Fig.~\ref{fig:21_sim_iref_vs_T_skewed_process_22nm}(a) displays that $I_{REF}$ spans from 2.02~nA in (FF, SF) to 1.02~nA in (SS, FS) without trimming, leading to process variations of \mbox{+34.1$\%$}$\:$/$\:$\mbox{-33.4$\%$}. However, trimming reduces variations to \mbox{+0.2$\%$}$\:$/$\:$\mbox{-0.3$\%$}, approximately 10$\times$ better than in 0.11~$\mu$m, which is consistent with the fact that the \mbox{22-nm} design uses three additional calibration bits for $I_{REF}$. Adding more bits to the \mbox{0.11-$\mu$m} design would be possible, but only at the expense of significant area overhead for the binary-weighted current mirror. Then, regarding the TC, it ranges from 19 to 210~ppm/$^\circ$C without trimming, and is reduced from 14 to 36~ppm/$^\circ$C after it, while being below 50~ppm/$^\circ$C in all corners [Fig.~\ref{fig:21_sim_iref_vs_T_skewed_process_22nm}(b)]. The effect of the TC trimming is to bring $V_{B2}$ CTAT slope closer to its value in the SLVT TT corner, with slightly larger (resp. lower) values in the fast (resp. slow) nMOS LVT corners [Fig.~\ref{fig:21_sim_iref_vs_T_skewed_process_22nm}(c)].
Next, considering MC simulations, without trimming, mismatch leads to a $(\sigma/\mu)$ of 2.55~$\%$ which is 3$\times$ larger than in 0.11~$\mu$m due to the smaller dimensions of the pMOS current mirror. Eq. (\ref{eq:var_iref}) indeed predicts a variability of 3.78~$\%$ from which 3.65~$\%$ come from the mirror mismatch. Process variations lead to a $(\sigma/\mu)$ of 10.70~$\%$ similar to 0.11~$\mu$m [Fig.~\ref{fig:22_sim_iref_vs_T_mc_22nm}(a)]. $I_{REF}$ TC suffers most from mismatch, with a \mbox{49-ppm/$^\circ$C} median and a \mbox{172-ppm/$^\circ$C} 99th percentile, than from process variations, with a \mbox{28-ppm/$^\circ$C} median and a \mbox{62-ppm/$^\circ$C} 99th percentile [Fig.~\ref{fig:22_sim_iref_vs_T_mc_22nm}(c)]. After TC trimming and no $I_{REF}$ trimming, $I_{REF}$ $(\sigma/\mu)$ slightly increases while process variations are reduced, in the same fashion as in 0.11~$\mu$m [Fig.~\ref{fig:22_sim_iref_vs_T_mc_22nm}(b)]. The TC now has a median around 22~ppm/$^\circ$C in all simulation types, with 99th percentiles at 33, 28, and 42~ppm/$^\circ$C, for mismatch, process, and combined effects.
Lastly, Fig.~\ref{fig:18_sim_startup}(b) highlights a nominal \mbox{99-$\%$} startup time of 0.94~ms, which spans from 0.04 and 371.2~ms in extreme corners.

\section{Measurement Results}
\label{sec:5_measurement_results}
\subsection{Measurement Testbench}
\label{subsec:5A_measurement_testbench}
Out of the four designs in Section~\ref{subsec:3C_final_implementation}, the \mbox{22-nm} design with TC and $I_{REF}$ trimming has been fabricated as part of the CERBERUS microcontroller unit (MCU) [Fig.~\ref{fig:23_microphotograph_layout}(b)]. Eleven dies have been measured, using the setup shown in Fig.~\ref{fig:24_meas_setup}. It consists of a host PC controlling an Espec \mbox{SH-261} climatic chamber for the temperature sweep, and two Keithley K2540 source measure units (SMUs) for the supply voltage sweep. On the one hand, SMU 1 measures $I_{OUT} = 7I_{REF}$ with $V_{SD}$ = 0.9~V, and the current mirror producing $I_{OUT}$ only slightly increases $I_{REF}$ $(\sigma/\mu)$ by 0.36~$\%$. Besides, the leakage of the electrostatic discharge (ESD) protection diodes in the analog pad is at most 8.12~pA at 85$^\circ$C, i.e., only 0.077~$\%$ of $I_{OUT}$. On the other hand, SMU 2 measures the sum of $I_{OUT}$ and the supply current $I_{VDD}$. Besides, the calibration bits are managed by an on-chip \mbox{Cortex-M4}, interacting through GPIO with a \mbox{Nucleo-64} platform piloted by the host PC. To measure the startup time, SMU 1 is replaced by an \mbox{82-M$\Omega$} resistance $R_S$, in series with the \mbox{1-M$\Omega$} input resistance of the probe.

\subsection{Measurement of the \mbox{22-nm} \mbox{FD-SOI} Design}
\label{subsec:4B_measurement_of_the_22-nm_FD-SOI_design}
\begin{figure}[!t]
	\centering
	\includegraphics[width=.5\textwidth]{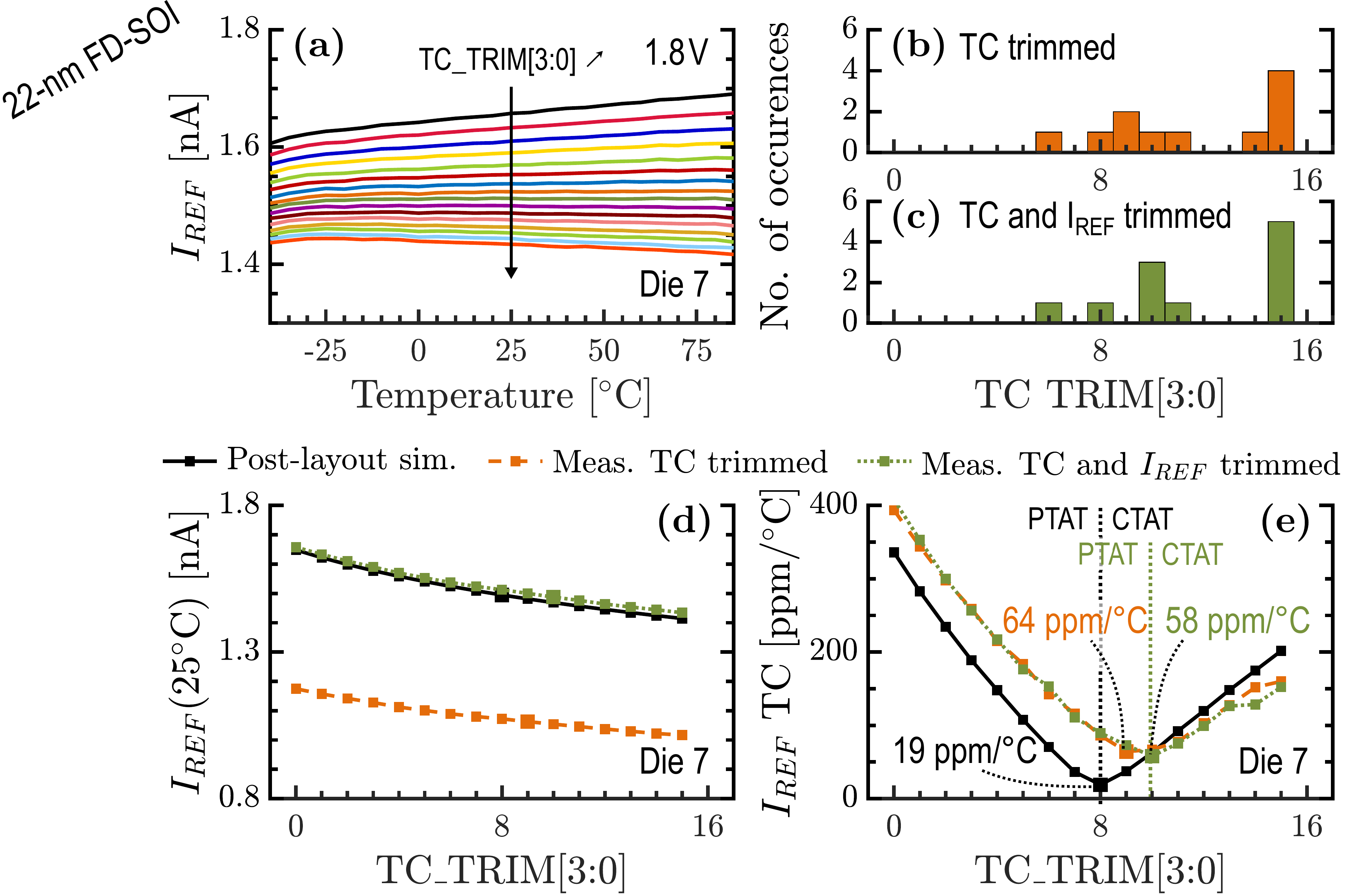}
	\caption{In \mbox{22-nm} \mbox{FD-SOI}, (a) measured temperature dependence of $I_{REF}$ at 1.8~V for all trimming codes, with $I_{REF}$ already trimmed. Histograms of the trimming code giving the minimum TC (b) before and (c) after $I_{REF}$ trimming. Impact of the TC trimming (d) on $I_{REF}$ at 25$^\circ$C and (e) on $I_{REF}$ TC. (b)(c) are for the 11 dies while (a)(d)(e) are for die 7.}
	\label{fig:25_meas_TC_calib_22nm}
\end{figure}
\begin{figure}[!t]
	\centering
	\includegraphics[width=.5\textwidth]{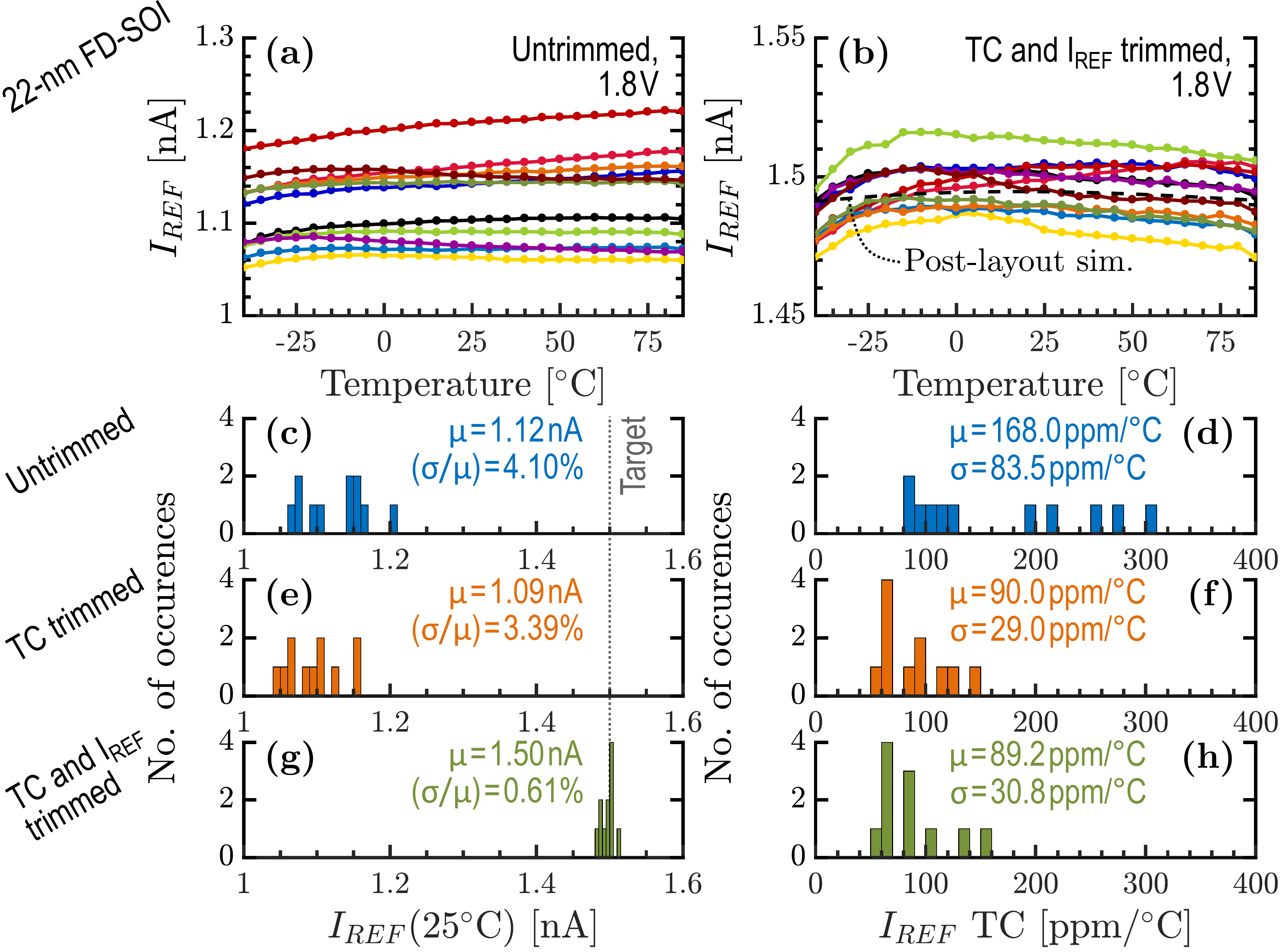}
	\caption{In \mbox{22-nm} \mbox{FD-SOI} and at 1.8~V, measured temperature dependence of $I_{REF}$ (a) without trimming, and (b) with TC and $I_{REF}$ trimming. Measured histograms of $I_{REF}$ at 25$^\circ$C [(c), (e) and (g)] and of $I_{REF}$ TC from -40 to 85$^\circ$C [(d), (f) and (h)], for the reference without trimming, with TC trimming, and with TC and $I_{REF}$ trimming.}
	\label{fig:26_meas_iref_vs_T_22nm}
\end{figure}
\begin{figure}[!t]
	\centering
	\includegraphics[width=.45\textwidth]{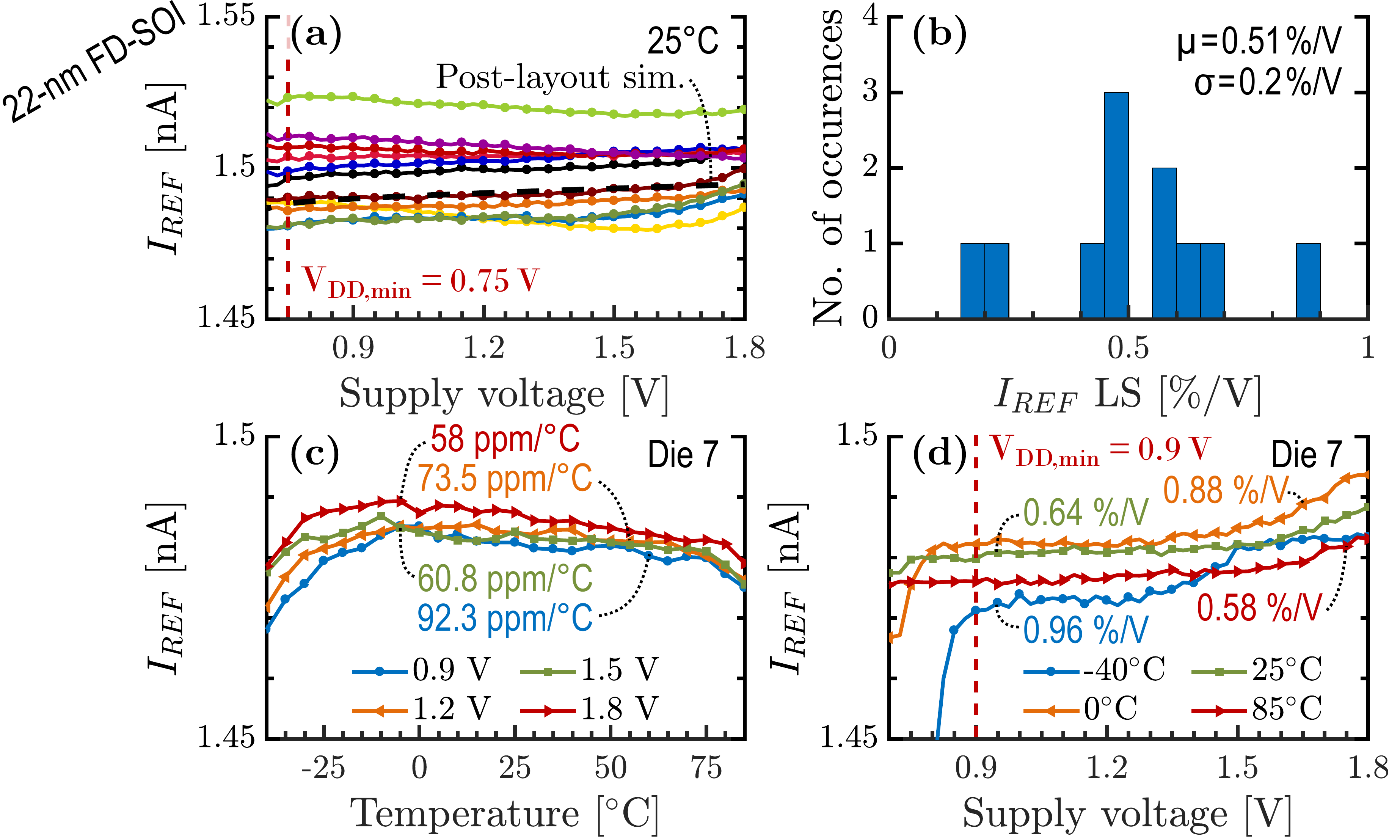}
	\caption{In \mbox{22-nm} \mbox{FD-SOI} and with TC and $I_{REF}$ trimmed, (a) measured supply voltage dependence at 25$^\circ$C and (b) histogram of LS from 0.75 to 1.8~V. For die 7, measured (c) temperature dependence at different supply voltages and (d) supply voltage dependence at different temperatures.}
	\label{fig:27_meas_iref_vs_vdd_22nm}
\end{figure}

Fig.~\ref{fig:25_meas_TC_calib_22nm}(a) illustrates that the temperature dependence of $I_{REF}$, with $I_{REF}$ already trimmed, changes from PTAT to CTAT for different TC trimming codes. Figs.~\ref{fig:25_meas_TC_calib_22nm}(b) and (c) show the histogram of the optimal code before and after $I_{REF}$ trimming, revealing only marginal changes and thereby confirming that trimming $I_{REF}$ does not impact its TC. This intuitively makes sense, as the TC of $I_{REF}$ depends on the TCR, and on the TCs of $V_{B2}$ and $\gamma_b^{*}$, but is not impacted by the change of resistance employed to trim $I_{REF}$. However, trimming the TC slightly modifies $I_{REF}$ [Fig.~\ref{fig:25_meas_TC_calib_22nm}(d)] due to a change of $V_{B2}$. The TC should thus be trimmed first, and $I_{REF}$ consequently adjusted to reach the target value. For the die showing the best TC, a minimum TC of 58~ppm/$^\circ$C is reached for a code of 0xA, compared to 19~ppm/$^\circ$C for 0x8 in the TT post-layout simulation [Fig.~\ref{fig:25_meas_TC_calib_22nm}(e)]. This gap originates from the behavior of $I_{REF}$ below -20$^\circ$C, as shown in Fig.~\ref{fig:26_meas_iref_vs_T_22nm}(b). Next, the temperature dependence of $I_{REF}$ is studied in Fig.~\ref{fig:26_meas_iref_vs_T_22nm} at the three stages of the trimming process, with Figs.~\ref{fig:26_meas_iref_vs_T_22nm}(a) and (b) representing the $I_{REF}$ vs. $T$ curves for the 11 dies without and with trimming, respectively. Before any trimming, $I_{REF}$ is 1.12~nA on average, with a $(\sigma/\mu)$ of 4.1~$\%$ and a TC of 168~ppm/$^\circ$C [Figs.~\ref{fig:26_meas_iref_vs_T_22nm}(c) and (d)], suggesting that the dies might originate from a slow SLVT and fast LVT nMOS process corner [Fig.~\ref{fig:22_sim_iref_vs_T_mc_22nm}(a)]. After trimming the TC, $(\sigma/\mu)$ diminishes to 3.39~$\%$ while the TC decreases to 90~ppm/$^\circ$C, with a standard deviation cut from 83.5 to 29~ppm/$^\circ$C [Figs.~\ref{fig:26_meas_iref_vs_T_22nm}(e) and (f)]. Finally, after complete trimming, $I_{REF}$ is close to the \mbox{1.5-nA} design point, with a $(\sigma/\mu)$ reduced to 0.61~$\%$ and an \mbox{89.2-ppm/$^\circ$C} average TC [Figs.~\ref{fig:26_meas_iref_vs_T_22nm}(g) and (h)]. These results correspond to a trimming based on the complete temperature profile from -40 to 85$^\circ$C, but a two-point trimming at -35 and 65$^\circ$C gives similar results as it only marginally increases the mean TC from 89.2 to 89.4~ppm/$^\circ$C. In both cases, we see a clear improvement compared to the \mbox{168-ppm/$^\circ$C} TC of the untrimmed reference, which could be even larger in other process corners, as suggested in Fig.~\ref{fig:21_sim_iref_vs_T_skewed_process_22nm}(b). The TC differs from post-layout simulations due to the decrease of $I_{REF}$ below \mbox{-20$^\circ$C}, which is most likely due to the nMOS switches in the TC trimming circuit limiting the current flowing through $M_{5Vi}$. Moreover, Fig.~\ref{fig:27_meas_iref_vs_vdd_22nm}(a) shows the supply voltage dependence of $I_{REF}$ for the 11 trimmed dies, with $V_{DD,\textrm{min}}$ = 0.75~V and a \mbox{0.51-$\%$/V} average LS [Fig.~\ref{fig:27_meas_iref_vs_vdd_22nm}(b)] in line with the \mbox{0.41-$\%$/V} simulated value.
\begin{figure}[!t]
	\centering
	\includegraphics[width=.45\textwidth]{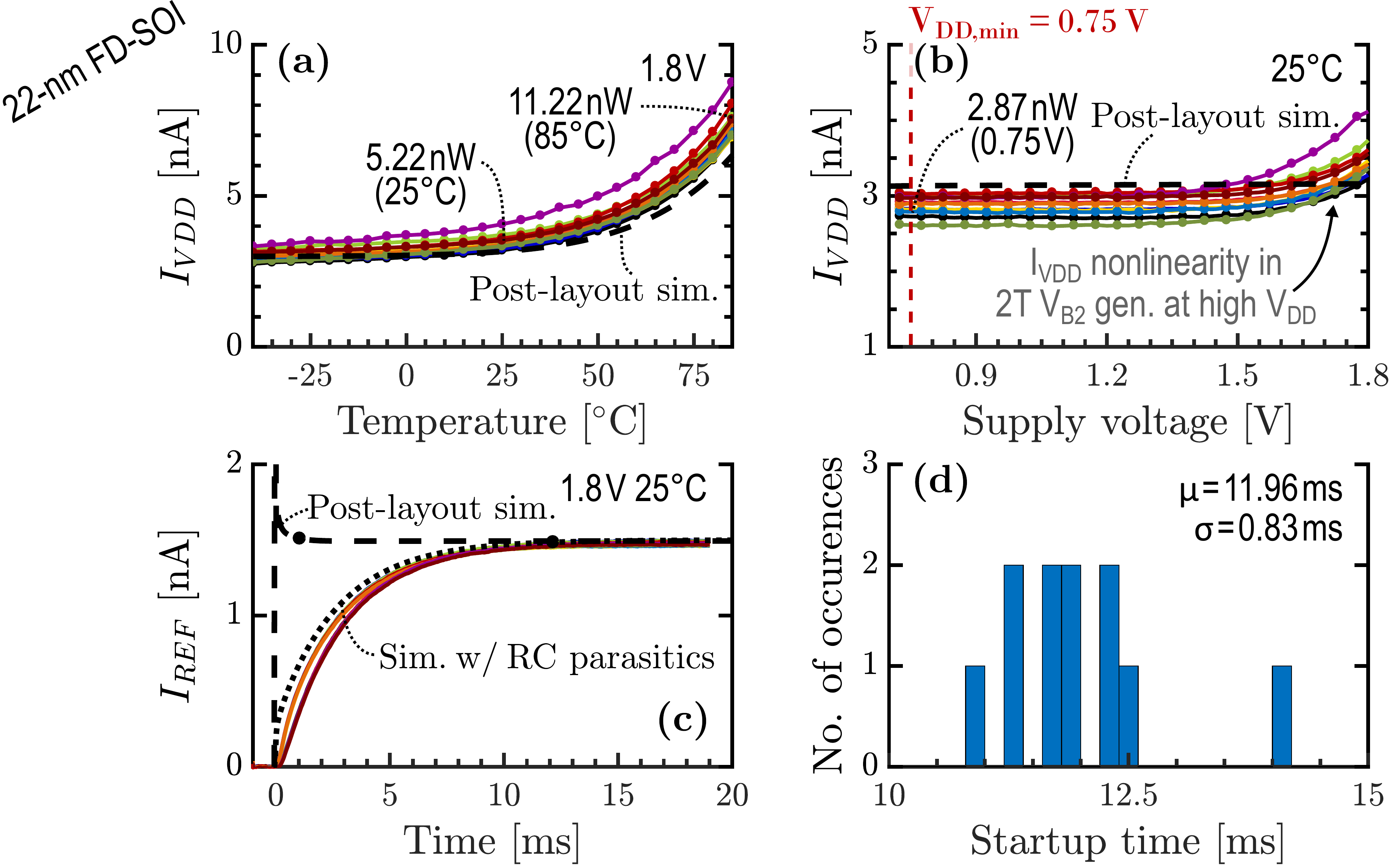}
	\caption{In \mbox{22-nm} \mbox{FD-SOI}, measured dependence of the supply current (a) to temperature at 1.8~V, and (b) to supply voltage at 25$^\circ$C. (c) Measured startup waveforms and (d) histogram of the \mbox{99-$\%$} startup time at 1.8~V 25$^\circ$C.}
	\label{fig:28_meas_ivdd_startup_22nm}
\end{figure}
\setlength{\tabcolsep}{2pt}
\begin{table*}[!t]
\centering
\caption{Comparison table of temperature-independent nA-range current references.}
\vspace{-0.15cm}
\label{table:soa_nanoamp_range}
\scalebox{.675}{%
\begin{threeparttable}
\begin{scriptsize}
\begin{tabular}[t]{l|ccccccccc|cccccccccc@{\hskip 5pt}|cc}
\toprule
& Far & Cordova & Santamaria & Agarwal & Aminzadeh & Mahmoudi & Bruni & Huang & Yang & De Vita & Dong & Ji & Wang & Wang & Huang & Lee & Chang & Shetty & Lefebvre & \multicolumn{2}{c}{\textbf{Lefebvre}}\\
& \cite{Far_2015} & \cite{Cordova_2017} & \cite{Santamaria_2019} & \cite{Agarwal_2022} & \cite{Aminzadeh_2022} & \cite{Mahmoudi_2022} & \cite{Bruni_2023} & \cite{Huang_2023} & \cite{Yang_2023} & \cite{DeVita_2007} & \cite{Dong_2017} & \cite{Ji_2017} & \cite{Wang_2019_VLSI} & \cite{Wang_2019_TCAS} & \cite{Huang_2020} & \cite{Lee_2020} & \cite{Chang_2022} & \cite{Shetty_2022} & \cite{Lefebvre_2023} & \multicolumn{2}{c}{\textbf{This work}}\\
\midrule
Publication & ROPEC & ISCAS & ISCAS & TCAS-II & AEU &  & CAE & AEU & AEU & ISCAS & ESSCIRC & ISSCC & VLSI-DAT & TCAS-I & TCAS-II & JSSC & JJAP & TCAS-I & JSSC & \multicolumn{2}{c}{JSSC} \\
Year & 2015 & 2017 & 2019 & 2022 & 2022 & 2022 & 2023 & 2023 & 2023 & 2007 & 2017 & 2017 & 2019 & 2019 & 2020 & 2020 & 2022 & 2022 & 2023 & \multicolumn{2}{c}{2024}\\
\cmidrule(lr){21-22}
Type of work & \multicolumn{9}{c}{\textbf{Simulations}} & \multicolumn{10}{c}{\textbf{Silicon measurements}} & Sim. & Meas.\\
\cmidrule(lr){2-10} \cmidrule(lr){11-20}
Samples & N/A & N/A & N/A & N/A & N/A & N/A & N/A & N/A & N/A & 20 & 16 & 10 & 10 & 16 & 10 & 10 & 3 & 10 & 20 & N/A & 11\\
\midrule
Technology & 0.18$\mu$m & 0.18$\mu$m & 0.18$\mu$m & 0.18$\mu$m & 0.18$\mu$m & 0.13$\mu$m & 0.18$\mu$m & 0.18$\mu$m & 0.18$\mu$m & 0.35$\mu$m & 0.18$\mu$m & 0.18$\mu$m & 0.18$\mu$m & 0.18$\mu$m & 0.18$\mu$m & 0.18$\mu$m & 90nm & 0.13$\mu$m & 22nm & 0.11$\mu$m & 22nm\\
$I_{REF}$ [nA] & 14 & 10.9 & 2.7 & 5.6 & 6.7 & 6.6 & 6.3 & 8.9 & 1.96 & 9.1 & 35 & 6.7 & 6.5 & 9.8 & 11.6 & 1 & 1.3 & 1.9 & 1.25/0.9\tnote{$\star$} & 5.03 & 1.12/1.50\tnote{$\dagger$}\\
Power [nW] & 150 & 30.5 & 26 & 9.5 & 51 & 3.7 & 3.3 & 0.05 & 9.2 & 54.8 & 1.02 & 9.3 & 15.8 & 28 & 48.6 & 4.5/14 & 8.6 & 10.2 & 7.8/5.8\tnote{$\star$} & 13.88 & 2.87\\
 & $@$1V & $@$0.9V & $@$2V & $@$0.55V & $@$1V & $@$0.4V & $@$0.6V & $@$0.8V & $@$0.55V & $@$1.5V & $@$1.5V & $@$N/A & $@$0.85V & $@$0.7V & $@$0.8V & $@$1.5V & $@$0.75V & $@$0.85V & $@$0.9V & $@$0.85V & $@$0.75V\\
Area [mm$^2$] & \textcolor{ECS-Blue}{\textbf{0.0102}} & \textcolor{ECS-Blue}{\textbf{0.01}} & \textcolor{ECS-Blue}{\textbf{0.0093}} & 0.032 & \textcolor{ECS-Red}{\textbf{0.46}} & \textcolor{ECS-Blue}{\textbf{0.0021}} & \textcolor{ECS-Blue}{\textbf{0.0018}} & \textcolor{ECS-Blue}{\textbf{0.008}} & \textcolor{ECS-Blue}{\textbf{0.0033}} & 0.035 & 0.0169 & \textcolor{ECS-Red}{\textbf{0.055}} & \textcolor{ECS-Red}{\textbf{0.062}} & \textcolor{ECS-Red}{\textbf{0.055}} & \textcolor{ECS-Red}{\textbf{0.054}} & \textcolor{ECS-Red}{\textbf{0.332}} & 0.0175 & 0.0163 & 0.0132 & \textcolor{ECS-Blue}{\textbf{0.00954}} & \textcolor{ECS-Blue}{\textbf{0.00214}}\\
\midrule
Supply range [V] & 1 -- 3.3 & 0.9 -- 1.8 & 2 -- 3.63\tnote{$\diamond$} & 0.55 -- 1.9 & 1.1 -- 1.8 & 0.4 -- 1.6 & 0.6 -- 1.8 & 0.8 -- 1.8 & 0.55 -- 1.8 & 1.5 -- 4 & 1.5 -- 2.5 & 1.3 -- 1.8 & 0.85 -- 2 & 0.7 -- 1.2 & 0.8 -- 2 & 1.5 -- 2 & 0.75 -- 1.55 & 0.85 -- 2 & 0.9 -- 1.8 & 0.85 -- 1.2 & 0.75 -- 1.8\\
LS [$\%$/V] & \textcolor{ECS-Blue}{\textbf{0.1}} & \textcolor{ECS-Blue}{\textbf{0.54}} & \textcolor{ECS-Red}{\textbf{8.9}}\tnote{$\diamond$} & \textcolor{ECS-Blue}{\textbf{0.022}} & \textcolor{ECS-Blue}{\textbf{0.03}} & 2.7 & \textcolor{ECS-Red}{\textbf{12.1}} & 1.39 & \textcolor{ECS-Blue}{\textbf{0.2}} & \textcolor{ECS-Blue}{\textbf{0.57}} & 3 & 1.16 & 4.15 & \textcolor{ECS-Blue}{\textbf{0.6}} & 1.08 & 1.4 & \textcolor{ECS-Blue}{\textbf{0.15}} & 4 & \textcolor{ECS-Blue}{\textbf{0.26}}/\textcolor{ECS-Blue}{\textbf{0.39}}\tnote{$\star$} & 2.84 & \textcolor{ECS-Blue}{\textbf{0.51}}\\
\midrule
Temp. range [$^\circ$C] & 0 -- 70 & -20 -- 120 & -40 -- 125 & -30 -- 70 & -40 -- 120 & -40 -- 120 & -40 -- 120 & 0 -- 125 & 0 -- 100 & 0 -- 80 & -40 -- 120 & 0 -- 110 & -10 -- 100 & -40 -- 125 & -40 -- 120 & -20 -- 80 & 0 -- 120 & -40 -- 120 & -40 -- 85 & -40 -- 85 & -40 -- 85\\
TC [ppm/$^\circ$C] & \textcolor{ECS-Blue}{\textbf{20}} & \textcolor{ECS-Blue}{\textbf{108}} & 309 & 256 & \textcolor{ECS-Blue}{\textbf{40.3}} & 308 & 219 & \textcolor{ECS-Blue}{\textbf{139}} & \textcolor{ECS-Blue}{\textbf{96.8}} & \textcolor{ECS-Blue}{\textbf{44}} & 282 & \textcolor{ECS-Red}{\textbf{680}}/283\tnote{$\dagger$} & \textcolor{ECS-Blue}{\textbf{157}} & \textcolor{ECS-Blue}{\textbf{150}} & \textcolor{ECS-Blue}{\textbf{169}} & 289/265\tnote{$\triangleright$} & \textcolor{ECS-Blue}{\textbf{53}}/394\tnote{$\star$} & 530/\textcolor{ECS-Red}{\textbf{822}}\tnote{$\star$} & \textcolor{ECS-Blue}{\textbf{203}}/565\tnote{$\star$} & \textcolor{ECS-Blue}{\textbf{65}} & 168/\textcolor{ECS-Blue}{\textbf{89}}\tnote{$\dagger$}\\
TC type & Typ. & $\mu$ & $\mu$ & $\mu$ & $\mu$ & $\mu$ & Typ. & $\mu$ & $\mu$ & $\mu$ & $\mu$ & $\mu$ & $\mu$ & $\mu$ & $\mu$ & $\mu$ & $\mu$ & $\mu$ & $\mu$ & $\tilde{x}$ & $\mu$\\
\midrule
$I_{REF}$ var. & \multirow{2}{*}{N/A} & 15.8 / & \multirow{2}{*}{N/A} & +55.4 / & \multirow{2}{*}{N/A} & \multirow{2}{*}{N/A} & +129.1 / & +2.8 / & \multirow{2}{*}{8.7} & \multirow{4}{*}{2.16} & \multirow{2}{*}{$\pm$4.7} & \multirow{2}{*}{N/A} & \multirow{2}{*}{N/A} & +11.7 / & +17.6 / & \multirow{2}{*}{N/A} & \multirow{4}{*}{21.1} & \multirow{2}{*}{N/A} & +9.9 / & +3.64 / & +0.21 /\\
(process) [$\%$] & & 11.6\tnote{$\dagger$} & & -28.5\tnote{$\diamond$} &  &  & -61.8\tnote{$\diamond$} & -13.9\tnote{$\diamond$} &  &  &  &  &  & -8.7\tnote{$\diamond$} & -10.3\tnote{$\diamond$} &  &  &  & -9.5 & -1.41\tnote{$\triangleleft$} & -0.54\tnote{$\triangleleft$}\\
$I_{REF}$ var. & \multirow{2}{*}{5.8} & \multirow{2}{*}{N/A} & \multirow{2}{*}{\textcolor{ECS-Red}{\textbf{20.3}}} & \multirow{2}{*}{\textcolor{ECS-Red}{\textbf{10.4}}} & \multirow{2}{*}{\textcolor{ECS-Blue}{\textbf{0.7}}} & \multirow{2}{*}{\textcolor{ECS-Red}{\textbf{6.1}}} & \multirow{2}{*}{N/A} & \multirow{2}{*}{2.6} & \multirow{2}{*}{1.7} &  & \multirow{2}{*}{1.6} & \multirow{2}{*}{4.07/1.19\tnote{$\star$}} & \multirow{2}{*}{3.33} & \multirow{2}{*}{1.6} & \multirow{2}{*}{4.3} & \multirow{2}{*}{1.26/\textcolor{ECS-Blue}{\textbf{0.25}}\tnote{$\dagger$}} & & \multirow{2}{*}{\textcolor{ECS-Red}{\textbf{15.6}}} & \multirow{2}{*}{\textcolor{ECS-Red}{\textbf{6.39}}/\textcolor{ECS-Red}{\textbf{9.20}}\tnote{$\star$}} & \multirow{2}{*}{1.32} & \multirow{2}{*}{4.10/\textcolor{ECS-Blue}{\textbf{0.61}}\tnote{$\dagger$}}\\
(mismatch) [$\%$] &  &  &  &  &  &  &  &  &  &  &  &  &  &  &  &  &  &  &  &  &\\
\midrule
Trimming & No & TC (6b) & No & No & No & TC (12b) & No & TC (6b) & No & No & No & $I_{REF}$ & No & TC (5b) & TC (4b) & $I_{REF}$ & $I_{REF}$ (4b) & $I_{REF}$ (4b) & No & TC (5b), & TC (4b), \\
& & & & & & & & & & & & & & & & & & & & $I_{REF}$ (5b) & $I_{REF}$ (8b)\\
Spec. components & No & \textcolor{ECS-Red}{\textbf{ZVT}} & No & Res. & Res., \textcolor{ECS-Red}{\textbf{BJT}} & LVT, HVT & \textcolor{ECS-Red}{\textbf{ZVT}} & \textcolor{ECS-Red}{\textbf{ZVT}}, HVT & No & No & No & Res., \textcolor{ECS-Red}{\textbf{BJT}} & No & Res., I/O & Res. & HVT & \textcolor{ECS-Red}{\textbf{ZVT}}, HVT & Res., HVT & HVT & Res., HVT, I/O & Res., LVT\\
\midrule
FoM$_1$ & \multirow{2}{*}{\textcolor{ECS-Blue}{\textbf{0.0029}}} & \multirow{2}{*}{\textcolor{ECS-Blue}{\textbf{0.0077}}} & \multirow{2}{*}{0.0174} & \multirow{2}{*}{0.0819} & \multirow{2}{*}{0.1159} & \multirow{2}{*}{\textcolor{ECS-Blue}{\textbf{0.0040}}} & \multirow{2}{*}{\textcolor{ECS-Blue}{\textbf{0.0025}}} & \multirow{2}{*}{\textcolor{ECS-Blue}{\textbf{0.0089}}} & \multirow{2}{*}{\textcolor{ECS-Blue}{\textbf{0.0032}}} & \multirow{2}{*}{0.0193} & \multirow{2}{*}{0.0298} & \multirow{2}{*}{0.1415} & \multirow{2}{*}{0.0887} & \multirow{2}{*}{0.0499} & \multirow{2}{*}{0.0570} & \multirow{2}{*}{0.9595} & \multirow{2}{*}{0.0575} & \multirow{2}{*}{0.0835} & \multirow{2}{*}{0.0597} & \multirow{2}{*}{\textcolor{ECS-Blue}{\textbf{0.0050}}} & \multirow{2}{*}{\textcolor{ECS-Blue}{\textbf{0.0015}}}\\
$[$ppm/$^\circ$C$^2 \times$mm$^2]$ &  &  &  &  &  &  &  &  &  &  &  &  &  &  &  &  &  &  &  &  &\\
FoM$_2$\tnote{$\ddagger$} & \multirow{2}{*}{3.061} & \multirow{2}{*}{2.398} & \multirow{2}{*}{9.017} & \multirow{2}{*}{7.896} & \multirow{2}{*}{\textcolor{ECS-Blue}{\textbf{1.917}}} & \multirow{2}{*}{2.698} & \multirow{2}{*}{\textcolor{ECS-Blue}{\textbf{1.195}}} & \multirow{2}{*}{\textcolor{ECS-Blue}{\textbf{0.0078}}} & \multirow{2}{*}{8.261} & \multirow{2}{*}{2.208} & \multirow{2}{*}{\textcolor{ECS-Blue}{\textbf{0.034}}} & \multirow{2}{*}{3.571} & \multirow{2}{*}{4.082} & \multirow{2}{*}{3.711} & \multirow{2}{*}{5.532} & \multirow{2}{*}{\textcolor{ECS-Red}{\textbf{24.733}}} & \multirow{2}{*}{\textcolor{ECS-Red}{\textbf{28.961}}} & \multirow{2}{*}{\textcolor{ECS-Red}{\textbf{32.447}}} & \multirow{2}{*}{\textcolor{ECS-Red}{\textbf{32.365}}} & \multirow{2}{*}{\textcolor{ECS-Blue}{\textbf{1.688}}} & \multirow{2}{*}{\textcolor{ECS-Blue}{\textbf{1.816}}}\\
$[$ppm/$^\circ$C$^2]$ &  &  &  &  &  &  &  &  &  &  &  &  &  &  &  &  &  &  &  &  &\\
\bottomrule
\end{tabular}
\end{scriptsize}
\vspace{0.1cm}
\begin{footnotesize}
\begin{tablenotes}
	\item[$\star$] Simulated and measured values.
	\item[$\dagger$] Before and after trimming.
	\item[$\diamond$] Estimated from figures.
	\item[$\triangleright$] For 25 and 2.5 minutes between two calibrations.
	\item[$\triangleleft$] After trimming.
	\item[$\ddagger$] $\textrm{FoM}_2 = \dfrac{\textrm{TC}}{(T_{\textrm{max}}-T_{\textrm{min}})} \times \dfrac{I_{VDD}}{I_{REF}}$, as used in \cite{Ballo_2022}.
\end{tablenotes}
\end{footnotesize}
\end{threeparttable}%
}
\end{table*}
Then, $I_{REF}$ temperature dependence is characterized at various supply voltages, revealing a consistent behavior, albeit the TC is slightly degraded from 58 to 92.3~ppm/$^\circ$C as $V_{DD}$ is reduced from 1.8 to 0.9~V [Fig.~\ref{fig:27_meas_iref_vs_vdd_22nm}(c)]. Additionally, Fig.~\ref{fig:27_meas_iref_vs_vdd_22nm}(d) characterizes the supply voltage dependence of $I_{REF}$ at different temperatures, and indicates that a $V_{DD,\textrm{min}}$ of 0.9~V would be required to operate down to \mbox{-40$^\circ$C}, but also that the LS deteriorates from 0.58 to 0.96~$\%$/V as temperature decreases from 85 to \mbox{-40$^\circ$C}. These curves highlight an increase of $I_{REF}$ above 1.5~V, which could be linked to drain-induced barrier lowering (DIBL). Moreover, the current consumption $I_{VDD}$ in Fig.~\ref{fig:28_meas_ivdd_startup_22nm}(a) is 2$I_{REF}$ = 3~nA at low temperature as it is dominated by the PCR, consisting of two branches drawing the same current, but increases exponentially with temperature when the 2T body bias generator and its replica start to prevail, leading to a \mbox{2.1-$\times$} increase from 25 to 85$^\circ$C. The dependence of $I_{VDD}$ to the supply voltage in Fig.~\ref{fig:28_meas_ivdd_startup_22nm}(b) presents a sharper increase at high $V_{DD}$ compared to $I_{REF}$, likely originating from the 2T body bias generator. An average power consumption of 2.87~nW is reached at 0.75~V and 25$^\circ$C. At last, regarding the \mbox{99-$\%$} startup time, an average value of 11.96~ms 11.4$\times$ larger than the simulated value is obtained [Fig.~\ref{fig:28_meas_ivdd_startup_22nm}(d)]. In simulation, when considering a \mbox{33-pF} capacitance at the drain of the pMOS transistor generating $I_{OUT}$, in parallel with the \mbox{82-M$\Omega$} resistance of $R_S$, a startup time of 12.15~ms is reached [Fig.~\ref{fig:28_meas_ivdd_startup_22nm}(c)], confirming that PCB parasitics are the source of the observed discrepancy.

\section{Comparison to the State of the Art}
\label{sec:6_comparison_to_the_state_of_the_art}
\begin{figure}[!t]
	\centering
	\includegraphics[width=.45\textwidth]{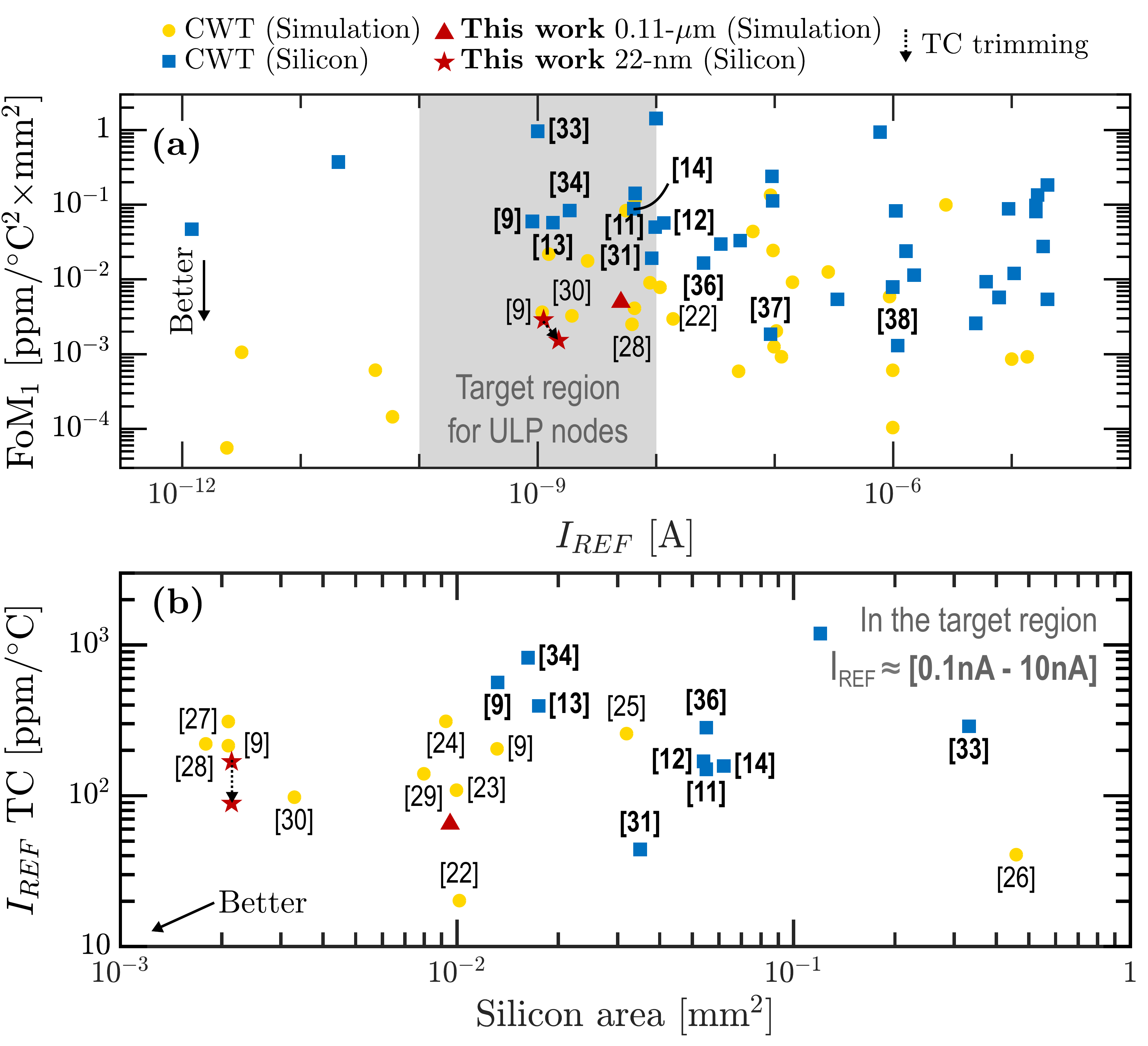}
	\caption{(a) Figure of merit combining temperature dependence and silicon area as a function of $I_{REF}$, and (b) trade-off between TC and area in the target region, based on the state of the art of current references.}
	\label{fig:29_comparison_to_soa}
\end{figure}

In this section, we compare our work to the state of the art of simulated and fabricated nA-range CWT current references, detailed in Table~\ref{table:soa_nanoamp_range} and illustrated in Fig.~\ref{fig:29_comparison_to_soa}. To better understand the performance of existing current references, we introduce a figure of merit (FoM) combining the temperature dependence of the current with the area occupied by the reference, i.e.,
\begin{equation}
	\textrm{FoM}_1 = \frac{\textrm{TC}}{(T_{\textrm{max}}-T_{\textrm{min}})} \times \textrm{Area} \quad \textrm{[ppm/$^\circ$C$^2 \times$ mm$^2$]}\textrm{.}\label{eq:FoM}
\end{equation}
It shares strong similarities with the one used in \cite{CamposDeOliveira_2017, Shetty_2022}, but puts less emphasis on the temperature range and does not include power consumption. In addition, it does not normalize performance, as the tradeoff between TC and silicon area is complex to model and depends on many factors. In our comparison, we focus on fabricated current references, and hence limit our discussion to the proposed \mbox{22-nm} design. Besides, we do not normalize the silicon area to the technology node as a fair normalization is not straightforward for analog circuits. First, Fig.~\ref{fig:29_comparison_to_soa}(a) indicates that trimming the TC of $I_{REF}$ improves the FoM of our \mbox{22-nm} design by 1.9$\times$ thanks to a TC reduction from 168 to 89~ppm/$^\circ$C. Compared to the closest fabricated competitor \cite{DeVita_2007}, our design offers a 12.9$\times$ FoM reduction explained by a 16.4$\times$ area reduction and an increased temperature range, despite a 2$\times$ larger TC. Moreover, \cite{Wang_2019_VLSI, Wang_2019_TCAS, Huang_2020} achieve an acceptable TC around 150~ppm/$^\circ$C within a silicon area above 0.05~mm$^2$, due to the use of large subthreshold transistors \cite{Wang_2019_VLSI} or resistors \cite{Wang_2019_TCAS, Huang_2020}. \cite{Chang_2022, Shetty_2022, Lefebvre_2023} exhibit areas between 0.01 and 0.02~mm$^2$, dominated by gate-leakage transistors \cite{Chang_2022}, resistors \cite{Shetty_2022}, or a subthreshold $\beta$-multiplier \cite{Lefebvre_2023}. Nevertheless, their area efficiency is counterbalanced by their relatively large TC above 400~ppm/$^\circ$C. \cite{Kayahan_2013} uses a pMOS with a $V_G$ tracking $V_T$, and a source degeneration resistor to achieve temperature compensation. It occupies a 0.0053-mm$^2$ area but consumes 28.5~$\mu$W, thereby limiting its interest. \cite{Lee_2020} employs a current DAC periodically trimmed by a high-precision duty-cycled current reference. It achieves a 265~ppm/$^\circ$C TC, but at the expense of a significant 0.332-mm$^2$ silicon footprint. Finally, \cite{Chouhan_2016} and \cite{Lefebvre_2022} are interesting architectures in terms of TC and area which lie outside of the target region. \cite{Chouhan_2016} relies on a $\beta$-multiplier with a triode transistor as $V$-to-$I$ converter, and produces a \mbox{92.2-nA} current with a \mbox{177-ppm/$^\circ$C} TC and \mbox{0.0013-mm$^2$} silicon area, but suffers from a large 6.1-$\%$ variability due to mismatch. \cite{Lefebvre_2022} biases a resistor with the threshold voltage difference between two transistors of different $V_T$ types, thereby generating a \mbox{1.1-$\mu$A} \mbox{38-ppm/$^\circ$C} current within a \mbox{0.0043-mm$^2$} silicon area. However, the limited area is achieved thanks to the fact that resistors are well suited to the generation of a $\mu$A current.

\section{Conclusion}
\label{sec:7_conclusion}
In this work, we presented a nA-range CWT peaking current reference, biasing a resistor with the $\Delta V_T$ between two subhtreshold transistors, one of them being forward body biased to decrease its $V_T$. The body bias is generated by a 2T voltage reference, with a replica to suppress the leakage of parasitic diodes at high temperature. In addition, the proposed reference does not require a startup circuit, and includes two simple mechanisms to trim $I_{REF}$ and its TC, so as to maintain performance across process corners. It requires high-density polysilicon resistors and transistors of two different $V_T$ types, which are common features of most technology nodes today. Then, we proposed a thorough sizing methodology and validated the reference with post-layout simulations in a bulk and an FD-SOI technology, to demonstrate that the body effect can indeed be employed in both of these technologies. Finally, we fabricated the proposed reference in \mbox{22-nm} \mbox{FD-SOI}, and measured a \mbox{1.5-nA} current across 11 dies with average TC and LS of 89~ppm/$^\circ$C and 0.51~$\%$/V, respectively. It consumes 2.87~nW at 0.75~V and occupies an area of 0.00214~mm$^2$. This results in a 12.9$\times$ FoM improvement compared to the closest fabricated competitor in the nA range. Further work could focus on extending the temperature range to 125$^\circ$C while avoiding a sharp increase of the power consumed by the 2T voltage reference.


%

\appendices
\section{Analytical Expression of $I_{REF}$ Variability}
\label{sec:8_appendix}
\indent To establish an analytical expression linking the variability of $I_{REF}$ to the transistor dimensions, we can rely on Pelgrom's law \cite{Pelgrom_1989}, and, under the assumption that the dominant source of mismatch is the threshold voltage, which is a common assumption in weak/moderate inversion \cite{Vancaillie_2003}, on the $I_{DS}$ mismatch between two transistors taken from \cite{Kinget_2005}. To compute the variability of $I_{REF}$ due to mismatch, we employ the following expression
\begin{equation}
	I_{REF} = \frac{\left(V_{T01}-V_{T02}\right) + \gamma_b^{*} V_{B2} + nU_T\ln\left(1+\frac{\Delta I_{DS}}{I_{DS}}\right)}{R}\textrm{,}
\end{equation}
which is a more accurate version of (\ref{eq:iref_proposed}) taking into account the difference of $V_{T0}$ and the current imbalance between $M_{1-2}$. Considering $R$, $\gamma_b^{*}$, $n$ and $U_T$ as constants, the variance of $I_{REF}$ in the proposed CWT PCR can be computed as
\begin{IEEEeqnarray}{RCL}
	\sigma^2(I_{REF}) & = & \frac{\sigma^2\left(\Delta V_{T0,1-2}\right)}{R^2} + \left(\frac{\gamma_b^{*}}{R}\right)^2 \sigma^2\left(V_{B2}\right)\IEEEnonumber\\
	& & + \left(\frac{nU_T}{R}\right)^2 \sigma^2\left(\ln\left(1+\frac{\Delta I}{I}\right)\right)\textrm{,}\label{eq:var_iref}
\end{IEEEeqnarray}
with
\begin{IEEEeqnarray}{RCL}
	\sigma^2(\Delta V_{T0,1-2}) & = & \frac{2 A_{V_{Tn}}}{W_1 L_1}\textrm{,}\label{eq:var_dvt}\\
	\sigma^2(V_{B2}) & = & A_{V_{Tn}}\left[\frac{1}{W_5 L_5} + \left(\frac{n_5}{n_6}\right)^2\frac{1}{W_6 L_6}\right]\textrm{.}\label{eq:var_vb2}
\end{IEEEeqnarray}
The random variable $X = \left(1+\frac{\Delta I}{I}\right)$ has a normal distribution with an expected value of one and a variance $\sigma^2_{\Delta I/I}$ given by \cite{Kinget_2005}. The third term in (\ref{eq:var_iref}) can thus be obtained as the variance of a log-normal distribution expressed as
\begin{equation}
	\sigma^2\left(\ln\left(X\right)\right) = \left[\exp\left(\sigma^2_{\Delta I/I}\right)-1\right]\exp\left(2 + \sigma^2_{\Delta I/I}\right)\textrm{,}\\
\end{equation}
using the common expression linking $\sigma^2\left(\ln\left(X\right)\right)$ to the parameters of the normal distribution $X$, and simplifies to
\begin{equation}
\sigma^2\left(\ln\left(X\right)\right) \approx \sigma^2_{\Delta I/I} \exp(2)\textrm{}
\end{equation}
as $\sigma^2_{\Delta I/I} \ll 1$.

\section*{Acknowledgments}
The authors would like to thank Pierre G\'{e}rard for the measurement testbench, El\'{e}onore Masarweh for the microphotograph, and ECS group members for their proofreading.

\ifCLASSOPTIONcaptionsoff
  \newpage
\fi



%
\bibliographystyle{IEEEtran}
\bibliography{Lefebvre_JSSC_2023_CERBERUS_current_ref}

%

\vspace{-1cm}
\begin{IEEEbiography}[{\includegraphics[width=1in,height=1.25in,clip,keepaspectratio]{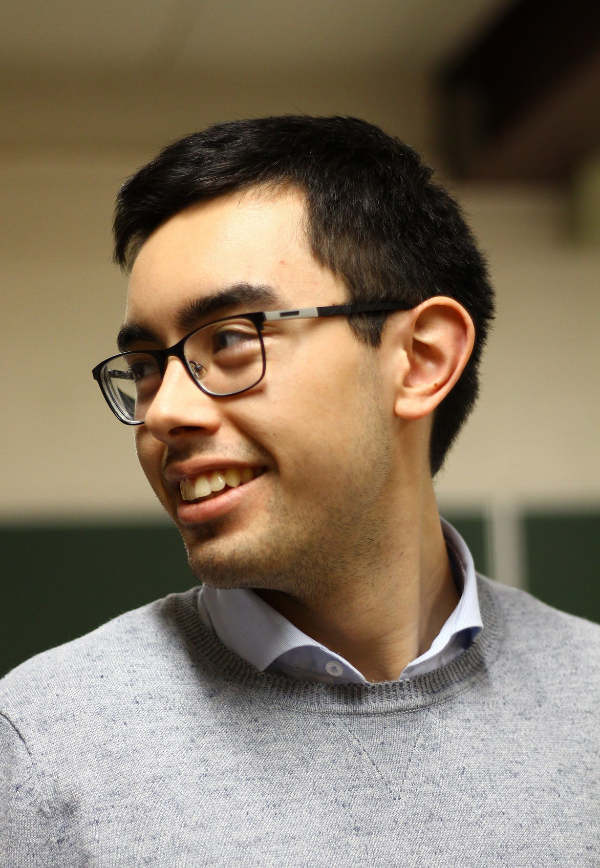}}]{Martin Lefebvre} (Graduate Student Member, IEEE) received the M.Sc. degree (summa cum laude) in Electromechanical Engineering and the Ph.D. degree from the Universit\'e catholique de Louvain (UCLouvain), Louvain-la-Neuve, Belgium, in 2017 and 2024, respectively. His Ph.D. thesis, focusing on area-efficient and temperature-independent current references for the Internet of Things, was supervised by Prof. David Bol. His current research interests include hardware-aware machine learning algorithms, low-power mixed-signal vision chips for embedded image processing, and ultra-low-power current reference architectures. He serves as a reviewer for various IEEE journals and conferences including JSSC, TCAS-I, TCAS-II, TVLSI, and ISCAS.
\end{IEEEbiography}

\vspace{-1cm}
\begin{IEEEbiography}[{\includegraphics[width=1in,height=1.25in,clip,keepaspectratio]{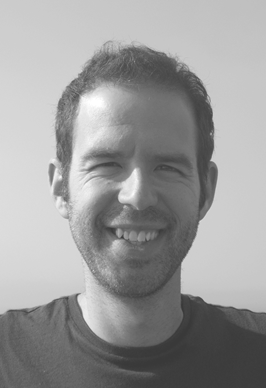}}]{David Bol} (Senior Member, IEEE) is an Associate Professor at UCLouvain. He received the Ph.D. degree in Engineering Science from UCLouvain in 2008 in the field of ultra-low-power digital nanoelectronics. In 2005, he was a visiting Ph.D. student at the CNM, Sevilla, and in 2009, a post-doctoral researcher at intoPIX, Louvain-la-Neuve. In 2010, he was a visiting post-doctoral researcher at the UC Berkeley Lab for Manufacturing and Sustainability, Berkeley. In 2015, he participated to the creation of e-peas semiconductors spin-off company. Prof. Bol leads the Electronic Circuits and Systems (ECS) group focused on ultra-low-power design of integrated circuits for environmental and biomedical IoT applications including computing, power management, sensing and wireless communications. He is actively engaged in a social-ecological transition in the field of information and communication technologies (ICT) research with a post-growth approach. Prof. Bol has authored more than 150 papers and conference contributions and holds three delivered patents. He (co-)received four Best Paper/Poster/Design Awards in IEEE conferences (ICCD 2008, SOI Conf. 2008, FTFC 2014, ISCAS 2020) and supervised the Ph.D. thesis of Charlotte Frenkel who received the 2021 Nokia Bell Scientific Award and the 2021 IBM Innovation Award for her Ph.D. He serves as a reviewer for various IEEE journals and conferences and presented several keynotes in international conferences. 
\end{IEEEbiography}




\end{document}